\pdfoutput=1
\documentclass[11pt,twoside,a4paper,cmspaper,final,collab]{cms-tdr}
\def\svnVersion{aeb5e30}\def\svnDate{2023/07/15}\def\cmsCernNoTag{CERN-EP-2022-142}\def\cmsCernDate{\today}\def\cmsMessage{Published in the Journal of High Energy Physics as \href{http://dx.doi.org/10.1007/JHEP07(2023)091}{\doi{10.1007/JHEP07(2023)091}.}}
\begin{document}\cmsNoteHeader{HIG-19-016}

\newcommand{\Zee}{\ensuremath{\PZ \to \Pe\Pe }\xspace}
\newcommand{\Zmm}{\ensuremath{\PZ \to \PGm\PGm}\xspace}
\newcommand{\Zmmg}{\ensuremath{\PZ \to \PGm\PGm\PGg}\xspace}
\newcommand{\mH}{\ensuremath{m_{\PH}}\xspace}
\newcommand{\mgg}{\ensuremath{m_{\PGg\PGg}}\xspace}
\newcommand{\Hgg}{\ensuremath{\PH\to\PGg\PGg}\xspace}
\newcommand{\Hbb}{\ensuremath{\PH\to\Pb\Pb}\xspace}
\newcommand{\Htautau}{\ensuremath{\PH\to\PGt\PGt}\xspace}
\newcommand{\HWW}{\ensuremath{\PH\to\PW\PW}\xspace}
\newcommand{\HWWSt}{\ensuremath{\PH\to\PW\PW^{\ast}}\xspace}
\newcommand{\HZZfl}{\ensuremath{\PH\to\PZ\PZst\to4\Pell}\xspace}
\newcommand{\ggF}{\ensuremath{\Pg\Pg\mathrm{F}}\xspace}
\newcommand{\VBF}{VBF\xspace}
\newcommand{\VH}{\ensuremath{\PV\PH}\xspace}
\newcommand{\ttH}{\ensuremath{\PQt\PQt\PH}\xspace}
\newcommand{\gamgam}{\ensuremath{\PGg\PGg}\xspace}
\newcommand{\gamplusjet}{\ensuremath{\PGg{+}\text{jet}}\xspace}
\newcommand{\cme}{\ensuremath{\sqrt{s} = 13\TeV\xspace }}
\newcommand{\NNLOPS}{\textsc{nnlops}\xspace}
\newcommand{\fb}{\ensuremath{\unit{fb}}\xspace}
\newcommand{\ptj}{\ensuremath{\pt^{\mathrm{j}}}\xspace}
\newcommand{\etaj}{\ensuremath{\eta^{\mathrm{j}}}\xspace}
\newcommand{\sigMoM}{\ensuremath{\sigma_{m}/m}\xspace}
\newcommand{\sigMD}{\ensuremath{\sigma_{m}^{\mathrm{D}}}\xspace}
\newcommand{\taucj}{\ensuremath{\tau_{\mathrm{C}}^{\mathrm{j}}}\xspace}
\newcommand{\njets}{\ensuremath{n_{\text{jets}}}\xspace}
\newcommand{\mjj}{\ensuremath{m^{\mathrm{jj}}}\xspace}
\newcommand{\Detajj}{\ensuremath{\Delta\eta^{\mathrm{jj}}}\xspace}
\newcommand{\phietaS}{\ensuremath{\phi_{\eta}^{\ast}}\xspace}
\newcommand{\yj}{\ensuremath{y_{\mathrm{j}}}\xspace}
\newcommand{\yh}{\ensuremath{y_{\PH}}\xspace}
\newcommand{\etaZepp}{\ensuremath{\overline{\eta}_{\mathrm{j}_{1}\mathrm{j}_{2}}-\eta_{\PGg\PGg}}\xspace}
\newcommand{\ptgamgam}{\ensuremath{\pt^{\PGg\PGg}}\xspace}
\newcommand{\DetajOjT}{\ensuremath{\Delta\eta_{\mathrm{j}_{1}\mathrm{j}_{2}}}\xspace}
\newcommand{\ptjO}{\ensuremath{\pt^{\mathrm{j}_{1}}}\xspace}
\newcommand{\DphiggjOjT}{\ensuremath{\Delta\phi_{\PGg\PGg,\mathrm{j}_{1}\mathrm{j}_{2}}}\xspace}
\newcommand{\DphijOjT}{\ensuremath{\Delta\phi_{\mathrm{j}_{1},\mathrm{j}_{2}}}\xspace}
\newcommand{\sigFid}{\ensuremath{\sigma_{\text{fid}}}\xspace}
\newcommand{\Igen}{\ensuremath{\mathcal{I}_{\text{gen}}^{\PGg}}\xspace}
\newcommand{\nbj}{\ensuremath{n_{\PQb\text{jets}}}\xspace}
\newcommand{\resultsCaption}{The observed differential fiducial cross section values are shown as black points with the vertical error bars showing the full uncertainty, the horizontal error bars show the width of the respective bin. The grey shaded areas visualize the systematic component of the uncertainty. The coloured lines denote the predictions from different setups of the event generator. All of them have the HX=VBF+VH+ttH component from \MGvATNLO in common, which is displayed in violet without uncertainties. The red lines show the sum of HX and the ggH component from \MGvATNLO reweighted to match the \NNLOPS prediction. For the blue lines no \NNLOPS reweighting is applied and the green lines take the prediction for the ggH production mode from \POWHEG. The hatched areas show the uncertainties in theoretical predictions on both the \ggF and HX components. Only effects coming from varying the set of PDF replicas, the \alpS value, and the renormalization and factorization scales that impact the shape are taken into account here, the total cross section is kept constant at the value from Ref.~\cite{deFlorian:2016spz}. The given $p$-values are calculated for the nominal SM prediction and the bottom panes show the ratio to the same prediction. If the last particle-level bin expands to infinity is explicitly marked on the plot together with the normalization of this bin.}
\newcommand{\resultsCaptionShort}{The content of each plot is described in the caption of Fig.~\ref{fig:resultsPtNjetsCosTsRapi}.}
\newcommand{\resultsCaptionShortOne}{The content of this plot is described in the caption of Fig.~\ref{fig:resultsPtNjetsCosTsRapi}.}

\providecommand{\cmsTable}[1]{\resizebox{\textwidth}{!}{#1}}

\cmsNoteHeader{HIG-19-016}
\title{Measurement of the Higgs boson inclusive and differential fiducial production cross sections in the diphoton decay channel with \texorpdfstring{$\Pp\Pp$}{pp} collisions at \texorpdfstring{$\sqrt{s} = 13\TeV$}{sqrt(s) = 13 TeV}}

\date{\today}

\abstract{
  The measurements of the inclusive and differential fiducial cross sections of the Higgs boson decaying to a pair of photons are presented. The analysis is performed using proton-proton collisions data recorded with the CMS detector at the LHC at a centre-of-mass energy of 13\TeV and corresponding to an integrated luminosity of 137\fbinv. The inclusive fiducial cross section is measured to be $\sigFid=73.4_{-5.3}^{+5.4}\stat_{-2.2}^{+2.4}\syst\fb$, in agreement with the standard model expectation of $75.4\pm4.1\fb$. The measurements are also performed in fiducial regions targeting different production modes and as function of several observables describing the diphoton system, the number of additional jets present in the event, and other kinematic observables. Two double differential measurements are performed. No significant deviations from the standard model expectations are observed.}

\hypersetup{
pdfauthor={CMS Collaboration},
pdftitle={Measurement of the Higgs boson inclusive and differential fiducial production cross sections in the diphoton decay channel with pp collisions at sqrt(s) = 13 TeV},
pdfsubject={CMS},
pdfkeywords={CMS, Higgs, differential, fiducial, diphoton}}

\maketitle

\section{Introduction}\label{intro}
The discovery of the Higgs boson (\PH) by the ATLAS and CMS Collaborations in 2012~\cite{HiggsAtlas, HiggsCMS, HiggsCMS1} marked the beginning of an 
extensive program of measurements aimed at determining its properties and testing their compatibility 
with the standard model (SM) predictions. The four dominant Higgs boson production mechanisms in proton-proton ($\Pp\Pp$) collisions are gluon-gluon fusion (\ggF), vector boson fusion (\VBF), Higgs strahlung off a vector boson (\VH) and Higgs boson production associated with two top quarks (\ttH). The \ggF production mode is the one with the largest cross section, exceeding the other cross sections by roughly one order of magnitude. 

The Higgs boson production cross section in $\Pp\Pp$ collisions has been measured via an array of different decay channels, namely \Hgg, \HZZfl, \Hbb, \Htautau, and \HWW, at centre-of-mass energies of $7\TeV$, $8\TeV$, and most recently $13\TeV$~\cite{ATLAS:2016neq, ATLAS:2014yga, CMS:2015qgt, ATLAS:2014xzb, CMS:2015zpx, ATLAS:2016vlf, CMS:2016ipg, ATLAS:2017qey, CMS:2017dib, CMS:2018piu, ATLAS:2018pgp}.

Higgs boson production is explored using several approaches. The simplified template cross section (STXS)~\cite{deFlorian:2016spz} approach separates the Higgs boson production modes and additionally defines bins based on certain observables aimed at minimizing the theory uncertainty of the measurement. The inclusive and differential measurements of the fiducial cross sections instead aim at providing a set of model-independent results. 

The ATLAS Collaboration has used the \HZZfl~\cite{ATLAS:2020rej}, \Hbb~\cite{ATLAS:2020fcp}, \Htautau~\cite{ATLAS:2022yrq}, \HWWSt~\cite{ATLAS:2022ooq} and \Hgg~\cite{ATLAS:2022vkf, ATLAS:2022tnm} channels to measure the Higgs boson production cross section within the STXS framework while exploiting the full integrated luminosity available from the LHC data-taking period that lasted during the years 2016--2018. The CMS Collaboration has published STXS Higgs boson production cross section results using the \HZZfl~\cite{CMS:2021ugl} and \Hgg~\cite{HIG-19-015} final states using the data taken during the period 2016--2018.

The fiducial and differential approach for Higgs boson production cross section measurements was used by the ATLAS Collaboration to publish results in the \HZZfl~\cite{ATLAS:2020wny} and \Hgg~\cite{ATLASHggDiff} channels while using the data set collected from 2015--2018. With the data collected during the same period by the CMS Collaboration, fiducial differential Higgs boson production cross section measurements using the \HZZfl~\cite{CMS:2021ugl}, \HWW~\cite{CMS:2020dvg} and \Htautau~\cite{HIG-20-015} channels have been reported. The data set collected in 2016 has been used by the CMS Collaboration for the most recent Higgs boson production cross section measurement with the differential fiducial approach exploiting the \Hgg final state~\cite{HIG-17-025}.  

The extensive set of differential, fiducial Higgs boson production cross section measurements for \Hgg presented in this paper builds upon previous analyses performed by the CMS Collaboration~\cite{HIG-17-025, CMS:2015qgt}.

In the SM, the branching fraction for the Higgs boson decaying to a pair of photons is computed to be only about 0.23\%~\cite{deFlorian:2016spz}. Nevertheless, thanks to the high precision of the diphoton invariant mass reconstruction and the fully reconstructed final state, the \Hgg decay channel provides the most comprehensive measurements of the Higgs boson differential production cross sections.

The analysis presented in this paper uses the data recorded by the CMS experiment during the LHC Run~2, corresponding to an integrated luminosity of 137\fbinv. A new correction scheme is applied to the output of the simulation of electromagnetic showers that reduces the impact of the leading experimental systematic uncertainties. Several other improvements in the CMS reconstruction and calibration are exploited as well.

The analysis strategy closely follows the two previous CMS measurements at \cme~\cite{HIG-17-025} and $\sqrt{s}=8\TeV$~\cite{CMS:2015qgt} and it is designed to provide a measurement of the \Hgg cross section as independent as possible from the theoretical assumptions used for its extraction. 
To this end, the definition of the fiducial region mimics the detector acceptance and the events are categorized at the reconstruction level using a per-event mass resolution estimator.
The large data sets collected so far allow for an approximate doubling of the number of bins for most observables compared with the previous publications and therefore the most precise measurement of the $\Pp\Pp\to\PH+\PX$, \Hgg fiducial cross section to date. The measurements presented in this paper can be seen as a complementary and less model dependent addition to the STXS measurements described in Ref.~\cite{HIG-19-015}.

The fiducial differential cross sections are extracted in observables of the diphoton system, the transverse momentum (\pt) of the jets produced in association with the diphoton system, the dijet system containing the two leading-\pt jets, and with respect to several other event-level observables. For the first time, the differential cross section is also measured as a function of the angular observable \phietaS, as defined in Eq.~\eqref{eq:phiStar}, which complements the diphoton \pt measurement at low \pt values, and \taucj, defined in Eq.~\eqref{eq:tauC}, which is the jet \pt weighted by a function of its rapidity. Furthermore, two double-differential measurements are performed in bins of the diphoton \pt and the number of jets, and in bins of the diphoton \pt and \taucj.

The paper is structured as follows. An overview of the CMS detector is given in Section~\ref{sec:cms_detector}. The data and simulation samples used to perform this analysis are described in Section~\ref{sec:data_samples}. Section~\ref{sec:event_reco} summarizes the reconstruction of \Hgg candidate events and contains a detailed description of a newly developed simulation correction method. The selection criteria for events entering this analysis are shown in Section~\ref{sec:event_selection} and their categorization in Section~\ref{sec:categorization}. An overview of the observables as a function of which the \Hgg cross section is measured, and the definitions of the fiducial phase spaces are given in Section~\ref{sec:observables}. In Section~\ref{sec:statistical_analysis}, the statistical procedure used to extract the results from data is presented, while the relevant systematic uncertainties are given in Section~\ref{sec:systematic_uncertainties}. The results are reported in Section~\ref{sec:results} and the paper is summarized in Section~\ref{sec:summary}. Tabulated results are provided in the HEPData record for this analysis~\cite{hepdata}.

\section{The CMS detector}\label{sec:cms_detector}
The central feature of the CMS apparatus is a superconducting solenoid of 6\unit{m} internal diameter, 
providing a magnetic field of 3.8\unit{T}. 
Within the solenoid volume are a silicon pixel and strip tracker, recording the tracks of charged particles up to $\abs{\eta}<2.5$, 
a lead tungstate crystal electromagnetic calorimeter (ECAL), 
and a brass and scintillator hadron calorimeter (HCAL), 
each composed of a barrel and two endcap sections. 
The ECAL consists of 75\,848 lead tungstate crystals, 
which provide coverage in pseudorapidity $\abs{\eta} < 1.48 $ in the barrel region (EB) 
and $1.48 < \abs{\eta} < 3.0$ in the two endcap regions (EE). 
Preshower detectors consisting of two planes of silicon sensors interleaved with 
a total of three radiation lengths of lead are located in front of each EE detector.
Forward calorimeters extend the pseudorapidity coverage up to $\abs{\eta}=5$. 
Muons are detected in gas-ionization chambers embedded in the steel flux-return yoke outside the solenoid with a coverage up to $\abs{\eta}<2.4$. The excellent resolution of the photon energy provided by the ECAL allows for a relatively large signal significance and therefore a precise Higgs boson cross section measurement. 

Events of interest are selected using a two-tiered trigger system~\cite{CMSTRG}. 
The first level, composed of custom hardware processors, 
uses information from the calorimeters and muon detectors to select events 
at a rate of around 100\unit{kHz} within a fixed latency of about 4\mus.
The second level, known as the high-level trigger (HLT), 
consists of a farm of processors running a version of the full event reconstruction software 
optimized for fast processing, and reduces the event rate to around 1\unit{kHz} before data storage.

A more detailed description of the CMS detector, 
together with a definition of the coordinate system used and the relevant kinematic variables, 
can be found in Ref.~\cite{Chatrchyan:2008zzk}.

\section{Data samples and simulated events}\label{sec:data_samples}
The analysis uses data corresponding to the integrated luminosities of 36.3\fbinv, 41.5\fbinv, and 59.4\fbinv collected in 2016, 2017, and 2018, respectively. 
The integrated luminosities of these data-taking periods are individually known 
with uncertainties in the 1.2--2.5\% range~\cite{CMSlumi2016,CMSlumi2017,CMSlumi2018}, 
while the total (2016--2018) integrated luminosity has an uncertainty of 1.6\%, 
the improvement in precision reflecting the (uncorrelated) time evolution of some systematic effects. 
In this section, the data and simulated event samples for all three years are described, highlighting the differences between the years.

Events from pp collisions are selected using a diphoton trigger at the HLT with asymmetric \pt
thresholds on the leading and subleading photon of 30~(18) and 30~(22)\GeV in 2016 (2017--2018) data, respectively.
The invariant mass of the photon pair is required to be $\mgg>90\GeV$, while each photon is also required to pass loose requirements on the calorimetric isolation and on the shape of the electromagnetic shower of the photon candidates.
The trigger efficiency is measured on \Zee events using the "tag-and-probe" technique~\cite{TAG-PROBE}. These events
are collected with a single-electron trigger.
The efficiency measured in data, in bins of \pt, the energy fraction in the centre of the electromagnetic cluster \RNINE~\cite{EGMpaper}, and $\eta$, is used to weight the simulated events to replicate the trigger
efficiency observed in data. The \RNINE~variable is defined as the fraction of energy deposited in three by three crystals matrix around the seed $E_{3{\times}3}$ with respect the total uncorrected energy $E_{\text{SC,raw}}$ in the supercluster:
\begin{linenomath*}\begin{equation*}
	\RNINE=\frac{E_{3{\times}3}}{E_{\text{SC,raw}}}\textnormal{.}
\end{equation*}\end{linenomath*}
The \RNINE~variable is useful to identify photons and electrons that started showering in the material upstream of the ECAL.

Simulated signal samples, corresponding to the four production modes with the largest cross section (\ggF, \VBF, \VH, \ttH) are generated using \MGvATNLO (version 2.6.5) at next-to-leading order (NLO) accuracy~\cite{AMCAT} in perturbative quantum chromodynamics (QCD).
Events produced via the gluon fusion mechanism are weighted as a function of the Higgs boson \pt
and the number of jets in the event, to match the prediction from the next-to-NLO including parton showering event generator (\NNLOPS) program~\cite{NNLOPS1, NNLOPS2, NNLOPS3}.
The parton-level samples are interfaced with \PYTHIA8 (version 8.240)~\cite{Pythia82} for parton showering and hadronization, 
with the \textsc{CUETP8M1}~\cite{CUETP8} (\textsc{CP5}~\cite{CP5}) tune used for the simulation of 2016 (2017--2018) data.
The signal cross sections and branching fractions recommended by the LHC Higgs Working Group~\cite{deFlorian:2016spz} are used.

The dominant source of background events in this analysis is from QCD diphoton (\gamgam) production.
The majority of the remaining background originates from \gamplusjet or dijet events, in which jets are misidentified as photons.

The diphoton background (\gamgam) is generated with the \SHERPA (version 2.2.4) generator~\cite{SHERPA22}. 
It includes the Born processes with up to three additional jets, as well as the box processes at leading order (LO) accuracy. The \gamplusjet and dijet background are simulated with \PYTHIA8. 
For the optimization of the event categorization, the \gamplusjet and \gamgam background samples are used. The training of the photon identification multivariate analysis (MVA) is performed with the \gamplusjet sample. As explained in Section~\ref{sec:statistical_analysis}, no simulated samples for the background are used in the measurement of the fiducial inclusive or differential cross sections.

Samples of \Zee, \Zmm, and \Zmmg simulated events are generated with \MGvATNLO at NLO precision and used for the derivation of energy scale, resolution, and photon selection correction factors.

\section{Event reconstruction}\label{sec:event_reco}
Physics objects are reconstructed using the particle-flow algorithm (PF)~\cite{ParticleFlow}, which aims to identify each individual particle (PF candidate) in an event, 
with an optimized combination of information from the various elements of the CMS detector. 

\subsection{Vertex identification}\label{subsec:vertex_choice}

Finding the correct primary vertex in the \Hgg final state when the Higgs boson was produced through gluon fusion faces the complication that photons (unless they convert into an electron-positron pair in the material upstream of the ECAL) do not leave any ionization signal in the tracker that can be used for finding their trajectory, and therefore the production vertex. On the other hand, assigning the right vertex is crucial for measuring the kinematic variables of the diphoton system and keeping a good diphoton mass resolution. It is found that if the reconstructed vertex is within 1\cm in distance along the beam axis of the true vertex, the angular component in the mass resolution is subdominant compared to the ECAL energy resolution.
To obtain the vertex position of an \Hgg event, a dedicated vertex finding algorithm has been developed based on a boosted decision tree (BDT)~\cite{CMS:2018piu}.
The BDT is trained on simulated events produced via gluon fusion, using track information from the objects recoiling against the diphoton system and tracks from electron-positron pairs originating from a converted photon if available. Its performance is validated in data with \Zmm events, in which the vertices are refitted omitting the muon tracks. The average vertex finding efficiency is 75--80\% across the years of data-taking~\cite{CMS:2018piu}.
\subsection{Photon reconstruction and identification}\label{subsec:photon_reco}

The reconstruction of a photon starts by clustering the energy deposited in the ECAL.	
The supercluster formation begins by finding the ECAL particle-flow clusters that locally record the largest energy above a given threshold.
Then the clusters spatially compatible with originating from the same prompt photon are merged into a supercluster. Photons are selected as superclusters that are not linked to any charged-particle trajectory associated to a primary interaction vertex. A variable supercluster size in $\phi$ recovers the energy of photons initiating an electromagnetic shower in the material upstream of the ECAL. 
A more detailed description of the photon reconstruction algorithm can be found in Ref.~\cite{EGMpaper}.

\subsubsection{Photon energy corrections}
The response of each ECAL crystal is calibrated individually and adjusted for time-dependent effects.
A semi-parametric regression is used to correct for the partial collection of the electromagnetic shower energy in the supercluster and effects from simultaneous soft $\Pp\Pp$ collisions (pileup), and to predict the per-photon energy resolution. This regression is implemented with a BDT and trained on simulated photons using several shower shape variables as input~\cite{CMS:2018piu}.

After this simulation-based correction, a time-dependent correction of the energy response measured in data to the one predicted by simulation is derived from \Zee events, where the photon reconstruction is applied to the electron candidates. The typical time granularity of this correction is of one LHC fill and is necessary to track the response evolution of the ECAL. In this way any residual energy drift in data is corrected as a function of time. The \Zee events are used to adjust the energy scale in data and the resolution in simulation: the energy of the electrons in data is corrected to have the peak of their invariant mass distribution matching the one in the Drell--Yan simulation; the energy in simulated electron pairs is instead smeared such that the width of their invariant mass spectrum matches the one observed in data. Further details about this procedure can be found in Ref.~\cite{HggMass}.

\subsubsection{Photon identification}
To discriminate prompt photons against those originating from the decay of neutral particles in jets, e.g. from \PGpz or \PGh mesons, a classifier implemented with a BDT, called the photon identification MVA, is trained on a \gamplusjet simulation sample. 
The BDT input consists of several variables associated with the shape of the supercluster associated with the photon candidate, and isolation sums, built from neutral electromagnetic components and charged hadrons, calculated for different vertex choices~\cite{EGMpaper}. For the calculation of the isolation from charged hadrons, it is necessary to consider different vertex choices, because for a given photon, the tracks (characterizing the charged hadrons) included in the isolation cone around it depend on the considered primary vertex. For photons detected in EE, the energy deposited into the preshower detector is considered by the photon identification MVA, in addition to the variables mentioned above.

\subsubsection{Correcting the photon identification MVA imperfect modelling}\label{subec:idMVAcorrections}
The photon identification MVA is the primary means of reducing the \gamplusjet and dijet QCD backgrounds in the analysis. Its imperfect modelling in simulation results in one of the most important sources of systematic uncertainty for this analysis and was the dominant one for the previous CMS differential fiducial \Hgg cross section measurement~\cite{HIG-17-025}.
The modelling inaccuracy comes from the imperfect modelling of the MVA input variables.
Among the reasons for this are the ever changing ECAL conditions due to radiation damage, and the imperfect knowledge of the amount of material traversed by a photon before reaching the ECAL. Together with a campaign of studies to address the root causes of the mismatch~\cite{EGMpaper}, an effective correction applied to simulation has been developed for the analyses with diphoton final states~\cite{HIG-19-015, HIG-19-013, HIG-19-018}.

A dedicated method has been devised to compute a per-photon correction that morphs the input variable distributions in simulation, and their correlations, to match data. The method, called chained quantile regression (CQR), is used for the cluster shape and isolation variables used as input for the photon identification MVA.
It is based on the multitarget regression algorithm described in Ref.~\cite{MTR} and implemented to correct these variables, conditional on the photon kinematic properties and the average spatial energy density of the event. 

The method originates from the quantile morphing technique, in which any continuous distribution can be morphed into any other continuous distribution, using the following algorithm:
\begin{itemize}
	\item[1.] Find the value of the cumulative distribution function (CDF) in simulation $F_{Y}^{\text{sim}}(y)$ for value $y$ of the variable $Y$ to be adjusted;
	\item[2.] Get the point of the cumulative distribution function of the target with the same value $F_{Y}^{\text{data}}(y_{C})=F_{Y}^{\text{sim}}(y)$;
	\item[3.] Assign the new value $y_{C}$ as the corrected value instead of $y$ in the distribution to be corrected.
\end{itemize}
This can be written as:
\begin{linenomath*}\begin{equation*}
	y_{C}={F_{Y}^{\text{data}}}^{-1}\left(F_{Y}^{\text{sim}}(y)\right).
\end{equation*}\end{linenomath*}
To be able to perform these steps, the CDFs ($F_{Y}^{\text{data}}$, $F_{Y}^{\text{sim}}$) of the distribution to be corrected and the target have to be known. 

For the purpose of correcting the set of input variables of the photon identification MVA, quantile morphing is extended to be conditional on the photon kinematic variables (\pt, $\eta$, $\phi$) and the average spatial event energy density $\rho$~\cite{CMS:2018piu}. The conditional shape of the CDFs in simulation and data are approximated using 21 quantile regressors (implemented with BDTs, called quantile BDTs from here), each predicting the quantile for a different value of the CDF.

\begin{figure}[tb]
	\centering
	\includegraphics[width=0.99\textwidth]{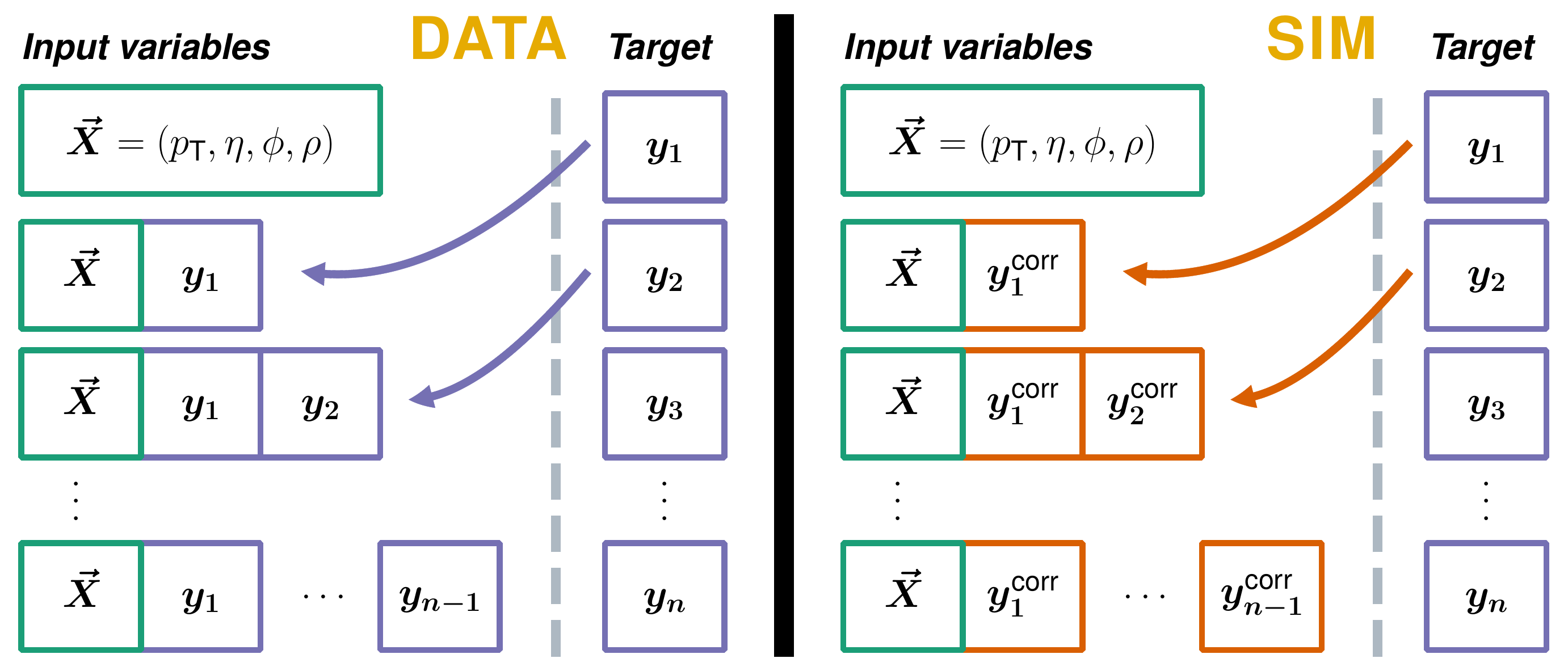}
	\caption{The chained approach for the set of input variables for the quantile BDTs. The input variables for the BDT for each variable being corrected are given. The variables to be corrected are used as the target, and the quantile is learned through the learning objective given in Eq.~\eqref{eq:quantLoss}. Within one group of variables ($y_{1}$, \dots, $y_{n}$), with nonnegligible correlations, an order is set. The quantile BDT for a given variable includes the prior set of variables, within this ordering, as additional inputs. For simulation (right), the additional input variables are corrected before using them as inputs for the quantile BDTs.}
	\label{fig:chainedVars}
\end{figure}

Apart from the kinematics of the photon and the event energy density, the photon identification MVA input variables can be sorted into three groups with nonnegligible correlations among the variables within them: the variables describing the shape of the electromagnetic shower, the isolation sums of charged candidates reconstructed by the PF algorithm, and the isolation sum of electromagnetic neutral objects. The quantile BDTs are trained using a set of input variables that depends on the variable to be corrected and the group it is assigned to. Within each group, a sequence of variables is defined and all variables appearing before the one to be corrected are added to the input variables used for the BDTs on top of the photon \pt, $\eta$, and $\phi$, and the event $\rho$.
The set of input variables for the quantile BDTs for the variables within one of the groups introduced above for simulation and data is illustrated in Fig.~\ref{fig:chainedVars}. 
This chained approach allows the correlations within the groups of variables to be corrected at the same time as their conditional shape. The order in which the variables are inserted in the chain has been optimized on the correction performance. The training objective of the quantile BDTs is defined as~\cite{quantileReg}: 
\begin{linenomath*}\begin{equation}\label{eq:quantLoss}
	q_{Y}\left(\tau\right)=\underset{u}{\mathrm{argmin}}\left\{\left(\tau-1\right)\int_{-\infty}^{u}\left(y-u\right)\rd F_{Y}\left(y\right)+\tau\int_{u}^{\infty}\left(y-u\right)\rd F_{Y}\left(y\right)\right\},
\end{equation}\end{linenomath*}
where $\tau$ corresponds to the CDF value for which the quantile $q_{Y}$ is computed. 

The training is performed independently in data and simulation using the \textsc{scikit-learn} package~\cite{scikit-learn} on electrons (reconstructed as photons) from simulated \Zee events and data collected using a single-electron trigger. The tag-and-probe method is then applied to data, while for simulation the reconstructed electrons are matched to their true generated counterparts before the tag-and-probe method is applied. This gives the unbiased sample of probe electrons needed to derive the conditional CDF shapes. 

The largest group of variables are the electromagnetic cluster shapes, for which the CQR approach introduced above is sufficient. The other groups of variables to be corrected include the photon and charged isolation sums which are discontinuous, preventing the immediate use of the CQR algorithm. 
The discontinuity is caused by detector thresholds applied to the energy of the constituents of the isolation sums, resulting in a fraction of events with exactly zero isolation energy. The isolation distributions present a peak at zero followed by a continuous tail. The number of events in the peak transforms into a plateau in the cumulative distribution function. Since the population of events with zero isolation sum is not necessarily the same in data and simulation, the codomains of the CDFs are not covering the same range of values, which is a necessary condition to apply the quantile morphing algorithm. 

A dedicated procedure has been developed to overcome this complication and still apply the quantile morphing to the tail parts of the distributions. The mitigation strategy differs between the two types of isolation sum: the photon isolation is corrected independently from all other variables, while the two charged isolation sums, which are correlated, are corrected simultaneously. 

\begin{figure}[!htb]
	\centering
	\includegraphics[width=0.7\textwidth]{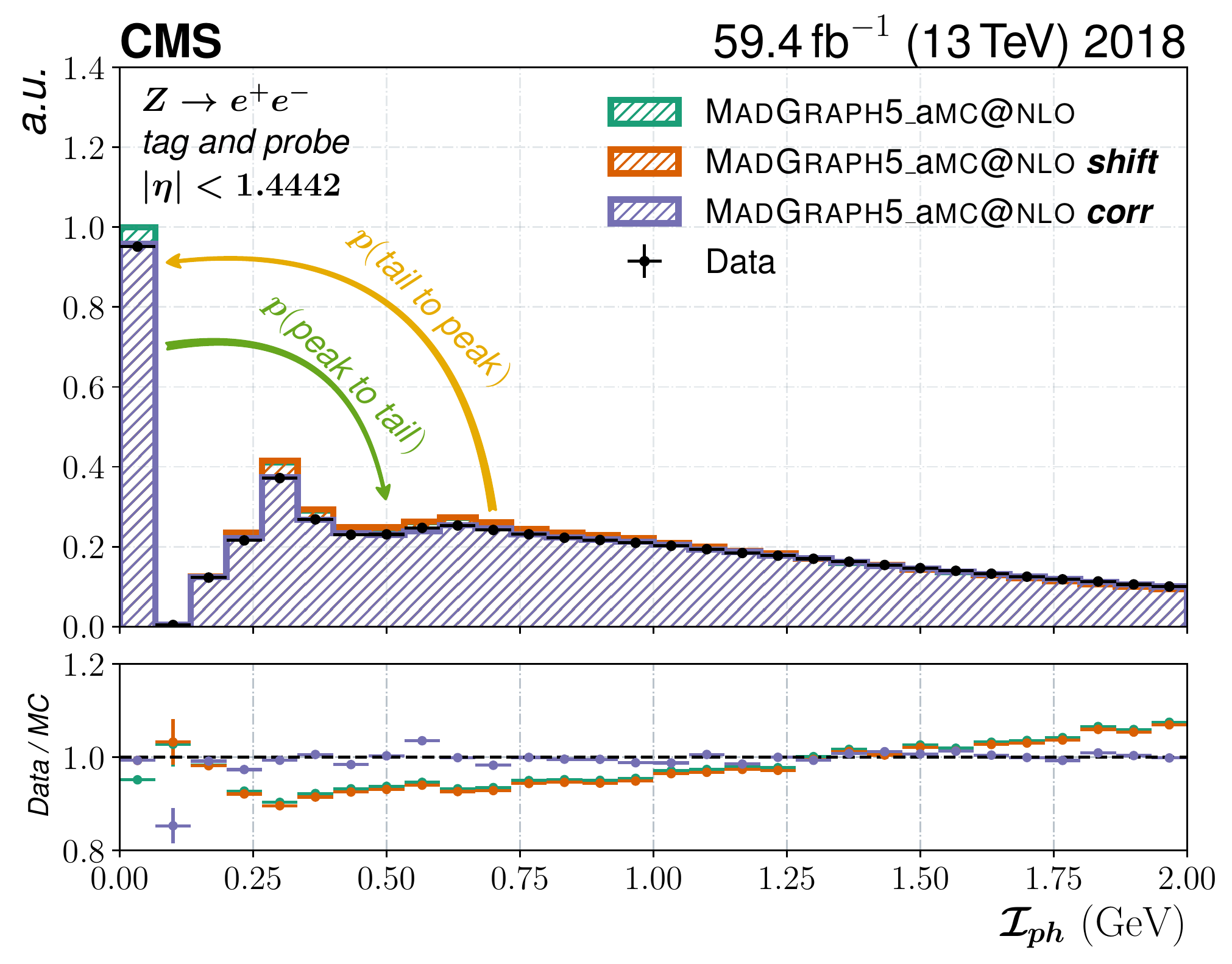}
	\caption{The upper pane shows the distribution of the photon isolation sum ($\mathcal{I}_{\text{ph}}$) in 2018 data (black dots) and simulation (coloured histograms). The green histogram shows the uncorrected distribution, the orange one the distribution after equalizing the number of events with zero isolation in simulation with data, and the purple one the distribution after applying the equalizing step and the CQR technique to its tail part. The arrows show the two ways events can be shifted. From peak to tail (green) and from tail to peak (yellow) with their respective probabilities $p(\text{peak to tail})$ and $p(\text{tail to peak})$. The lower pane shows the ratio of the three simulation distributions to the one from data. The distributions shown in this figure are taken from a set of events distinct from the ones used for the derivation of the correction method.}
	\label{fig:isoCorrFig}
\end{figure}

For the photon isolation sum, the first step is equalizing the number of events in the first bin of the isolation distribution between data and simulation using a stochastic procedure to move events from the peak (zero isolation) to the tail of the distribution and vice-versa. The events to be moved are randomly picked with a probability calculated from the relative population between the peak and the tail. 
The probability to have an isolation sum of zero is estimated with a BDT trained with the binary cross-entropy loss function conditionally on the photon kinematic variables and $\rho$ using the \textsc{xgboost} package~\cite{xgboost}. For all isolation variables, a newly assigned value for an event moved to the tail part (isolation greater than zero) is obtained by sampling from the tail part of the distribution conditional on the photon kinematic variables and $\rho$. A set of example distributions for the photon isolation sum, one being uncorrected, one after equalizing the number of events with zero isolation, and the third one after applying the equalizing step and the CQR technique to the continuous tail, together with an illustration of the equalizing step are shown in Fig.~\ref{fig:isoCorrFig}.

The charged isolation computed with respect to the vertex giving it its largest value, called worst charged isolation, is by construction larger than or equal to the charged isolation calculated with respect to the chosen vertex (hereafter called charged isolation). This leads to three possible categories: both charged isolation sums being exactly zero, the charged isolation being zero and the worst charged isolation being larger than zero, and both of them being larger than zero. A classifier implemented with a BDT, using (\pt, $\eta$, $\phi$, $\rho$) as input, is trained to distinguish between them. The events to be moved between the three categories are randomly picked with a probability calculated from the output of this classifier such that their populations in simulation match the ones observed in data. 

Finally, the continuous distributions of the two charged-isolation sums greater than zero are corrected using a CQR: the morphed charged-isolation is fed as an additional input to the quantile BDTs for the worst charged isolation, to correctly model their correlation.

\begin{figure}[!htb]
  \centering
    \includegraphics[width=0.49\textwidth]{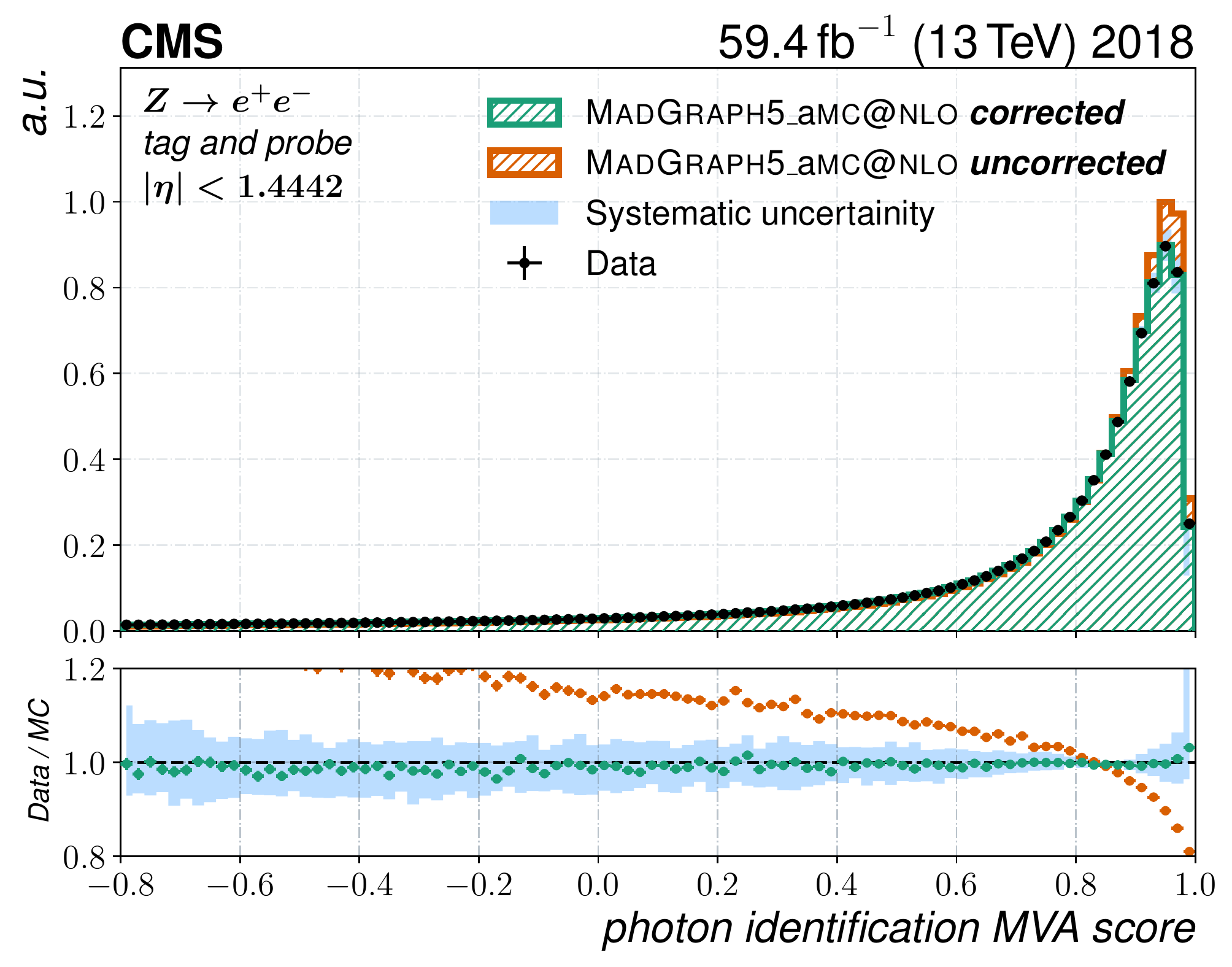}
    \includegraphics[width=0.49\textwidth]{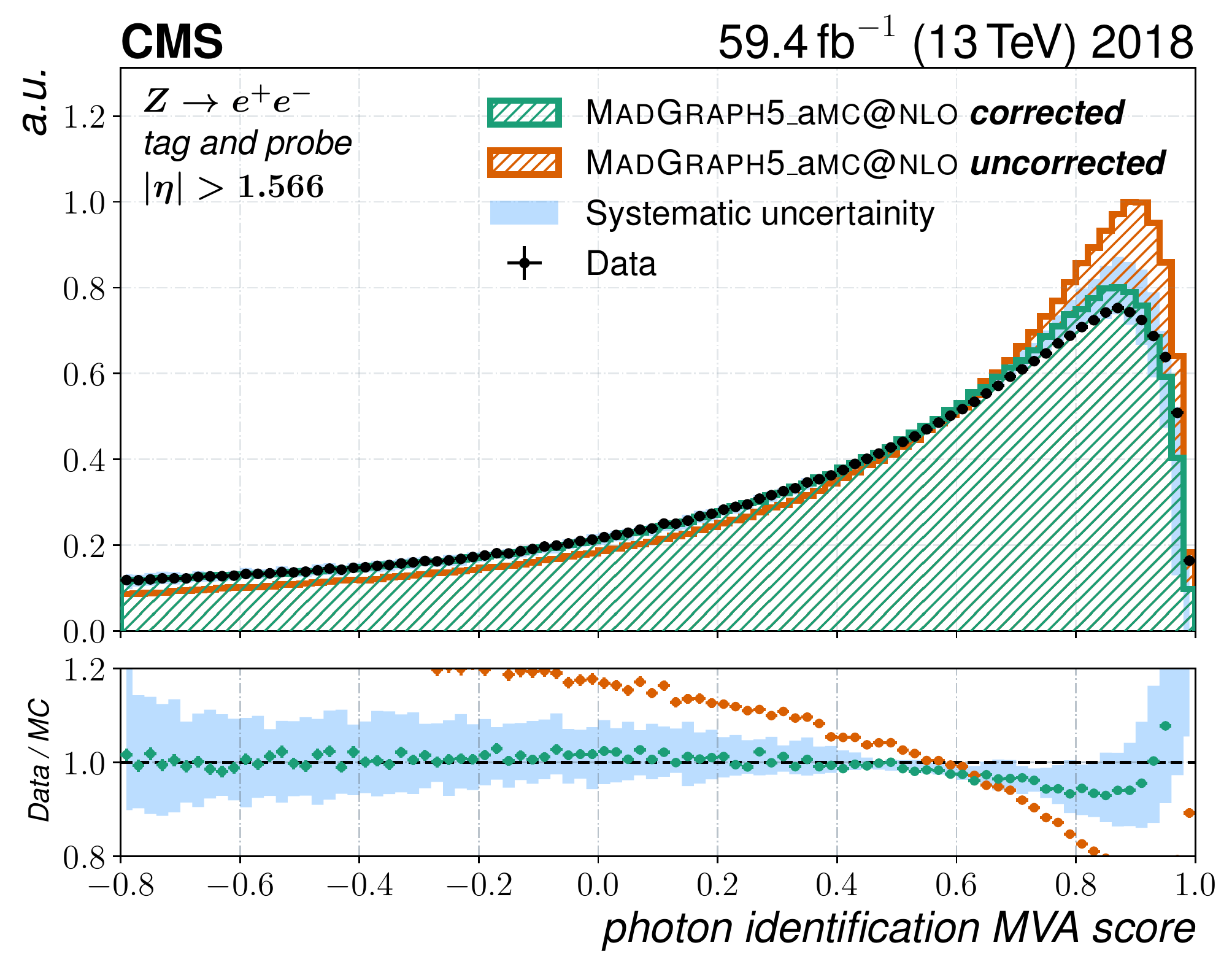}
  \caption{Distribution of the output of the photon identification MVA for the probe candidate in a \Zee tag-and-probe sample for data and the \MGvATNLO simulation. The electrons have been reconstructed as photons and a selection to reduce the number of misidentified electrons in data is applied. The simulated events have been reweighted with respect to \pt, $\eta$, $\phi$, and $\rho$ to match data in order to remove effects from mismodelled kinematic variables. Electrons that are detected in the barrel ($\abs{\eta}<1.4442$) or endcap ($\abs{\eta}>1.566$) part of the ECAL and the corresponding distributions are shown on the left or right, respectively. The blue band shows the systematic uncertainty assigned to the data simulation mismatch of the output of the photon identification MVA. The orange histogram and points in the upper and lower plots, respectively, show the photon identification MVA distribution and its ratio to data evaluated using the uncorrected version of its input variables in simulation.}
  \label{fig:phoIdMVA}
\end{figure}

After correcting the photon identification MVA input variables in simulation with the method just described, its output distribution matches the corresponding one in data within 1\% for the ECAL barrel and within 5\% for the ECAL endcap, as shown in Fig.~\ref{fig:phoIdMVA}. Consequently, the impact of the associated systematic uncertainty on the inclusive fiducial cross section measurement is significantly reduced from 5\%, for the previous analysis~\cite{HIG-17-025}, to 1.5\%, when adding the per-year contributions to the uncertainty in quadrature.

\subsection{Other objects}\label{subsec:other_ojects_reco}

Electrons are identified from a charged-particle track and the ECAL energy clusters consistent with this track's extrapolation to the ECAL. Additionally, possible bremsstrahlung photons emitted along the way through the tracker material are taken into account.
The energy of electrons is determined from a combination of the electron momentum 
at the primary interaction vertex as determined by the tracker, 
the energy of the corresponding ECAL cluster, 
and the energy sum of all bremsstrahlung photons spatially compatible 
with originating from the electron track.

Muons are identified from tracks in the central tracker consistent with either a track or several hits in the muon system, and associated with calorimeter deposits compatible with the muon hypothesis. Their momentum is obtained from the curvature of the corresponding track. 

The energy of charged hadrons is determined from a combination of their momentum 
measured in the tracker and the matching ECAL and HCAL energy deposits, 
corrected for the response function of the calorimeters to hadronic showers. 
Finally, the energy of neutral hadrons is obtained 
from the corresponding corrected ECAL and HCAL energies. 

Particle flow candidates are clustered to jets using the infrared and collinear-safe anti-\kt clustering algorithm~\cite{AntiKt, fastjet} with a distance parameter of 0.4.
In this process, the contribution from each calorimeter tower is assigned a momentum, 
the absolute value and the direction of which are given by the energy measured in the tower, 
and the coordinates of the tower. 
The raw jet energy is obtained from the sum of the PF candidate energies, 
and the raw jet momentum by the vectorial sum of the candidate momenta, 
which results in a nonzero jet mass. Additional proton-proton interactions within the same or nearby bunch crossings can contribute additional tracks and calorimetric energy depositions, increasing the apparent jet momentum. To mitigate this effect, tracks identified to be originating from pileup vertices are discarded and an offset correction is applied to correct for remaining contributions.
The raw jet energies are then corrected to establish a uniform relative response in $\eta$ and a calibrated absolute response in \pt. Jets originating from the hadronization of \PQb quarks are identified using a tight selection on the output of the \textsc{DeepJet} b~tagger~\cite{DeepJet}, corresponding to an efficiency of 56\% for 2016 and 62\% for 2017 and 2018. 

The missing transverse momentum vector \ptvecmiss is computed as the negative vector sum 
of the transverse momenta of all the PF candidates in an event, 
and its magnitude is denoted as \ptmiss~\cite{METperformance}. 
The \ptvecmiss is modified to account for corrections to the energy scale of the reconstructed jets in the event.

\section{Event selection}\label{sec:event_selection}
The selection criteria for photons forming the diphoton candidates that enter the analysis are designed to be tighter than the trigger thresholds based on \pt, isolation, and cluster shape variables~\cite{EGMpaper}. The details of this photon preselection are given in Ref.~\cite{HIG-19-015}. 

Each candidate in the photon pair must have a supercluster with $\abs{\eta}<2.5$, where the region $1.4442<\abs{\eta}<1.566$ is excluded. The transverse momentum relative to the invariant mass of the diphoton pair ($\pt/\mgg$) of the leading-\pt (subleading-\pt) photon has to be greater than $1/3$\,($1/4$). Both photons must satisfy a minimal score of the photon identification MVA. If an event contains more than one photon pair, the one with the highest \pt is selected.

Two selections of jets enter the analysis, one with $\abs{\eta}<2.5$, the other with $\abs{\eta}<4.7$. In both cases the jets are required to have $\pt>30\GeV$. The first selection is applied in differential measurements with respect to observables calculated when exactly one additional jet is present in the event, or with event-level observables, such as the number of jets and the number of \PQb-tagged jets. The looser $\eta$ criterion is used for observables taking into account variables involving two or more jets. Events with the tighter $\eta$ requirement benefit from a better jet energy resolution because of tracker information being available for jets with $\abs{\eta}<2.5$. To ensure a mutually exclusive selection of photon and jet candidates, each jet is required to have a $\Delta R>0.4$, which is defined as the quadratic sum of the differences in azimuthal angle and pseudorapidity between a photon candidate and the jet: 
\begin{linenomath*}\begin{equation}
	\Delta R=\sqrt{\Delta\phi^{2}+\Delta\eta^{2}}.
\end{equation}\end{linenomath*}
A jet reconstructed within $\abs{\eta}<2.4$, and satisfying the identification criteria of \PQb jets described in Section~\ref{subsec:other_ojects_reco}, is considered a \PQb jet at reconstruction level. At the particle level, a jet is considered as originating from a \PQb quark if, among its constituent hadrons, the decay products of at least one \PQb hadron are present.

Both electrons and muons are required to have $\pt>20\GeV$ and $\abs{\eta}<2.4$. They have to be separated by $\Delta R>0.35$ from the photon candidates. As with the photons, electrons are not reconstructed in the ECAL 
region $1.4442<\abs{\eta}<1.566$. To reject $\PZ{+}\PGg\to\Pep\Pem\PGg$ events with one electron misidentified as a photon, the invariant mass of any electron-photon pair cannot be within 5\GeV of the \PZ boson mass. The loose identification criteria defined in Ref.~\cite{EGMpaper} are applied to all electrons. Muons are selected when fulfilling a tight selection criterion \cite{CMS:2018rym} based on the quality of the track and the relative isolation, which is calculated as the sum of the transverse energy of charged and neutral hadrons, and photons, in a cone with radius parameter $\Delta R=0.4$ around the muon divided by the muon \pt.  
\section{Event categorization}\label{sec:categorization}
The photon pairs that enter this analysis are categorized according to the decorrelated mass resolution estimator, described below, and the photon identification MVA. The same boundary for the photon identification MVA selection is applied to the \pt-leading and \pt-subleading photons in events falling into any resolution category. These categorization criteria are designed to ensure minimal model dependence, while yielding the best possible analysis sensitivity. 
\subsection{Mass resolution estimator}\label{sec:sigmaMoM}
The mass resolution estimator \sigMoM for a photon pair is defined as the quadratic sum of the relative energy resolution of each photon as estimated by the photon energy regression BDT. The per-photon energy resolution in simulation is smeared to match the one observed in data using the \Zee decay as a reference.
The relative ECAL energy resolution ($\sigma_{E}/E$) improves with larger photon energies, for which the impact of noise and stochastic effects on the resolution decreases. This effect leads to a correlation between the mass resolution estimator (\sigMoM) and the invariant mass of the diphoton system. Such a correlation, combined with a selection on \sigMoM, can distort the shape of the background by depleting the low-mass region for high-resolution categories, creating a rising edge in the background shape that spoils the assumption of a monotonically falling background shape, which is necessary to model it using the discrete profiling method presented in Section~\ref{subsec:backgroundModel}.
To avoid this effect, the mass resolution estimator is decorrelated with respect to the diphoton invariant mass. This procedure is based on applying the quantile morphing algorithm described in Section~\ref{subec:idMVAcorrections} to the mass dependent \sigMoM spectrum and begins by dividing the \sigMoM distribution into mass bins. The \sigMoM distribution in each mass bin is then morphed to match the shape of the target, which is the \sigMoM distribution at a reference mass value of $\mgg=125\GeV$. In the following, the decorrelated relative mass resolution estimator is referred to as \sigMD.
More details on this procedure can be found in Ref.~\cite{CMS:2015qgt}.

\subsection{Decorrelated relative mass resolution categorization}
\begin{figure}[!ht]
	\includegraphics[width=0.49\textwidth]{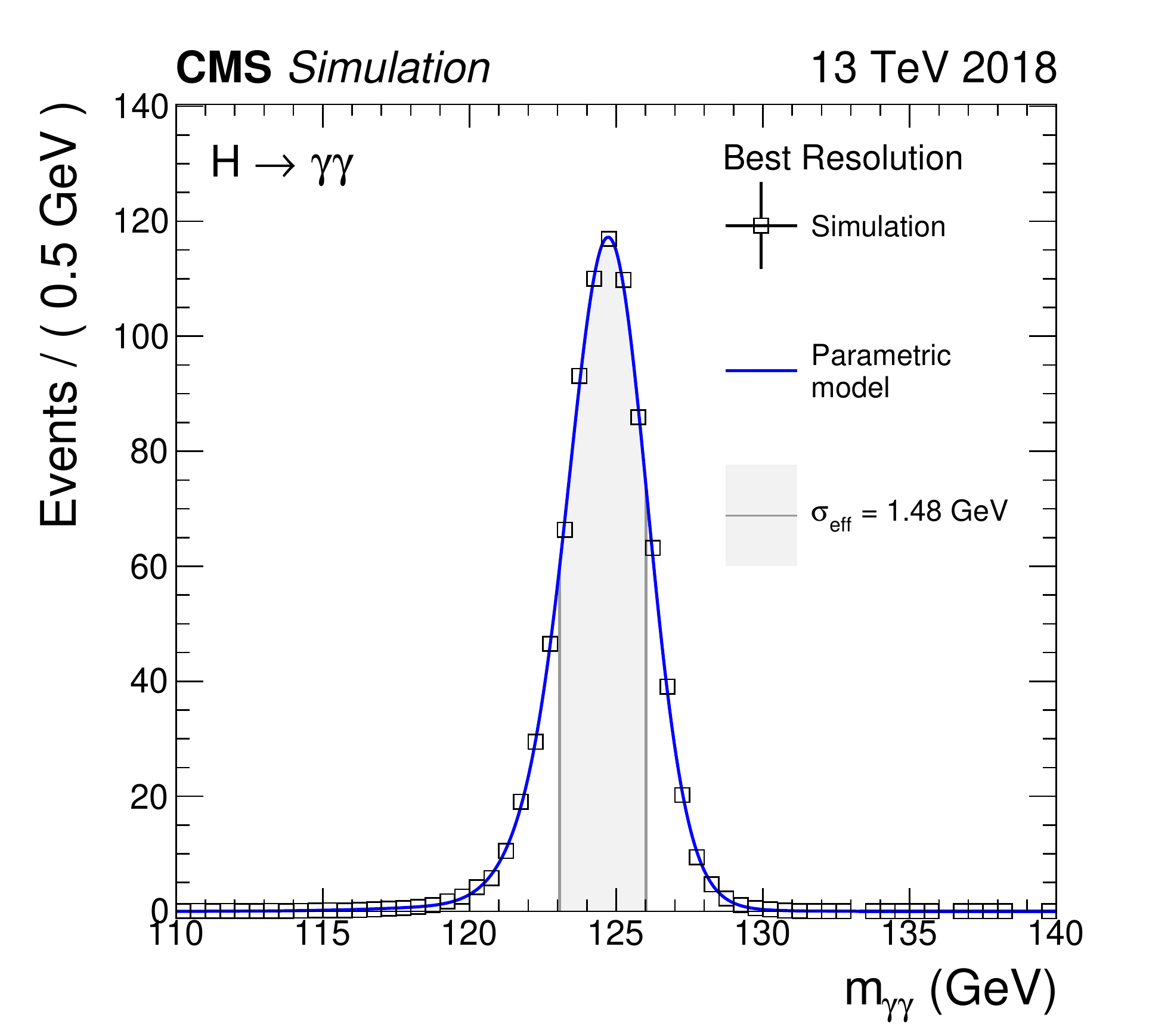}
	\includegraphics[width=0.49\textwidth]{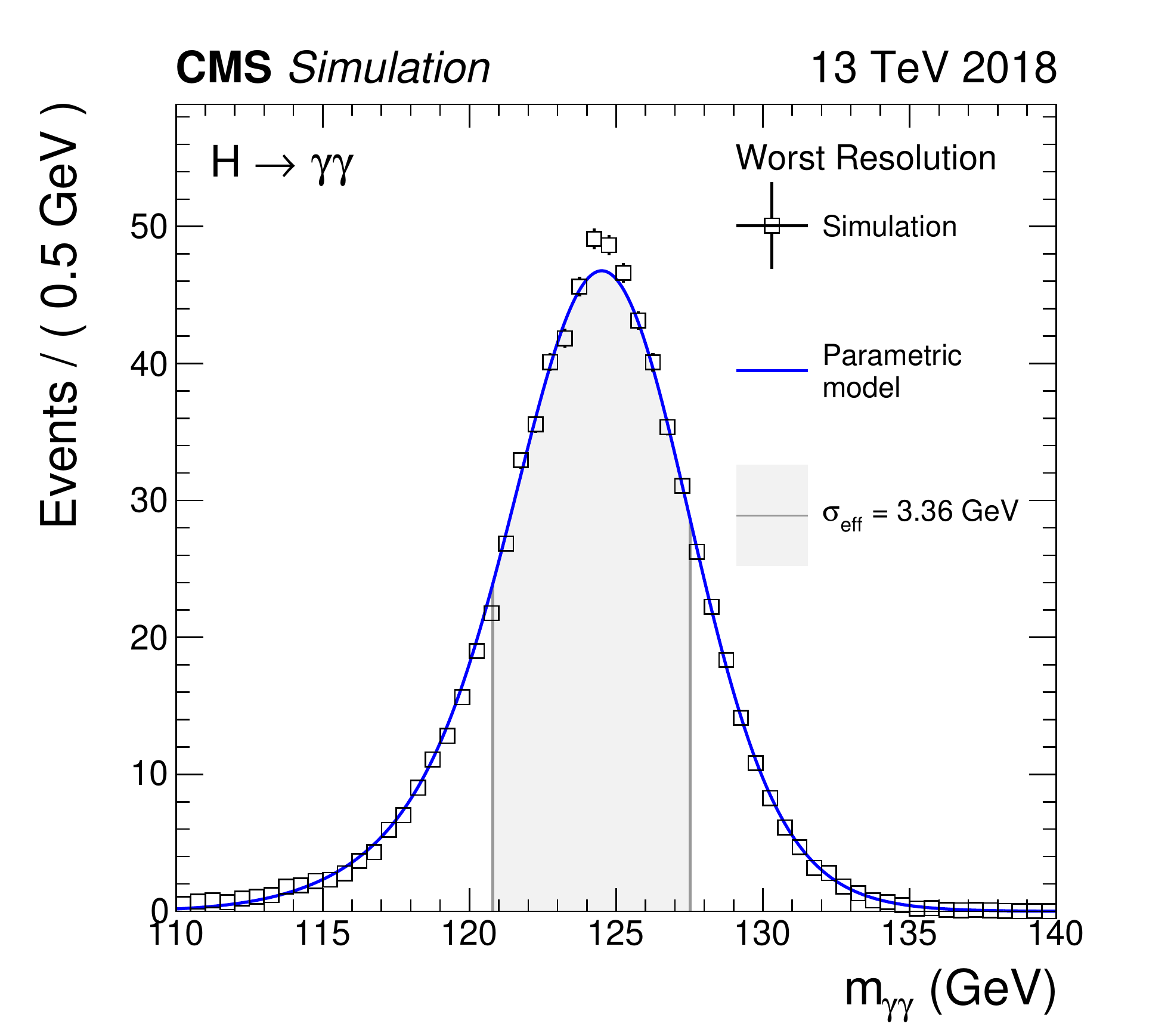}
	\centering
	\caption{The signal model distributions used in the fiducial cross section measurement for the best and worst resolution categories in 2018. The half width of the \mgg distribution region around its peak that contains 68.3\% of the total integral is denoted as $\sigma_{\text{eff}}$. The distributions shown here are taken from the signal simulation including the four dominant Higgs boson production mechanisms with a mass hypothesis of $\mH=125\GeV$. Details of the derivation of the signal model distributions are discussed in Section~\ref{subsec:signalModel}.}
	\label{fig:sigMod}
\end{figure}

The boundaries of the \sigMD categories are optimized simultaneously with the photon identification MVA selection, using the expected signal significance as a figure of merit. The optimization is performed using simulated samples of signal events at $\mH=125\GeV$, and diphoton and $\PGg{+}\text{jet}$ events to model the background, simulating different detector conditions for each year of data taking.
For all years, it is found that the optimal number of \sigMD categories is three. Photon pairs above the highest \sigMD boundary are discarded. The \sigMD boundaries roughly group the photon pairs by the ECAL regions where each of the two photons were detected. The overall efficiency of the photon identification criterion ranges from 75 to 80\% across the three years of data taking. The effect of the event categorization is to increase the signal significance by between $13$ and $18\%$ depending on the particular reconstruction level bin. 
A detailed breakdown, together with the efficiencies of the \sigMD categorization, is given in Table~\ref{tab:efficiencies}. Examples of the signal shape in different categories are shown in Fig.~\ref{fig:sigMod}.
\begin{table}[htbp]
	\centering
	\topcaption{Efficiencies of the photon identification MVA and \sigMD categories for events taken from the signal sample for all three years of data taking. The second row shows the efficiency of the photon identification MVA selection in the three \sigMD categories and for the full sample (Overall). The third row shows the efficiencies of the selections for the three \sigMD categories without the photon identification MVA selection applied. The forth row reports the effective width ($\sigma_{\text{eff}}$) of the Higgs boson signal in each category.
          The four dominant Higgs boson production modes considered for this analysis are included in the sample and $\mH=125\GeV$ is used. Only events satisfying the fiducial selection are included.}
	\begin{tabular}{lllll}
		&2016&&&\\\hline
		Resolution category&Best&Medium&Worst&Overall\\\hline
		Photon ID MVA efficiency&$84.7\%$&$81.7\%$&$70.3\%$&$77.3\%$\\
		Resolution category efficiency&$31.4\%$&$25.4\%$&$41.3\%$&\\
		$\sigma_{\text{eff}}$ (\GeV)&$1.46$&$1.98$&$2.80$&\\\hline
	\end{tabular}
	\begin{tabular}{lllll}
		\\
		&2017&&&\\\hline
		Resolution category&Best&Medium&Worst&Overall\\\hline
		Photon ID MVA efficiency&$90.4\%$&$83.9\%$&$73.0\%$&$80.3\%$\\
		Resolution category efficiency&$24.5\%$&$34.0\%$&$38.8\%$&\\
		$\sigma_{\text{eff}}$ (\GeV)&$1.31$&$1.87$&$2.66$&\\\hline
	\end{tabular}
	\begin{tabular}{lllll}
		\\
		&2018&&&\\\hline
		Resolution category&Best&Medium&Worst&Overall\\\hline
		Photon ID MVA efficiency&$86.8\%$&$78.8\%$&$63.5\%$&$75.3\%$\\
		Resolution category efficiency&$27.8\%$&$35.9\%$&$35.5\%$&\\
		$\sigma_{\text{eff}}$ (\GeV)&$1.48$&$2.13$&$3.36$&\\\hline
	\end{tabular}
	\label{tab:efficiencies}
\end{table}

\section{Observables and fiducial phase space definitions}\label{sec:observables}
The measurements of the inclusive and differential production cross sections are performed in fiducial phase spaces to reduce their model dependence.
A set of kinematic and identification selections is applied at the level of stable particles obtained from the event generator and parton shower simulation (particle-level) to the single photons and to the diphoton system, as summarized in Table~\ref{tab:phaseSpace}. The particle-level information does include QCD+EWK final state radiation effects but not the simulation of the interaction of the stable particles with the detector. The \Igen is defined as the sum of the energy of all stable hadrons produced in a cone of radius $\Delta R=0.3$ around the photon.
This requirement is introduced to mimic the photon identification MVA selection criterion at the reconstruction level.
\begin{table}[tbp]
	\centering
	\topcaption{Definition of the fiducial phase space. The labels 1, 2 refer to the \pt-ordered leading and subleading photon in the diphoton system. The variable \Igen is defined as the sum of the energy of all stable hadrons produced in a cone of radius $\Delta R=0.3$ around the photon.}
	\begin{tabular}{cc}
		\hline
		Observable&Selection\\\hline
		$\pt^{\PGg_1}/\mgg$&${>}1/3$\\
		$\pt^{\PGg_2}/\mgg$&${>}1/4$\\
		\Igen&${<}10\GeV$\\
		$\abs{\eta^{\PGg}}$&${<}2.5$\\\hline
	\end{tabular}
	\label{tab:phaseSpace}
\end{table}
In addition to the baseline fiducial phase space, other fiducial phase spaces are defined specifically to target certain observables of interest. There are several observables that are calculated with variables of the leading-\pt, or the leading- and subleading-\pt additional jets. All additional jets considered in this analysis are required to satisfy $\ptj>30\GeV$. If the measured observable requires one additional jet to be evaluated, the 1-jet phase space is used. It requires at least one particle-level jet being in the event and the leading-\pt jet to be within $\abs{\etaj}<2.5$. An observable that is calculated involving two additional jets is measured in the 2-jet phase space, for which the $\abs{\eta}$ selection is relaxed to $\abs{\etaj}<4.7$ for all jets. A region within the 2-jet phase space is designed to predominantly accept events originating from the vector boson fusion (VBF) Higgs boson production process and is called VBF-enriched. The additional requirements for events to enter this phase space region are: $\njets\geq2$, $\mjj>200\GeV$, and $\Detajj>3.5$, where \mjj is the invariant mass of the dijet system containing the two leading-\pt jets and \Detajj is the difference in pseudorapidity between these two jets. The selection for the VBF-enriched region is designed with loose cut-based kinematic criteria to ensure the model dependence of the measurement is kept as small as possible.

All differential cross section observables, along with the phase space regions in which they are measured, are given in Table~\ref{tab:obsBins}. The bin boundaries have been chosen such that the statistical uncertainty is similar for the cross section measured in each bin, with the goal of a per-bin statistical uncertainty of around 30\%. Where applicable, the bin boundaries have been unified with the ones for the ATLAS \Hgg differential cross section measurement~\cite{ATLASHggDiff}. The same bin boundaries are used for the corresponding reconstruction-level quantities. The measurements are performed binned in the transverse momentum and absolute rapidity of the diphoton system, which are sensitive to the Higgs boson production mechanism, the modelling of the QCD radiation, and the parton distribution functions (PDFs) of the proton. Two additional observables are defined for the diphoton system. One is the cosine of the polar angle in the Collins--Soper reference frame of the diphoton system $\cos\theta^{\ast}$~\cite{Collins:1977}, which is sensitive to the spin and CP properties of the diphoton resonance. The other is $\abs{\phietaS}$, designed to probe the low \pt region while minimizing the impact of experimental uncertainties. It is defined as \cite{Banfi:2010cf, boggia2017higgstools}
\begin{linenomath*}\begin{equation}\label{eq:phiStar}
	\phietaS = \tan\left(\frac{\phi_{\mbox{\text{acop}}}}{2}\right)\sin(\theta_{\eta}^{\ast}),
\end{equation}\end{linenomath*}
where the acoplanarity angle is related to the angle between the photons in the transverse plane, $\Delta\phi$, as $\phi_{\mbox{acop}} = \pi -\Delta\phi$, and $\theta_{\eta}^{\ast}$ is the scattering angle of the two photons with respect to the proton beam in the reference frame in which the two particles are back to back in the $(r,\theta)$ plane. Observables involving the two \pt-leading jets, such as their transverse momentum or absolute rapidity, are sensitive to the Higgs boson production in QCD.
In the system of the two photons originating from the Higgs boson and the leading jet, the difference in azimuthal angle and rapidity between the diphoton system and the jet are taken as observables. These specifically probe the properties of the VBF production mechanism. The \taucj observable is also considered. It is defined as~\cite{Gangal:2015}: 
\begin{linenomath*}\begin{equation}\label{eq:tauC}
	\taucj=\max_{k\in\text{jets}}\left(\frac{\sqrt{E_{\mathrm{k}}^{2}-p_{\mathrm{z},\mathrm{k}}^{2}}}{2\cosh\left(y_{\mathrm{k}}-\yh\right)}\right),
\end{equation}\end{linenomath*}
and calculated for the (up to) six largest \pt jets in the event, taking the maximum among them. It represents the transverse momentum of the jet weighted by a function that depends on the rapidity of the jet and smoothly suppresses the contribution from forward jets. In the denominator of Eq.~\eqref{eq:tauC}, the difference between the rapidity of the jet \yj and the rapidity of the Higgs boson \yh is taken. This means that \taucj is calculated in the frame where $\yh=0$.
Contrary to requiring an unweighted \ptj veto on jets, applying a kinematic selection using or binning events in \taucj for jets does not introduce extra logarithms (or minimizes their contribution) in the resummation region (low-\pt) of gluon fusion Higgs boson production cross section calculations including additional jets. Therefore, measuring the Higgs boson production cross section with respect to \taucj represents a test of QCD resummation~\cite{Gangal:2015}.

In the $\PGg\PGg\mathrm{j}_{1}\mathrm{j}_{2}$ system, the difference in $\phi$ and $\eta$ between the photon pair and the two jets, and between the two jets themselves, are the quantities of interest, where the $\eta$ difference between the photon pair and the jets is taken as the $\eta$ difference between the photon pair and the average $\eta$ of the two jets, defined as $\abs{\etaZepp}$ \cite{Rainwater:1996}. This observable serves as an additional probe of the properties of the VBF production mode. The invariant mass of the two leading-\pt jets is also considered. 

In the VBF-enriched phase space the cross section is measured with respect to the transverse momentum of the diphoton system and of the subleading jet, as well as the difference in $\phi$ between the dijet and the diphoton and the two leading jets themselves.

Two double-differential measurements of the \Hgg~cross section are performed: one with respect to the transverse momentum of the diphoton system and the number of jets, and the other one with respect to the diphoton transverse momentum and \taucj. 

Measurements are also performed with respect to the number of jets, the number of leptons, the number of b-tagged jets, as well as the missing transverse momentum in the \Hgg~event.

Finally, the \Hgg cross section is measured in dedicated regions of the fiducial phase space designed to loosely target specific production modes. Their selection criteria require the presence of one extra lepton, and $\ptmiss>100\GeV$ ($\PW\PH$) or $\ptmiss<100\GeV$ (complementary to $\PW\PH$). For the third region, targeting the \ttH production mode, events need to have at least one extra lepton and at least one b-tagged jet.

\begin{table}
  \centering
  \topcaption{Binning per observable of interest. The first block of rows of the table shows the observables measured in the baseline fiducial phase space, the second one observables involving one additional jet, and the third one involving two or more additional jets. In the fourth block observables for the VBF-enriched phase space are shown. Energy, invariant mass and momentum are in \GeV.}
  \cmsTable{
    \begin{tabular}{llllllllll}
      \hline
      Phase space region&Observable&\multicolumn{8}{c}{Bin boundaries}\\\hline
      \multirow{17}{*}{\parbox{3.0cm}{\textit{Baseline} \newline $\pt^{\PGg_1}/\mgg > 1/3$ \newline $\pt^{\PGg_2}/\mgg > 1/4$ \newline $\abs{\eta^\PGg} < 2.5$ \newline $\Igen<10\GeV$}}&\ptgamgam&$0$&$5$&$10$&$15$&$20$&$25$&$30$&$35$\\
      &&$45$&$60$&$80$&$100$&$120$&$140$&$170$&$200$\\
      &&$250$&$350$&$450$&$\infty$&&&&\\
      &\njets&$0$&$1$&$2$&$3$&$\geq4$&&&\\
      &$\abs{y^{\PGg\PGg}}$&$0.0$&$0.1$&$0.2$&$0.3$&$0.45$&$0.6$&$0.75$&$0.90$\\
      &&$2.5$&&&&&&&\\
      &$\abs{\cos(\theta^{\ast})}$&$0.0$&$0.07$&$0.15$&$0.22$&$0.35$&$0.45$&$0.55$&$0.75$\\
      &&$1.0$&&&&&&&\\
      &$\abs{\phietaS}$&$0.0$&$0.05$&$0.1$&$0.2$&$0.3$&$0.4$&$0.5$&$0.7$\\
      &&$1.0$&$1.5$&$2.5$&$4.0$&$\infty$&&&\\
      &$\ptgamgam,\njets=0$&$0$&$5$&$10$&$15$&$20$&$25$&$30$&$35$\\
      &&$45$&$60$&$\infty$&&&&&\\
      &$\ptgamgam,\njets=1$&$0$&$30$&$60$&$100$&$170$&$\infty$&&\\
      &$\ptgamgam,\njets>1$&$0$&$100$&$170$&$250$&$350$&$\infty$&&\\
      &\nbj&$0$&$1$&$\geq2$&&&&&\\
      &$n_{\text{leptons}}$&$0$&$1$&$\geq2$&&&&&\\
      &\ptmiss&$0$&$30$&$50$&$100$&$200$&$\infty$&&\\\hline
      \multirow{10}{*}{\parbox{3.0cm}{\textit{1-jet} \newline Baseline + ${\geq} 1$ jet \newline $\ptj>30\GeV$ \newline $\abs{\etaj}<2.5$}}&\ptjO&$30$&$40$&$55$&$75$&$95$&$120$&$150$&$200$\\
      &&$\infty$&&&&&&&\\
      &$\abs{y^{\mathrm{j}_{1}}}$&$0.0$&$0.3$&$0.6$&$0.9$&$1.2$&$1.6$&$2.0$&$2.5$\\
      &$\abs{\Delta\phi_{\PGg\PGg,\mathrm{j}_{1}}}$&$0.0$&$2.0$&$2.6$&$2.85$&$3.0$&$3.07$&$\pi$&\\
      &$\abs{\Delta y_{\PGg\PGg,\mathrm{j}_{1}}}$&$0.0$&$0.3$&$0.6$&$1.0$&$1.4$&$1.9$&$2.5$&$\infty$\\
      &\taucj&$<15$&$15$&$20$&$30$&$50$&$80$&$\infty$&\\
      &$\ptgamgam,\taucj<15~\GeV$&$0$&$45$&$120$&$\infty$&&&&\\
      &$\ptgamgam,15~\GeV\leq\taucj<25~\GeV$&$0$&$45$&$120$&$\infty$&&&&\\
      &$\ptgamgam,25~\GeV\leq\taucj<40~\GeV$&$0$&$120$&$\infty$&&&&&\\
      &$\ptgamgam,40~\GeV\leq\taucj$&$0$&$200$&$350$&$\infty$&&&&\\\hline
      \multirow{7}{*}{\parbox{3.0cm}{\textit{2-jets} \newline Baseline + ${\geq} 2$ jets \newline $\ptj>30\GeV$ \newline $\abs{\eta^{j}}<4.7$}}&$\pt^{\mathrm{j}_{2}}$&$30$&$40$&$65$&$90$&$150$&$\infty$&&\\
      &$\abs{y^{\mathrm{j}_{2}}}$&$0.0$&$0.6$&$1.2$&$1.8$&$2.5$&$3.5$&$5.0$&\\
      &$\abs{\DphijOjT}$&$0.0$&$0.5$&$0.9$&$1.3$&$1.7$&$2.5$&$\pi$&\\
      &$\abs{\DphiggjOjT}$&$0.0$&$2.0$&$2.7$&$2.95$&$3.07$&$\pi$&&\\
      &$\abs{\etaZepp}$&$0.0$&$0.2$&$0.5$&$0.85$&$1.2$&$1.7$&$\infty$&\\
      &\mjj&$0$&$75$&$120$&$180$&$300$&$500$&$1000$&$\infty$\\
      &$\abs{\DetajOjT}$&$0.0$&$0.7$&$1.6$&$3.0$&$5.0$&$\infty$&&\\\hline
      \multirow{4}{*}{\parbox{3.0cm}{\textit{VBF-enriched} \newline 2-jets + $\Detajj>3.5$ \newline $\mjj>200$\GeV }}&\ptgamgam&$0$&$30$&$60$&$120$&$200$&$\infty$&&\\
      &$\pt^{\mathrm{j}_{2}}$&$30$&$40$&$65$&$90$&$150$&$\infty$&&\\
      &$\abs{\DphijOjT}$&$0.0$&$0.5$&$0.9$&$1.3$&$1.7$&$2.5$&$\pi$&\\
      &$\abs{\DphiggjOjT}$&$0.0$&$2.0$&$2.7$&$2.95$&$3.07$&$\pi$&&\\\hline
    \end{tabular}
  }
  \label{tab:obsBins}
\end{table}

\section{Statistical analysis}\label{sec:statistical_analysis}
The events are categorized in bins of the observable of interest and in categories of the relative mass resolution estimator \sigMD, as described in Section~\ref{sec:categorization}. To unfold the bins defined at reconstruction-level to the particle-level bins, the signal simulation sample is further split into bins in the particle-level observable of interest.

The signal production cross section is extracted through a simultaneous extended maximum likelihood fit of parametric signal and background probability density functions to the diphoton invariant mass spectrum of events falling into each reconstruction-level bin further divided into \sigMD categories. The shape of the signal contribution is determined separately for each of the two-dimensional bins originating from dividing the events in reconstruction-level bins and particle-level bins, as described in Section~\ref{subsec:signalModel}.
This makes the signal model sensitive to the fraction of events from each particle-level bin being reconstructed in a reconstruction-level bin and category, thus performing the maximum likelihood unfolding of the detector response. The background shape is derived through the discrete profiling method~\cite{DiscreteProfilingMethod} in each reconstruction-level bin further split in categories of \sigMD, as described in Section~\ref{subsec:backgroundModel}.

The imperfect alignment of the fiducial phase space at the particle level and the reconstruction selection means that there are events that do not enter the fiducial phase space, but are selected at reconstruction level and vice versa. This effect has to be taken into account by the analysis since the theoretical predictions for the cross sections that are to be measured are calculated at the particle level. This is done by deriving a dedicated probability density function (pdf) for signal events originating outside the fiducial phase space at particle level, and taking them into account in the maximum likelihood fit as an additional signal-like component.

\begin{figure}
\centering
  \includegraphics[width=0.72\textwidth]{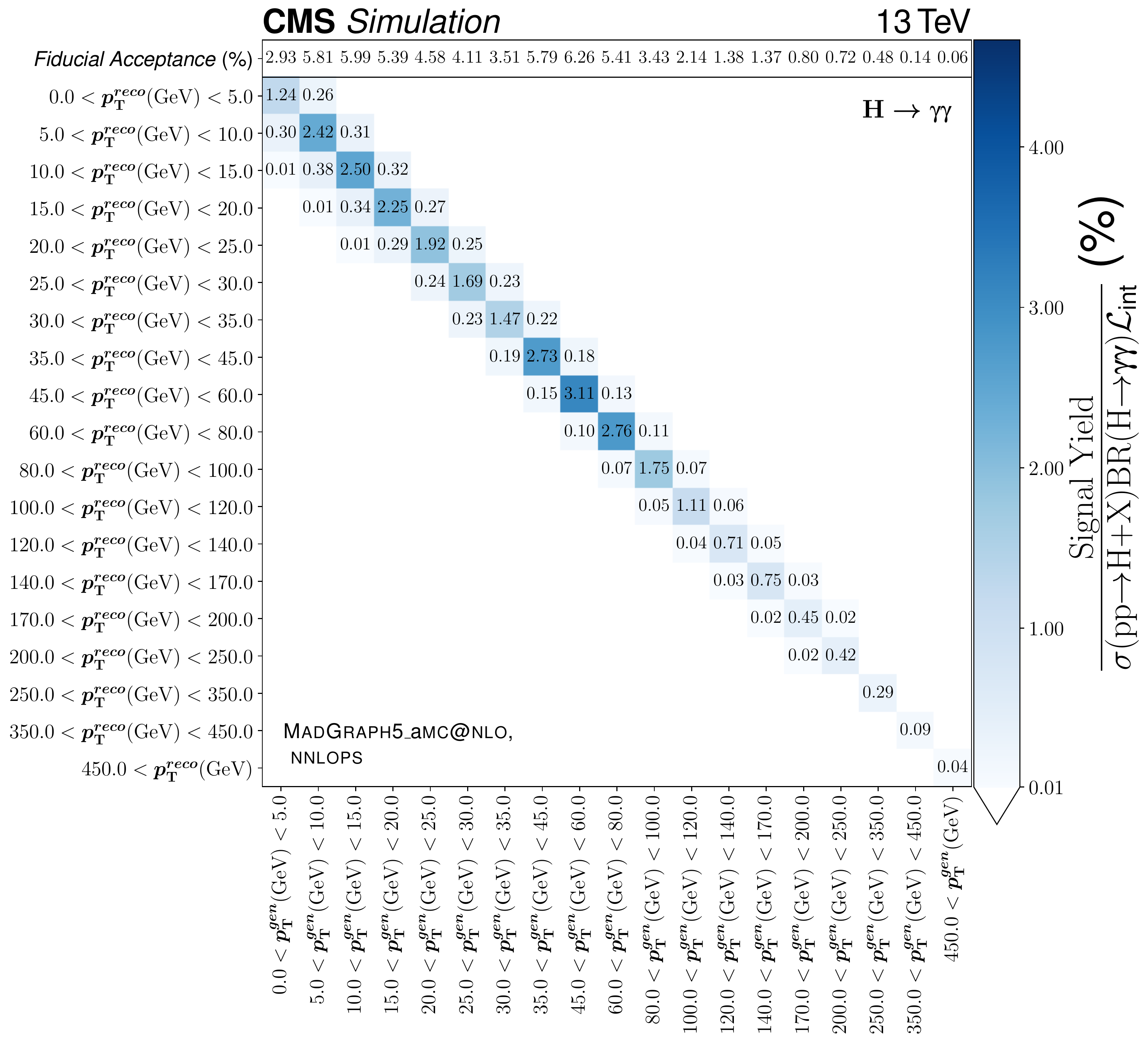}\\
  \includegraphics[width=0.62\textwidth]{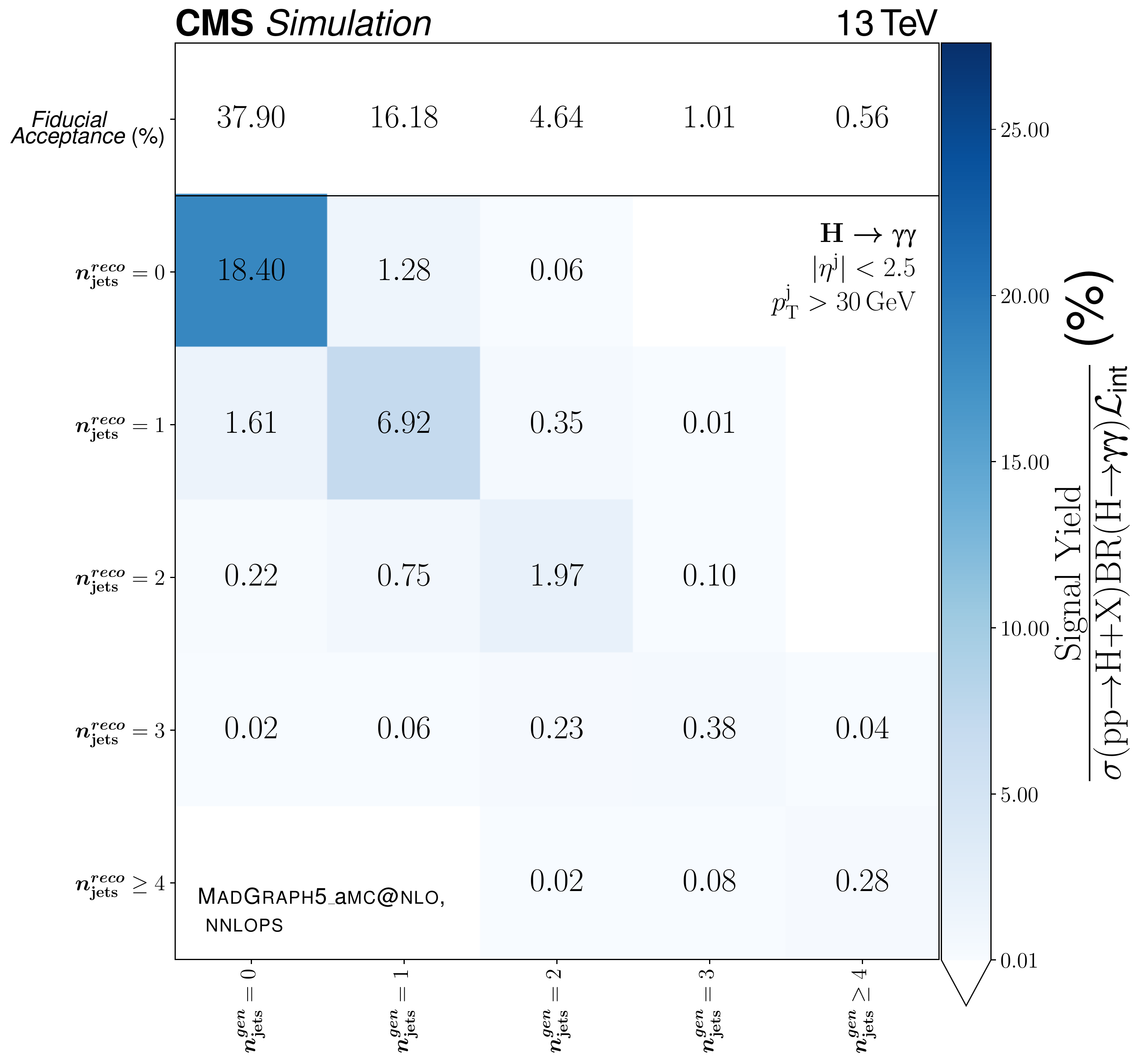}
  \caption{The event yields summed across all resolution categories divided by the total \Hgg cross section \cite{deFlorian:2016spz} multiplied by the integrated luminosity for the bins in the particle-level, reconstruction-level observables for the year 2018 for the observables \pt and \njets are shown. There is one column per particle-level bin and one row per reconstruction-level bin. The top row shows the predicted fiducial acceptance, \ie the per particle-level bin \Hgg cross section divided by the total \Hgg cross section. The values of the matrix $K$ in Eq.~\eqref{form:LikelihoodBin} can be computed by dividing, column by column, the values in each bin by the predicted fiducial acceptance reported in the top row. The version of \PYTHIA used here is 8.240 and the \MGvATNLO version is 2.6.5.}
  \label{fig:unfoldMatr}
\end{figure}

For observables that are measured in restricted regions of the phase space, the cross section of the events that enter the baseline fiducial phase space, but not the restricted region, is left floating in the fit and treated as an additional bin. Letting the cross section in this additional bin float and taking migrations into and out of it into account allows for a tighter constraint of the cross sections within the restricted phase space region. The efficiency multiplied with the acceptance per particle-level bin, reconstruction-level bin, and resolution category combination is shown in Fig.~\ref{fig:unfoldMatr} for the year 2018 and the observables \pt and \njets.

The likelihood in the $i$-th \sigMD category and $j$-th reconstruction-level observable bin can be written as:
\begin{linenomath*}\begin{gather}
\mathcal{L}_{ij}\left(\text{data}|\Delta\vec{\sigma}^{\text{fid}},\vec{n}_{\text{bkg}},\vec{\theta}_{\mathrm{S}},\vec{\theta}_{\mathrm{B}}\right)=\nonumber\\
\prod_{l=1}^{n_{\mgg}}\left(\frac{\sum_{k=1}^{n_{\mathrm{b}}}\Delta\sigma_{k}^{\text{fid}}K_{k}^{ij}\left(\vec{\theta}_{\mathrm{S}}\right)S_{k}^{ij}\left(\mgg^{l}|\vec{\theta}_{\mathrm{S}}\right)L^{i}+n_{\text{OOA}}^{ij}S_{\text{OOA}}^{ij}\left(\mgg^{l}|\vec{\theta}_{\mathrm{S}}\right)+n_{\text{bkg}}^{ij}B^{ij}\left(\mgg^{l}|\vec{\theta}_{\mathrm{B}}\right) }{n_{\text{sig}}^{ij}+n_{\text{bkg}}^{ij}}\right)^{n_{\text{ev}}^{lij}}.\label{form:LikelihoodBin}
\end{gather}\end{linenomath*}
The \sigMD categories also run over the three years of data taking, with three \sigMD categories per year, as presented in Section~\ref{sec:categorization}. In Eq.~\eqref{form:LikelihoodBin}, $L^{i}$ stands for the total analyzed integrated luminosity for the year with which the respective category is associated. The number of mass bins of the \mgg distribution is denoted with $n_{\mgg}$, and $n_{\mathrm{b}}$ stands for the number of reconstruction-level bins for a given variable. The parameters of interest are encoded in the vector $\Delta\vec{\sigma}^{\text{fid}}=(\Delta\sigma_{1}^{\text{fid}}, \dots, \Delta\sigma_{n_{\mathrm{b}}}^{\text{fid}})$, which contains the per particle-level bin fiducial cross sections multiplied by the branching fraction of the Higgs boson decaying into two photons. The particle-level cross sections are directly obtained from the maximum likelihood fit and are not constrained to be positive, as this would invalidate the method to extract the confidence intervals described below. 

The events originating from a particle-level bin $k$ are reconstructed in the reconstruction-level bin $j$, and category $i$. The magnitude of the migration is encoded in the detector response matrix $K_{k}^{ij}$. The pdfs in \mgg modelling the signal are denoted with $S_{k}^{ij}$ for the particle-level bin $k$ and reconstruction-level bin $j$ and category $i$. The corresponding background pdf for reconstruction-level bin $j$ and category $i$ is written as $B^{ij}$. The sum of the signal and background pdfs is used to fit the number of observed events $n_{\text{ev}}^{lij}$ in \mgg bin $l$ of reconstruction-level bin $j$ and category $i$. The numbers of signal and background events are denoted as $n_{\text{sig}}^{ij}$ and $n_{\text{bkg}}^{ij}$, respectively. For signal events that enter the reconstruction-level analysis selection, while originating from outside the fiducial phase space, the label $\text{OOA}$ (outside of acceptance) is used, where $n_{\text{OOA}}^{ij}$ is the number of events that fulfil this condition and $S_{\text{OOA}}^{ij}$ the pdf that describes their shape in \mgg. The number of events, $n_{\text{OOA}}^{ij}$, is fixed to the value predicted by the \MGvATNLO generator including the \NNLOPS reweighting for the gluon fusion production mode. Finally, the nuisance parameters associated with the signal and background model are encoded as $\vec{\theta}_{\mathrm{S}}$ and $\vec{\theta}_{\mathrm{B}}$, respectively. 

The complete extended likelihood can be written as:
\begin{linenomath*}\begin{equation}\label{form:Likelihood}
\mathcal{L}\left(\text{data}|\Delta\vec{\sigma}^{\text{fid}},\vec{n}_{\text{bkg}},\vec{\theta}_{\text{S}},\vec{\theta}_{\mathrm{B}}\right)=\prod_{i=1}^{n_{\text{cat}}}\prod_{j=1}^{n_{\mathrm{b}}}\left(\mathcal{L}_{ij}\text{Pois}\left(n_{\text{ev}}^{ij}|n_{\text{sig}}^{ij}+n_{\text{bkg}}^{ij}\right)\right)\textnormal{pdf}\left(\vec{\theta}_{\mathrm{S}}\right)\text{pdf}\left(\vec{\theta}_{\mathrm{B}}\right),
\end{equation}\end{linenomath*}
with $n_{\text{cat}}$ being the number of \sigMD categories across all years. Here, pdf stands for the probability density function of the nuisance parameters, while Pois denotes the Poisson probability distribution. From Eq.~\eqref{form:Likelihood} the fiducial cross section $\Delta\vec{\sigma}^{\text{fid}}$ can be extracted with a maximum likelihood fit. The mass of the Higgs boson $m_{\PH}$ is treated as a Gaussian constraint with a central value and width set according to the most precise CMS measurement of $\mH=125.38\pm0.14\GeV$~\cite{HggMass}.
This means that effectively the mass values are independent per observable, but all particle-level bins of one observable share the same value.

The negative log-likelihood ratio $q(\Delta\vec{\sigma})$ is used as the test statistic to obtain the uncertainties and correlation matrices~\cite{Asimov}:
\begin{linenomath*}\begin{equation}\label{form:testStat}
q\left(\Delta\vec{\sigma}\right)=-2\log\left(\frac{\mathcal{L}\left(\text{data}|\Delta\vec{\sigma},\hat{\vec{\theta}}_{\Delta\vec{\sigma}}\right)}{\mathcal{L}\left(\text{data}|\Delta\hat{\vec{\sigma}},\hat{\vec{\theta}}\right)}\right),
\end{equation}\end{linenomath*}
where the number of background events and both sets of nuisance parameters are written as $\theta=(n_{\text{bkg}},\vec{\theta}_{\mathrm{S}},\vec{\theta}_{\mathrm{B}})$; $\Delta\hat{\vec{\sigma}}$ and $\hat{\vec{\theta}}$ denote the best fit estimates of the cross sections and nuisance parameters, while $\hat{\vec{\theta}}_{\Delta\vec{\sigma}}$ denotes the best fit estimates of the nuisance parameters at a fixed value of $\Delta\vec{\sigma}$.

\subsection{Signal model}\label{subsec:signalModel}
The signal model is derived in a parametric form for each observable separately. The simulated \mgg~signal shape at reconstruction level, including efficiency corrections estimated with the help of control samples, is fit with a sum of up to four Gaussian pdfs in each bin of the particle-level observable, reconstruction-level observable bins, and reconstruction categories. The shapes for right and wrong vertices are fit individually and added together to give the full signal pdf per bin. A signal shape is derived for each of the three mass hypotheses with $m_{H}\in\left\{120,125,130\right\}\GeV$. For mass values between these points, the signal pdf parameters are linearly interpolated. The sample for deriving the signal model is obtained by combining the four dominating SM Higgs boson production modes simulated with the \MGvATNLO generator, reweighted to the \NNLOPS prediction for gluon fusion. Examples of signal pdfs used for the fiducial cross section measurement for different \sigMD categories are shown in Fig.~\ref{fig:sigMod}.

\subsection{Background model}\label{subsec:backgroundModel}
{\tolerance=500 The determination of the background model uses a strategy called the discrete profiling method~\cite{DiscreteProfilingMethod}. Since the exact form of the background is not known, a number of different functions to fit the smoothly falling shape of the \mgg~ background are tried, and their choice is treated as a discrete nuisance parameter in the maximum likelihood fit used to extract the differential cross section. The background fit is done in each reconstruction-level observable and reconstruction category bin over the range of $100\GeV<\mgg<180\GeV$.
The families of functions considered for $B\left(\mgg|\vec{\theta}_{\mathrm{B}}\right)$ are exponentials, power-law functions, Laurent polynomials, and polynomials in the Bernstein basis. A penalty term equal to the number of degrees of freedom is added to the double negative log-likelihood, hence allowing the fit to determine the number of degrees of freedom for all considered families. \par}

\section{Systematic uncertainties}\label{sec:systematic_uncertainties}
In the following, the sources of systematic uncertainty that are considered in the analysis are described. Since the discrete profiling method is used to derive both the shape and the normalization of the background, the systematic uncertainty related to the background description is included in the final results as a component of the statistical uncertainty.
Uncertainties that influence the shape of the signal model are implemented as nuisance parameters built into the signal model. Systematic variations that leave the shape of the \mgg distribution unaltered are taken into account as log-normal nuisances affecting only the predicted signal rate. If a systematic variation influences the event selection and thus changes the event yields in several categories, it is considered as a category migration systematic uncertainty, i.e. as a correlated log-normal uncertainty in the category yields. The systematic uncertainty arising from the imperfect modelling of the pileup distribution in the simulation is incorporated in to the uncertainty of the affected quantities (chiefly the photon MVA identification and vertex identification).
Uncertainties of a theoretical nature are fully correlated among all years of data taking. The overall normalization of the particle-level bins is kept constant while evaluating the effect of the theoretical systematic uncertainties. This means that their effect is only considered as causing event migrations between reconstruction level bins and categories within the particular year of data taking. All yield and simple log-normal systematic variations of an experimental nature are treated as independent across years. The shape systematic uncertainties related to the scale and smearing corrections of the photon energy are fully correlated among years.

The theoretical systematic uncertainties include:
\begin{itemize}
	\item \textit{PDF uncertainty:} this accounts for the imperfect knowledge of the parton distribution functions of the proton.
	These PDF uncertainties are derived through the relative yield variation per particle-level bin after reweighting the events with several alternative sets of weights. These weights are derived according to the prescriptions in \textsc{pdf4lhc}~\cite{Butterworth:2015oua} and \textsc{nnpdf3.1}~\cite{NNPDF31} for 2017 and 2018, and \textsc{nnpdf3.0}~\cite{NNPDF3} with the \textsc{MC2hessian}~\cite{MC2Hessian} procedure for 2016. This systematic uncertainty is treated as fully correlated between 2017 and 2018 and uncorrelated between 2016 and the other years of data taking. The largest effect on the cross section from reweighting across all sets of PDFs is at the per-mille level;
	\item \textit{\alpS uncertainty:} this uncertainty is related to the variation of the value of the strong force coupling constant used for the evaluation of the parton distribution functions, following the prescription in Ref.~\cite{Butterworth:2015oua}. The impact of the nuisance parameter associated with this uncertainty in the cross section is at the per-mille level, similar to the PDF uncertainty;
	\item \textit{Renormalization and factorization scales uncertainties:} the migration uncertainty between analysis bins caused by the scale variations are estimated by moving both scales by a factor of 2, excluding the $\left(2, 1/2\right)$ and $\left(1/2 ,2\right)$ combinations, as they are unphysical. The associated nuisance parameters can have an impact of up to 0.5\% on the measured cross section.
\end{itemize}
The experimental uncertainties affecting the signal shape include:
\begin{itemize}
	\item \textit{Photon energy scale and resolution:} this is the uncertainty related to the scale of the photon energy in data and its smearing correction in simulation~\cite{HggMass}. The variations are evaluated using \Zee~events by varying the \RNINE~distribution and the selection criteria. The statistical uncertainty from the \Zee sample used to derive the scale and smearing corrections is also taken into account. The uncertainty amounts to 0.05--0.15\% for most of the photons, while they can go up to 3\% in some categories.
	\item \textit{Photon energy scale nonlinearity:} the difference between the linearity of the energy scale in data and simulation is covered by this variation. It is estimated using Lorentz-boosted \Zee events and it amounts to 0.2\%.
	\item \textit{Cluster shape uncertainties:} the energy regression used to correct the photon energy, as well as the data-to-simulation corrections of the photon identification MVA input variables, are all evaluated on electrons from \PZ boson decays. Therefore an uncertainty in the shower shape variables that accounts for the difference between electrons and photons has to be taken into account in an analysis that considers photons as signal. The resulting uncertainty in the photon energy scale is 0.01--0.15\%, depending on the $\abs{\eta}$ and \RNINE~values of the photon \cite{CMS:2014afl}.
	\item \textit{Nonuniformity of light collection:} this uncertainty is related to the modelling of the light collection depending on the emission depth in the ECAL crystals. It amounts to 0.16--0.25\% for photons with $\RNINE>0.96$. For low-\RNINE~photons, the uncertainty is below 0.07\%~\cite{HggMass}.
	\item \textit{Modelling of the material upstream of ECAL:} the showering behaviour of photons is influenced by the material upstream of the ECAL. This systematic variation is associated with the uncertainty in the tracker material model. Its impact on the photon energy scale is evaluated by varying the amount of material in the simulation by 5\%. The uncertainty amounts to 0.02--0.05\% for central photons, increasing to 0.25\% for photons in the endcap \cite{CMS:2018piu}.
	\item \textit{Vertex assignment:} this variation covers the uncertainty in the fraction of events that have a reconstructed vertex that lies within $1\cm$ of the true vertex in data and simulation. The signal model includes a nuisance parameter that allows the number of events in which the vertex is at a distance of more than $1\cm$ from the true vertex to vary by ${\pm} 2\%$. The corresponding nuisance parameter can lead to variations of the cross section by a few parts per mille.
\end{itemize}
The systematic uncertainties that only affect the event yield are:
\begin{itemize}
	\item \textit{Integrated luminosity:} the uncertainty in the measured value of the integrated luminosity and its correlation across the years is applied following the CMS recommendations in Refs.~\cite{CMSlumi2016, CMSlumi2017, CMSlumi2018}. Its magnitude is 1.2\% for 2016, 2.3\% for 2017, and 2.5\% for 2018, while the overall uncertainty for the 2016--2018 period is 1.6\%.
	\item \textit{Per-photon energy resolution:} the impact on the per-category event yields from the per-photon energy resolution mismodelling in simulation is evaluated by shifting the $\sigma_{E}/E$ value for each photon up and down by 2\% in all events and repeating the full analysis. The magnitude of the per-event shift covers the data-to-simulation mismatch in the invariant mass resolution for \Zee events. This uncertainty has one of the largest systematic effects on the measurement, which can be up to 2\% in certain cross sections.
	\item \textit{Jet energy scale and smearing corrections:} the uncertainty in the jet energy scale is at the percent level and depends on the \pt and $\eta$ of the jet.
	It is propagated to the migrations between jet bins by varying the corrections within their uncertainties. The systematic uncertainties related to jet energy scale and smearing are partly correlated across years~\cite{JetsInRun2}.
	For observables involving jets, this uncertainty can impact the cross section measured as a function of those by up to 20\%.
	\item \textit{Trigger efficiency:} the trigger efficiency is measured with \Zee~ events using the tag-and-probe method. Its uncertainty affects bin migrations and has an effect of less than 1\%. An additional source of uncertainty is included to cover the gradual shift of the inputs to the ECAL L1 trigger timing for values of $\abs{\eta}>2$ in 2016 and 2017. This affected mostly jets but also photons. The measured cross sections are barely affected by this systematic uncertainty, its effect being below the permille level.
	\item \textit{Photon identification MVA score:} the input variables to the photon identification MVA are corrected in simulation to match data as described in Section~\ref{subec:idMVAcorrections}. This leads to a better agreement of the photon identification MVA score between simulation and data. The residual mismatch is covered by evaluating the uncertainty of the correction method and shifting the value of the photon identification BDT accordingly, as shown in Fig.~\ref{fig:phoIdMVA}. The uncertainty of the correction method is evaluated by varying the size of the data set used for its derivation. Finally, the uncertainty in the category yields is evaluated by propagating the variation of the photon identification MVA through the event selection.
	The nuisance parameters associated with this systematic variation have an effect of up to 1\% on the measured cross section. Summing these effects in quadrature yields an impact of 1.5\% on the inclusive fiducial cross section measurement.
	\item \textit{Photon preselection:} the uncertainty in the efficiency of the photon preselection is evaluated by varying the signal and background shapes that are used to compute the preselection efficiency. This is done in data and simulation using \Zee~ tag-and-probe events. The resulting variation is propagated to the corresponding scale factor. Its magnitude is below 1\%.
	\item \textit{Missing transverse momentum:} the uncertainty in the missing transverse momentum is computed by shifting the scale and resolution of the particles included in the computation of the missing transverse momentum. This leads to yield variations for bins in observables involving missing transverse momentum. This systematic uncertainty affects only the measurement of the \Hgg cross section differential in \ptmiss. Its impact on the cross section is negligible.
	\item \textit{Pileup jet identification:} the variation in the pileup jet identification score leads to bin migration in analysis bins related to jet variables. The resulting effect on the results presented here is negligible.
	\item \textit{Lepton identification:} the electron and muon identification efficiencies are varied within their uncertainties leading to bin migrations in lepton observables. These are propagated through the full analysis chain to calculate the related impact on the measured cross sections. This uncertainty only affects the measurement of the \Hgg cross section as a function of the number of leptons produced alongside the diphoton system. The magnitude of its effect is roughly 5\%.
	\item \textit{b tagging efficiency:} the \PQb jet tagging efficiency for the working point used in this analysis is varied within its uncertainty. This leads to bin migrations for observables involving \PQb jets and in the phase spaces with a requirement on the number of \PQb jets. This only affects the cross section measurement with respect to the number of \PQb tagged jets. The effect of the corresponding nuisance parameter on the cross section is between 5 and 20\%.
\end{itemize}

A summary of the dominant systematic impacts on the inclusive fiducial \Hgg cross section measurement are given in Table~\ref{tab:systImpacts}.

\begin{table}[htb]
	\centering
	\topcaption{Breakdown of the systematic uncertainties in the inclusive fiducial cross-section measurament. The impacts on the measured inclusive fiducial \Hgg cross section by varying the nuisance parameters for the dominating sources of systematic uncertainties by one standard deviation are given. The distinct contributions for systematic uncertainties that were split by category or year of data taking are added in quadrature for simplicity. Theoretical uncertainties summarizes the theoretical systematic uncertainties given above.}
	\begin{tabular}{lc}
		\hline
		Systematic uncertainty&Impact\\\hline
		{Per-photon energy resolution}&$2.0\%$\\
		{Integrated luminosity}&$1.5\%$\\
		{Photon identification MVA score}&$1.5\%$\\
		{Photon energy scale and resolution}&$1.0\%$\\
		{Photon preselection}&$0.6\%$\\
		{Theoretical uncertainties}&$0.5\%$\\
		{Vertex assignment}&$0.1\%$\\
		\hline
	\end{tabular}
	\label{tab:systImpacts}
\end{table}

\section{Results}\label{sec:results}
In this section, the results of the analysis are presented. The cross section measured for the fiducial region defined in Table~\ref{tab:phaseSpace} is:
\begin{equation}
\sigFid=73.4_{-5.3}^{+5.4}\stat_{-2.2}^{+2.4}\syst\fb=73.4_{-5.9}^{+6.1}\fb.
\end{equation}

\begin{figure}[btp]
\centering
  \includegraphics[width=0.55\textwidth]{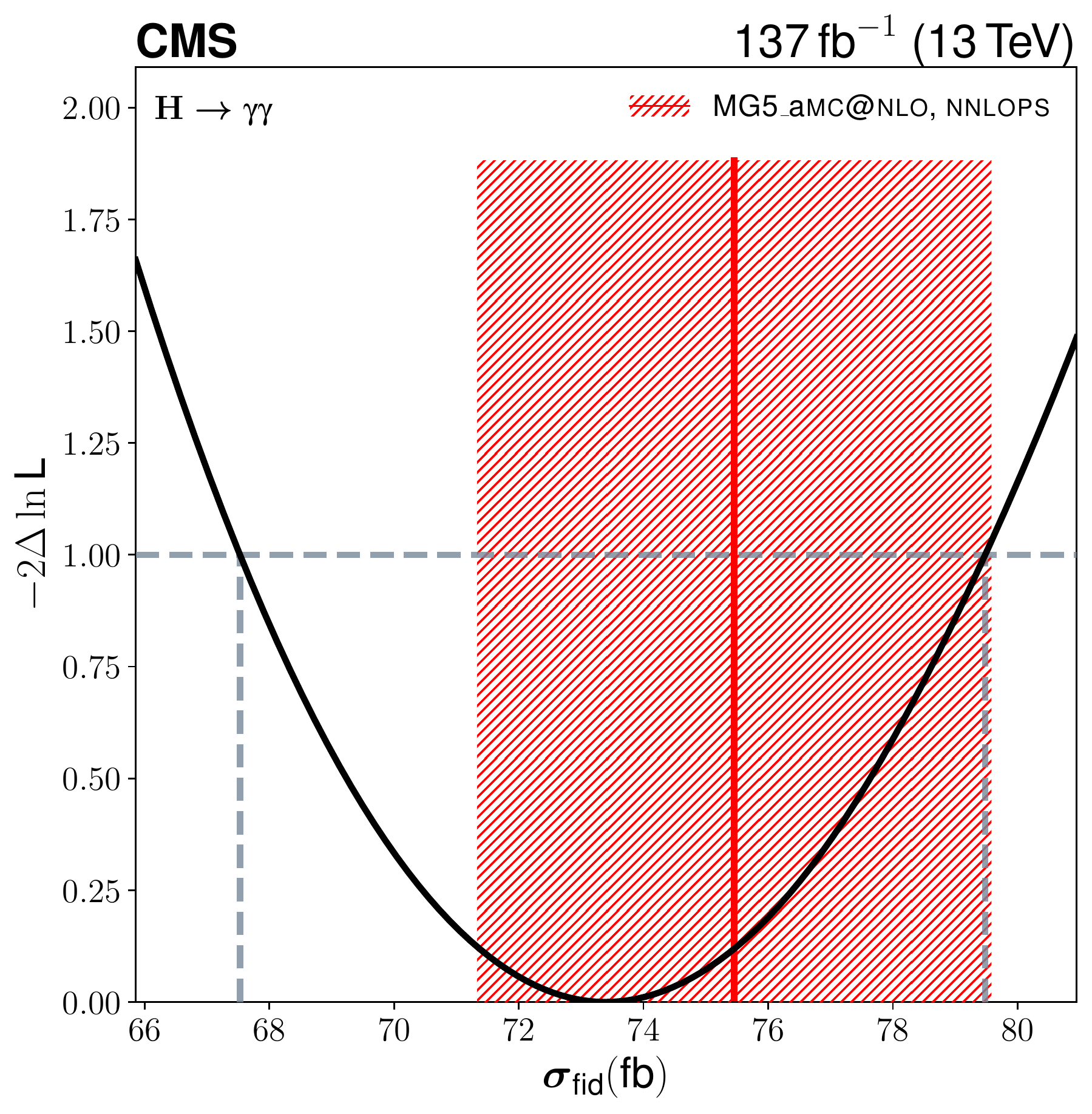}
  \caption{The black line shows the scan of $q\left(\Delta\vec{\sigma}\right)=-2\Delta\ln\mathrm{L}$ for the \Hgg~ cross section in the fiducial region. The red line shows the theoretical prediction for the SM, obtained with \MGvATNLO. Its uncertainty is shown as the hatched area.}
  \label{fig:resultsFid}
\end{figure}

\begin{figure}[t]
\centering
  \includegraphics[width=0.55\textwidth]{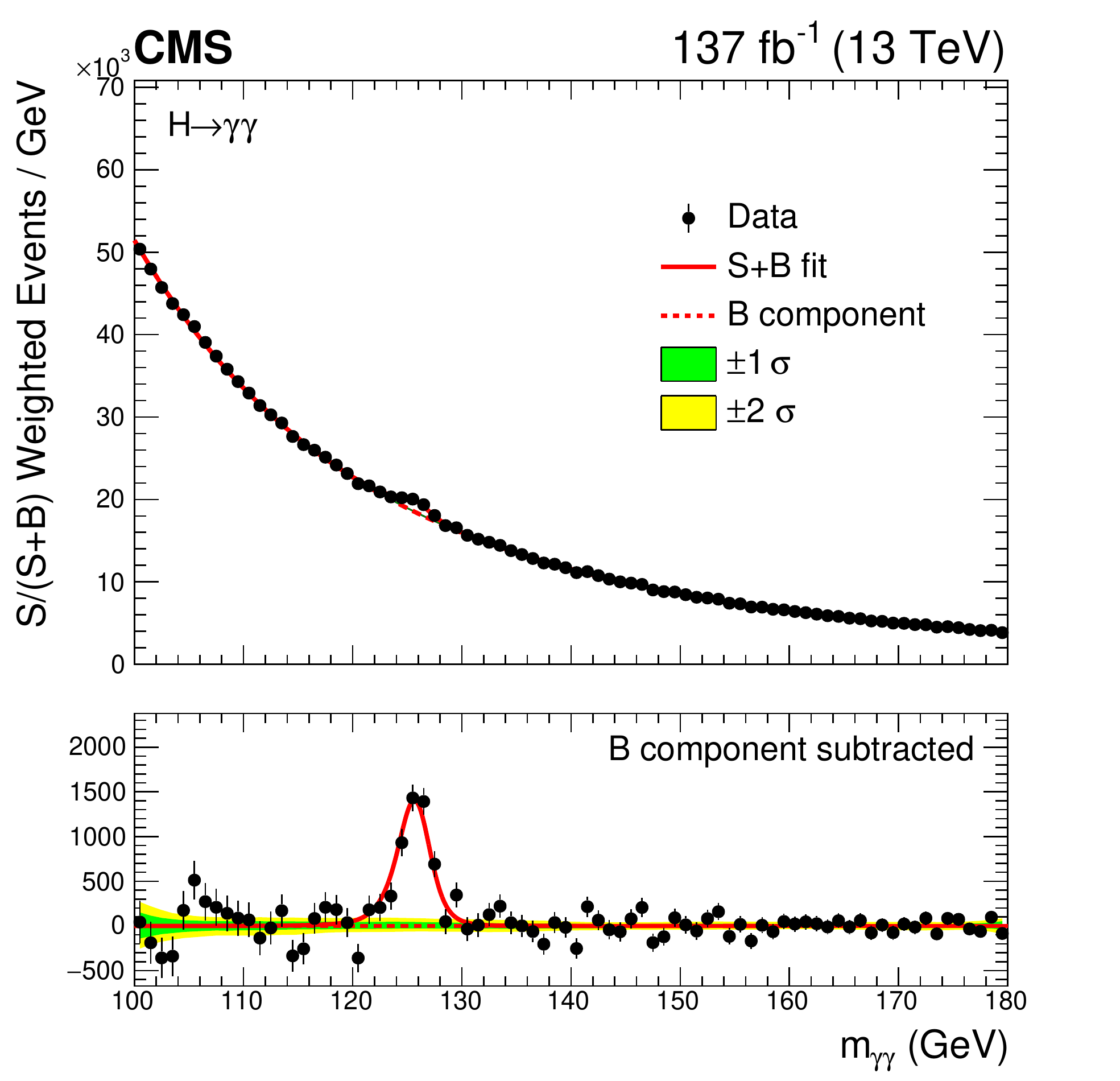}
  \caption{Diphoton invariant mass distribution with combining all categories used for the inclusive fiducial cross section measurement. The displayed \mgg histogram and signal+background hypothesis (red line) represent their sums across all categories weighted by their respective $S/(S+B)$ ratio. In the lower panel, the \mgg histogram subtracting the background component, as estimated by the background pdf, is shown.}
  \label{fig:mggHistSpB}
\end{figure}

The nominal theoretical predictions for all results presented in this section have been extracted from \MGvATNLO (version 2.6.5), reweighted to match the \NNLOPS prediction for gluon fusion and interfaced with \PYTHIA8 (version 8.240) using the CP5 tune.

Figure~\ref{fig:resultsFid} shows the likelihood scan for the fiducial cross section and the nominal theoretical prediction from \MGvATNLO reweighted to match the \NNLOPS prediction for gluon fusion. Figure~\ref{fig:mggHistSpB} shows the \mgg histogram with the hypotheses from the signal and background models at the best fit point. The events from all categories are added, weighted by the $S/(S+B)$ ratio for the respective category. The uncertainty in the measurement is dominated by the statistical component of about 7\%, while the systematic uncertainty amounts to about 3\%. The nominal theoretical prediction of the fiducial cross section is $75.4\pm4.1\fb$. The best fit value is consistent with the SM value within one standard deviation. The acceptance of the fiducial phase space with respect to the full phase space amounts to 60.3\% as predicted by the \MGvATNLO generator including the \NNLOPS reweighting for gluon fusion. The result for the \Hgg cross section in the full phase space can be found in Ref.~\cite{HIG-19-015}.

\begin{figure}[htbp]
\centering
  \includegraphics[width=\textwidth]{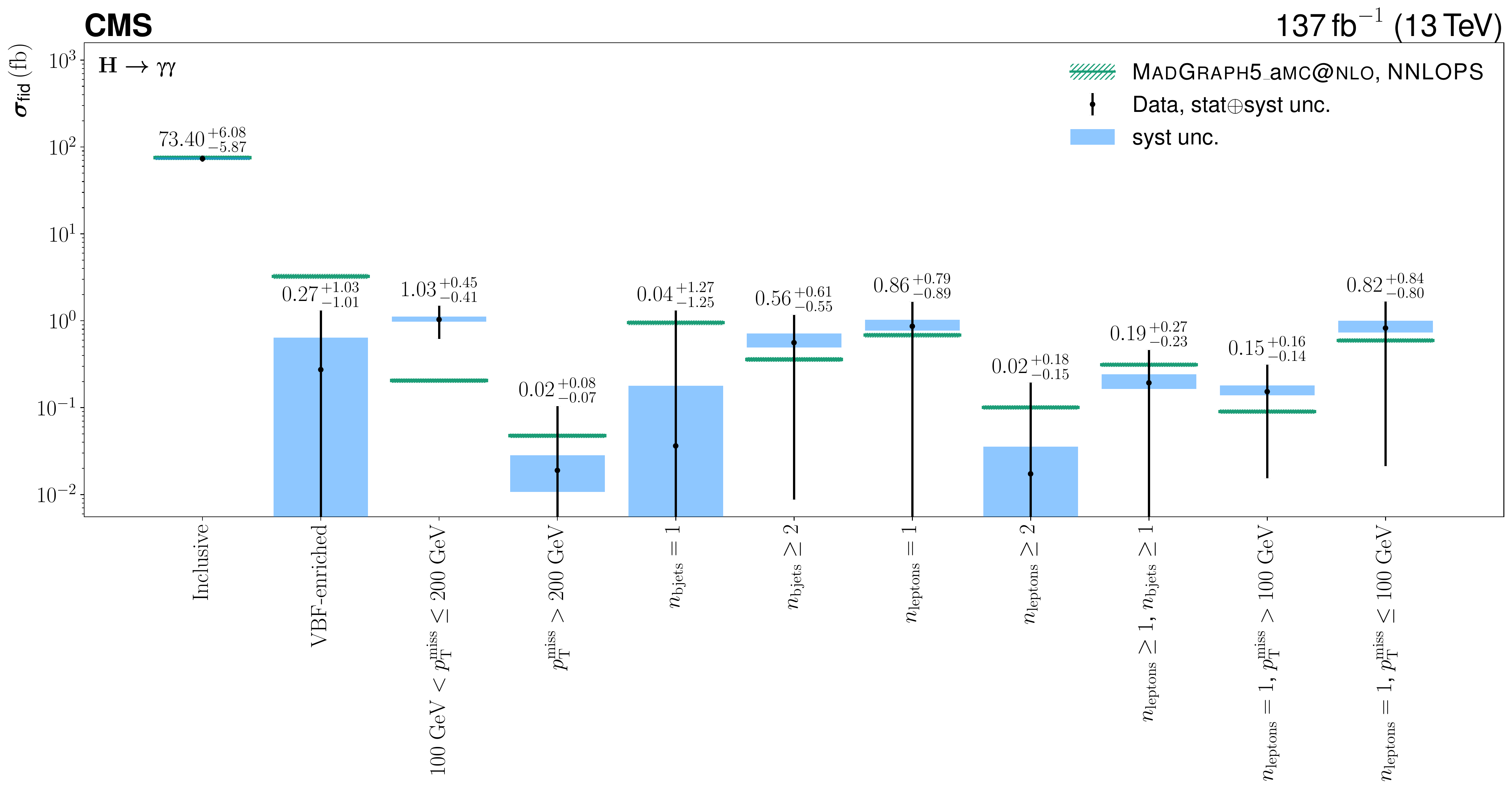}
  \caption{The \Hgg cross section in dedicated regions of the fiducial phase space. Their selection criteria on top of the fiducial requirements are indicated on the plot. The prediction from \MGvATNLO including the \NNLOPS reweighting, with its uncertainty from acceptance variation due to PDF, \alpS, and scale uncertainties, as well as cross section and branching fraction uncertainties, is shown. The systematic uncertainty in the measured value is shown as a blue band and the full systematic$\oplus$statistical uncertainty is shown as the error bar, where $\oplus$ stands for the sum in quadrature.}
  \label{fig:fidXSSummPlot}
\end{figure}

\begin{figure}
\centering
  \includegraphics[width=0.7\textwidth]{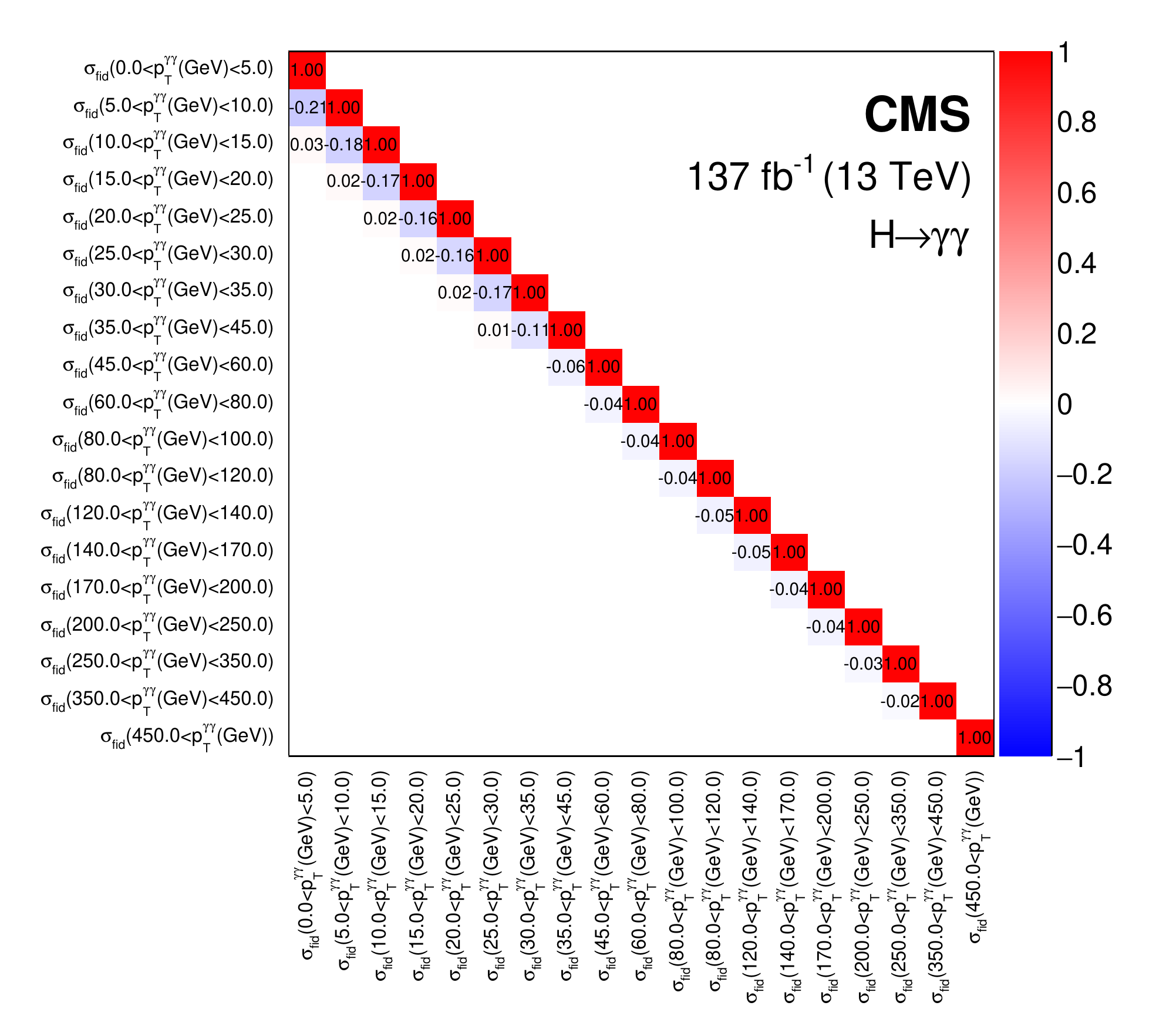}\\
  \includegraphics[width=0.6\textwidth]{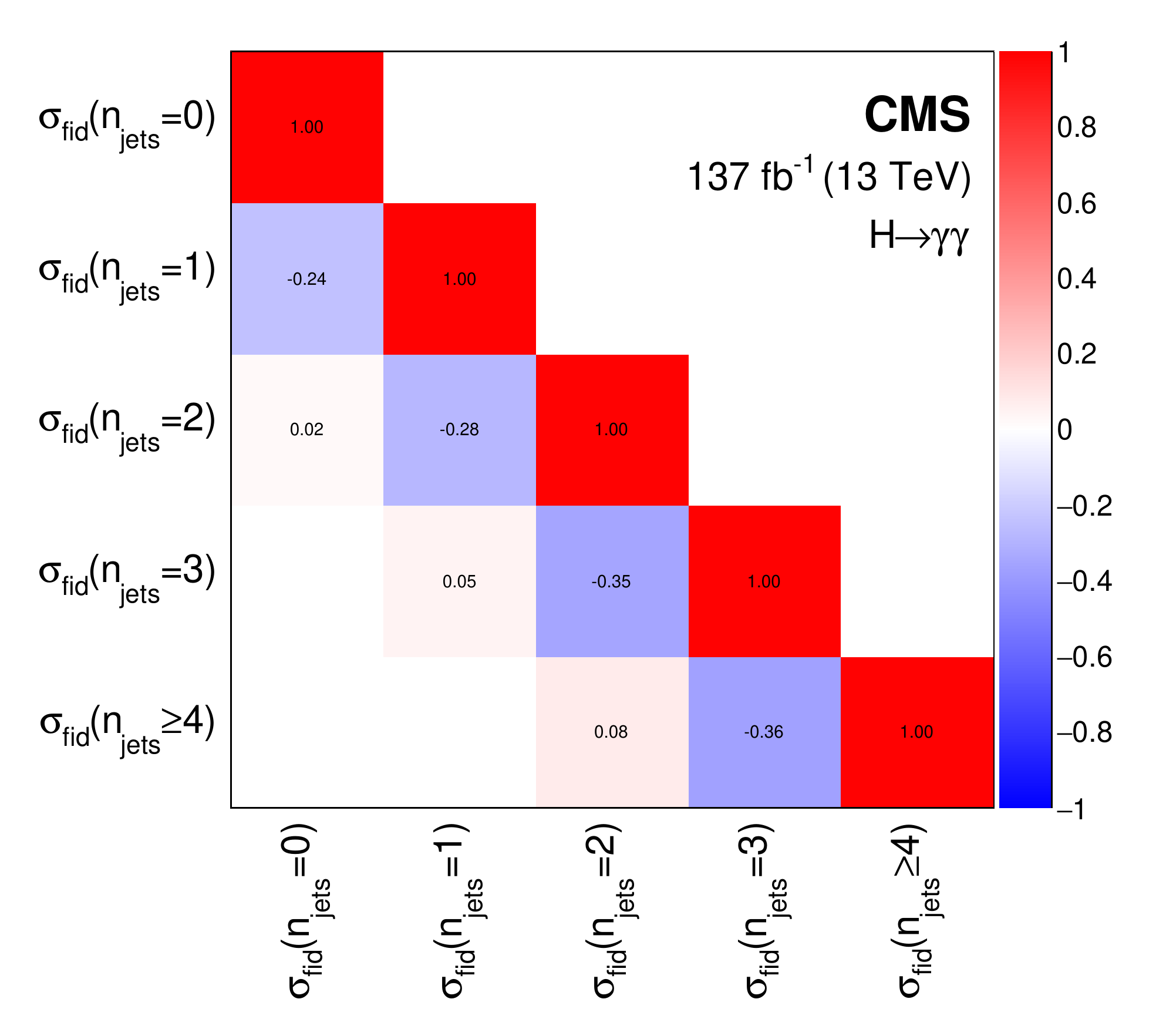}
  \caption{The correlation matrices for the cross sections \sigFid per particle-level bin for \ptgamgam (upper), and \njets (lower), as given in Table~\ref{tab:obsBins}, extracted from the simultaneous maximum likelihood fit for the cross sections.}
  \label{fig:corrMatrices}
\end{figure}

The uncertainty in the prediction of the fiducial inclusive \Hgg cross section, shown as the hatched area in Fig.~\ref{fig:resultsFid} and of the cross section in the dedicated phase space regions, shown as the hatched areas in Fig.~\ref{fig:fidXSSummPlot}, are the combination of several components. The dominating one of these is the uncertainty in the Higgs boson production cross section, with the second largest contribution coming from the uncertainty in the \Hgg branching fraction, both taken from~\cite{deFlorian:2016spz}. The remaining contributions come from the variation of the fiducial acceptance due to varying the set of PDF replicas~\cite{NNPDF31}, the value of \alpS by 0.002 around its nominal values of 0.118 and finally the variation of renormalization and factorization scales by a factor of 2, while excluding the $\left(2,1/2\right)$ and $\left(1/2,2\right)$ variations. The interference between the \Hgg signal and the continuous diphoton background~\cite{IntHggDipho} has not been taken into account for the predictions presented here. The measured and predicted cross sections for selected special regions of the fiducial phase space are shown in Fig.~\ref{fig:fidXSSummPlot}. Two examples of correlation matrices, calculated using Eq.~\eqref{form:testStat}, are shown in Fig.~\ref{fig:corrMatrices}. The correlation matrices for all other observables are available in the HEPData record for this analysis~\cite{hepdata}.

\begin{figure}[p]
\centering
  \includegraphics[width=0.48\textwidth]{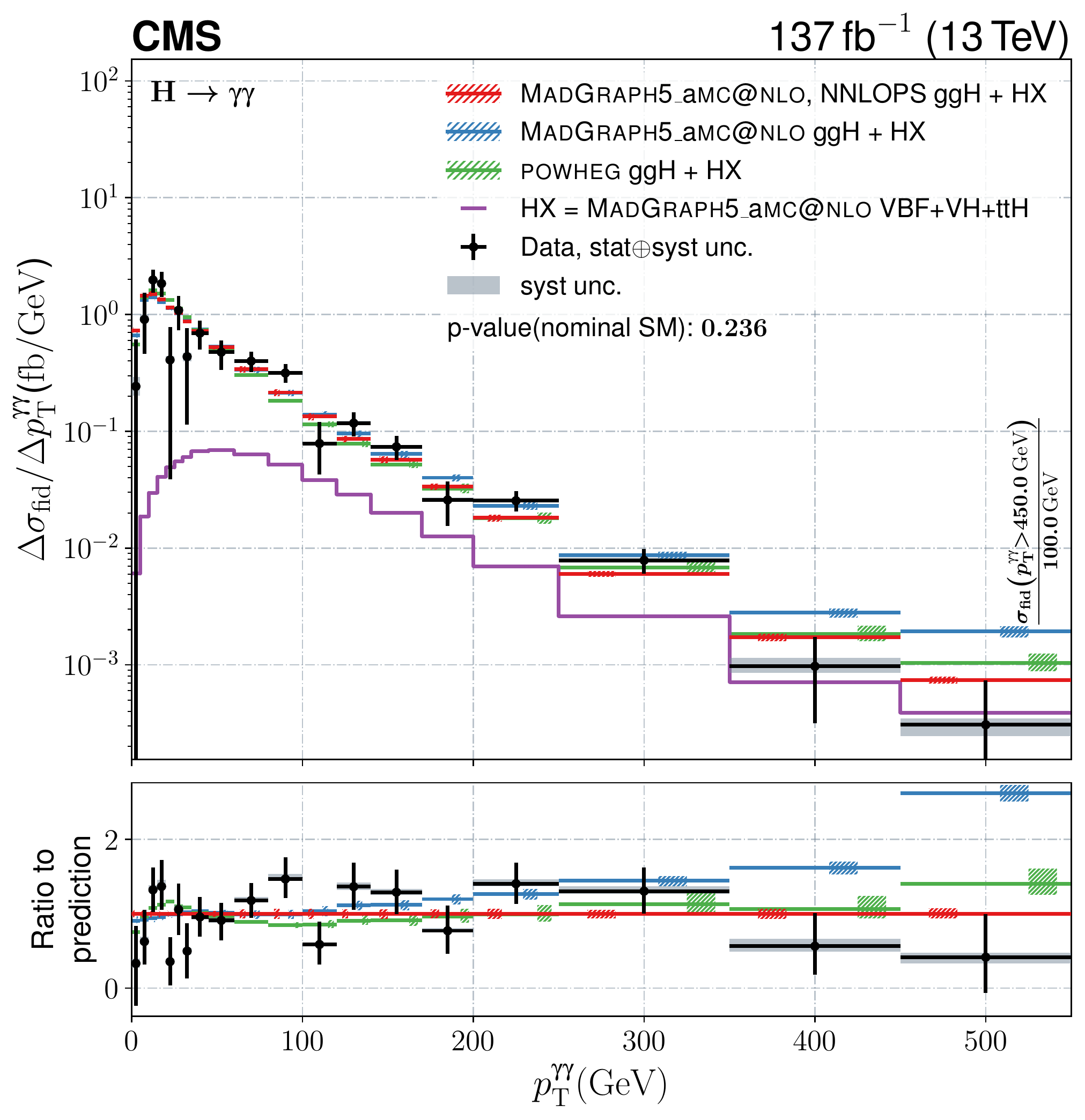}
  \includegraphics[width=0.48\textwidth]{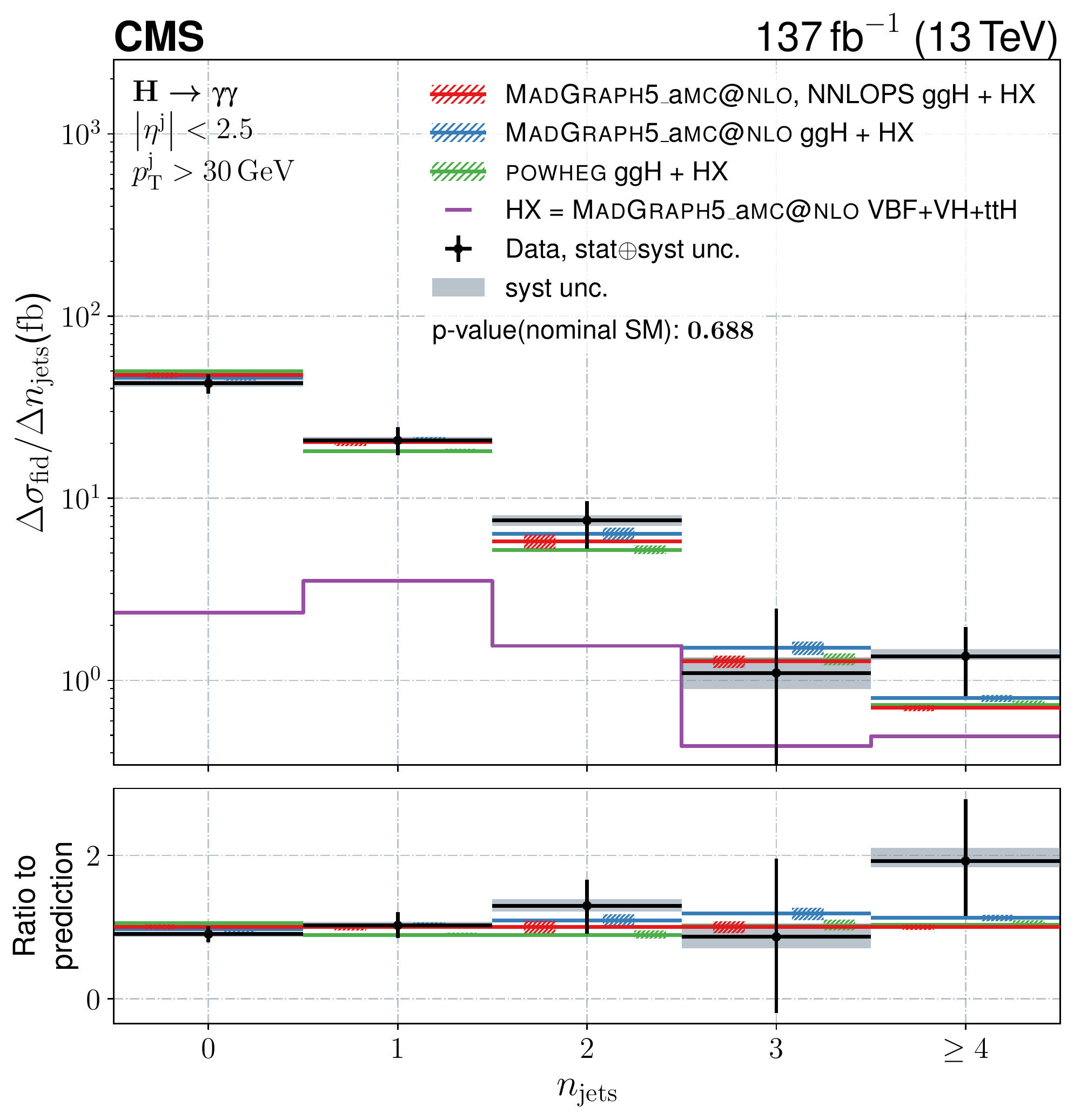}\\
  \includegraphics[width=0.48\textwidth]{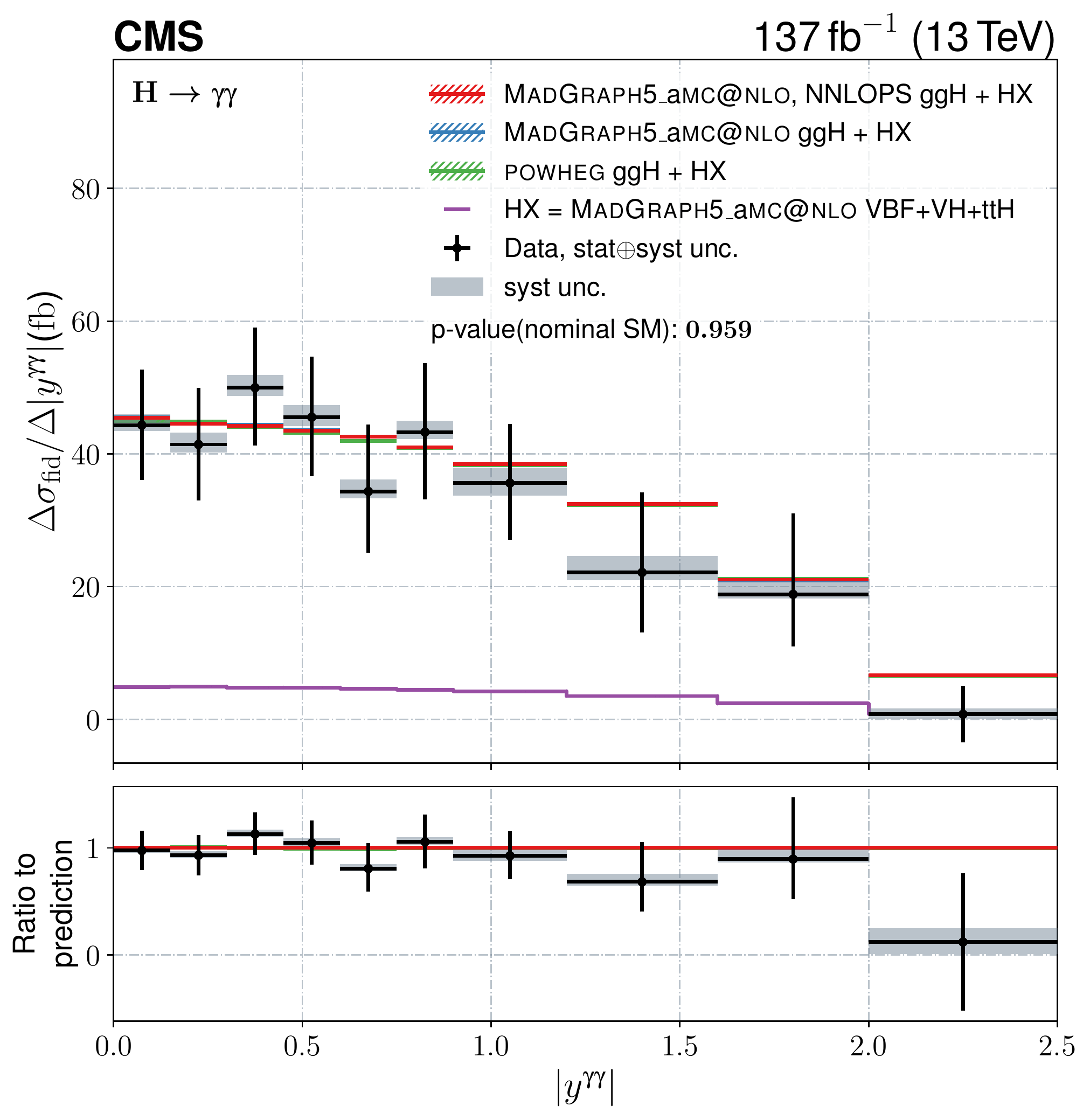}
  \includegraphics[width=0.48\textwidth]{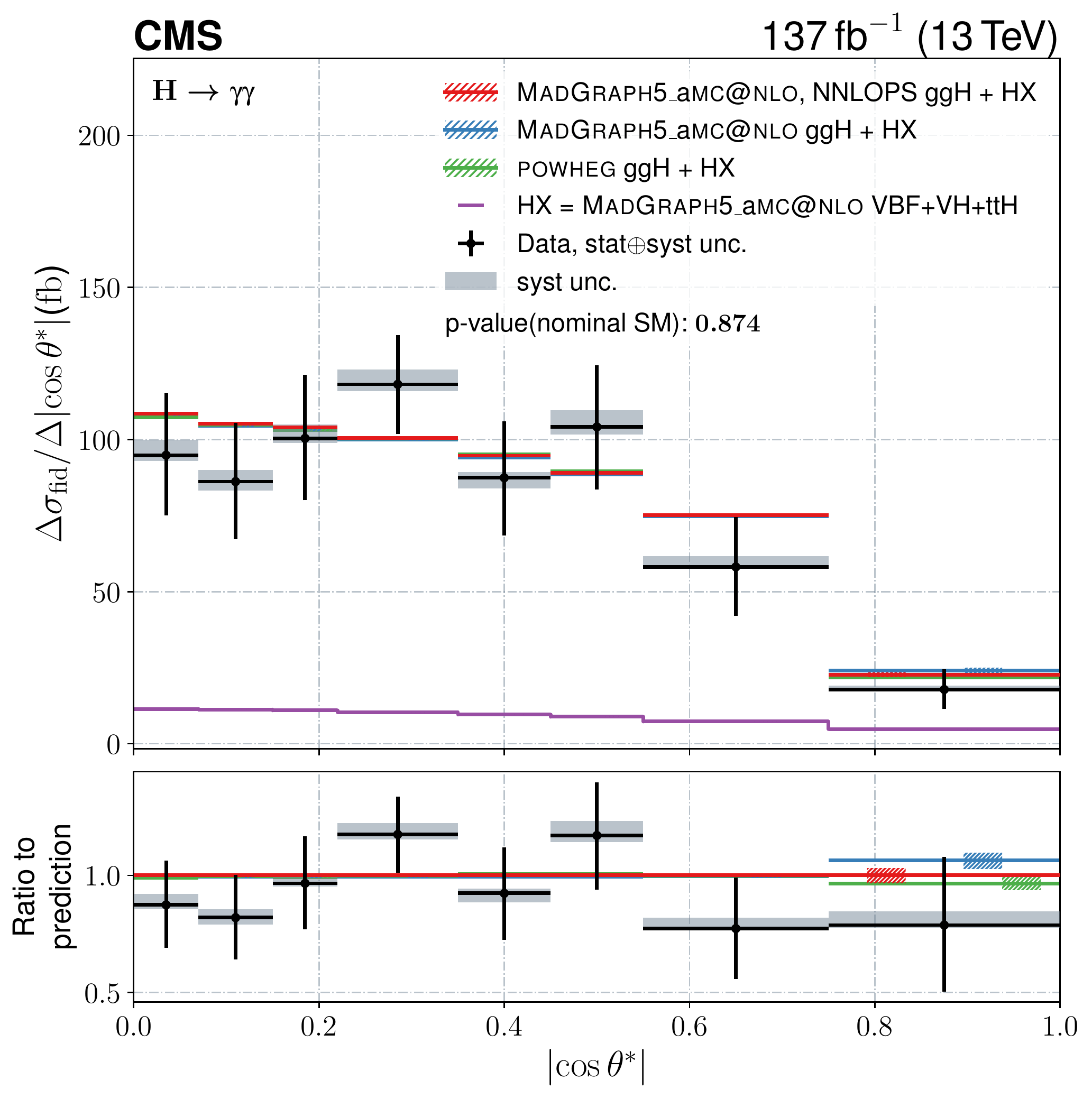}
  \caption{Differential fiducial cross sections for \ptgamgam, \njets, $\abs{y^{\PGg\PGg}}$, and $\abs{\cos\theta^{\ast}}$. \resultsCaption}
  \label{fig:resultsPtNjetsCosTsRapi}
\end{figure}

\begin{figure}
\centering
  \includegraphics[width=0.49\textwidth]{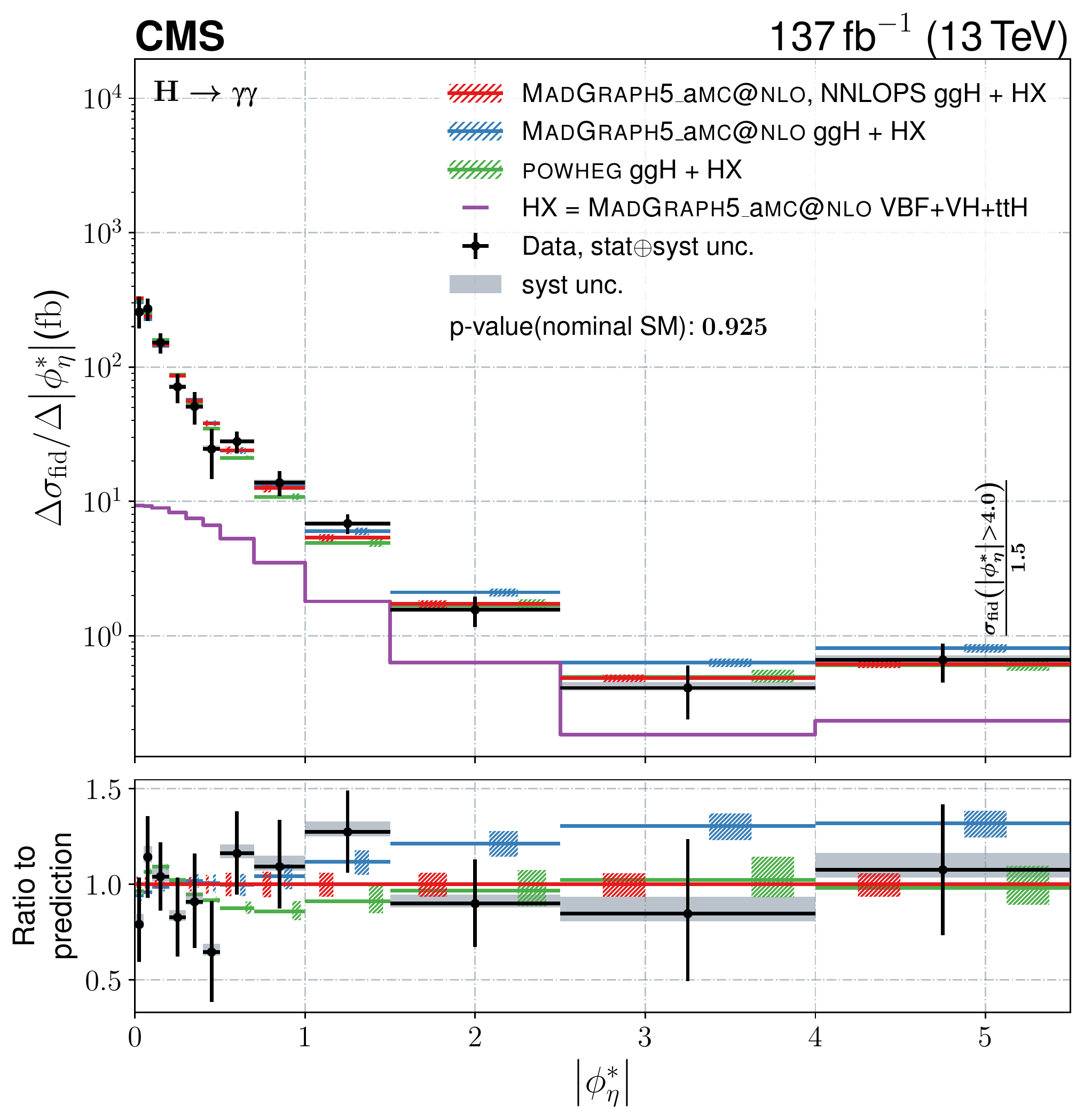}
  \includegraphics[width=0.49\textwidth]{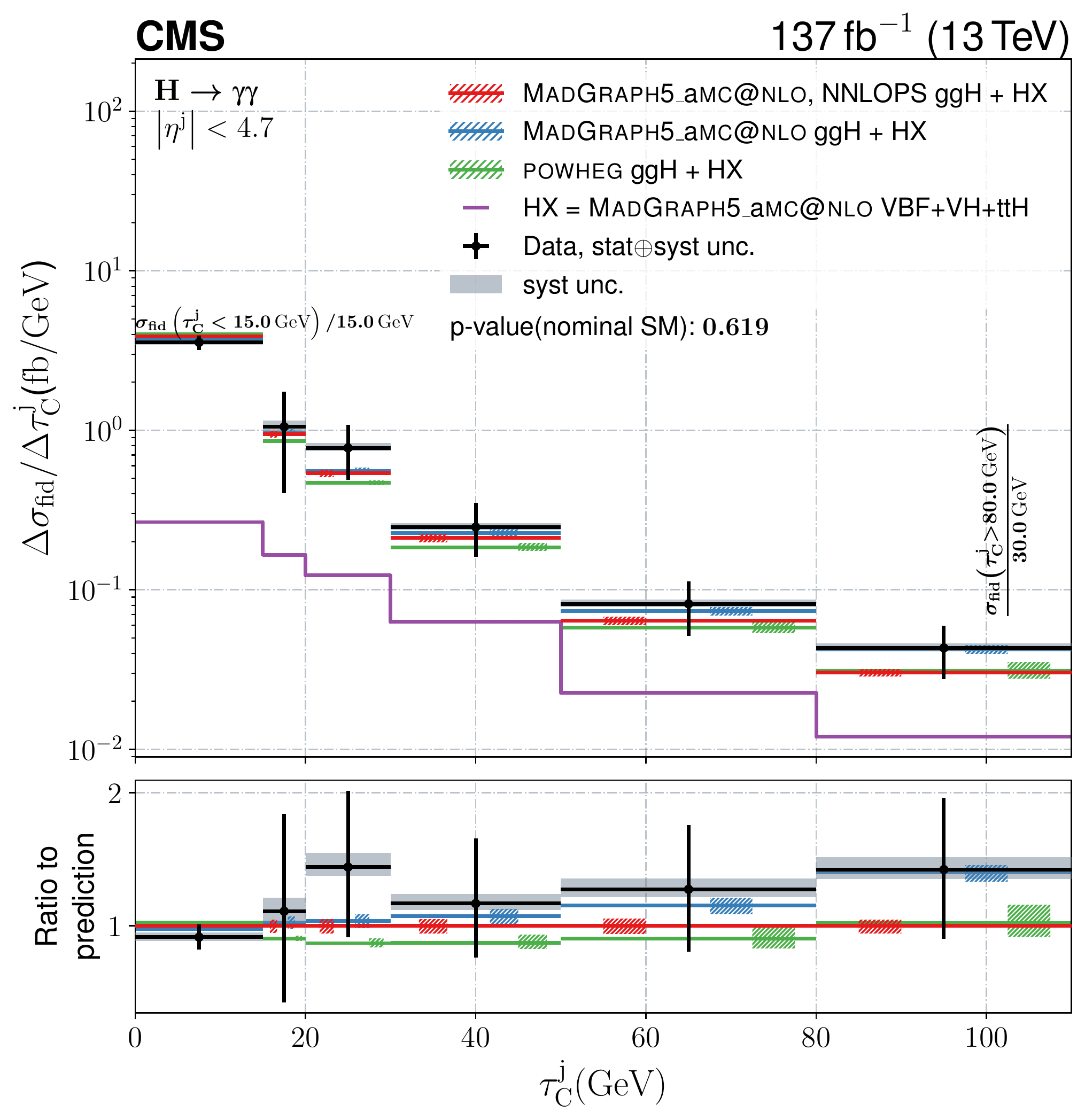}\\
  \includegraphics[width=0.49\textwidth]{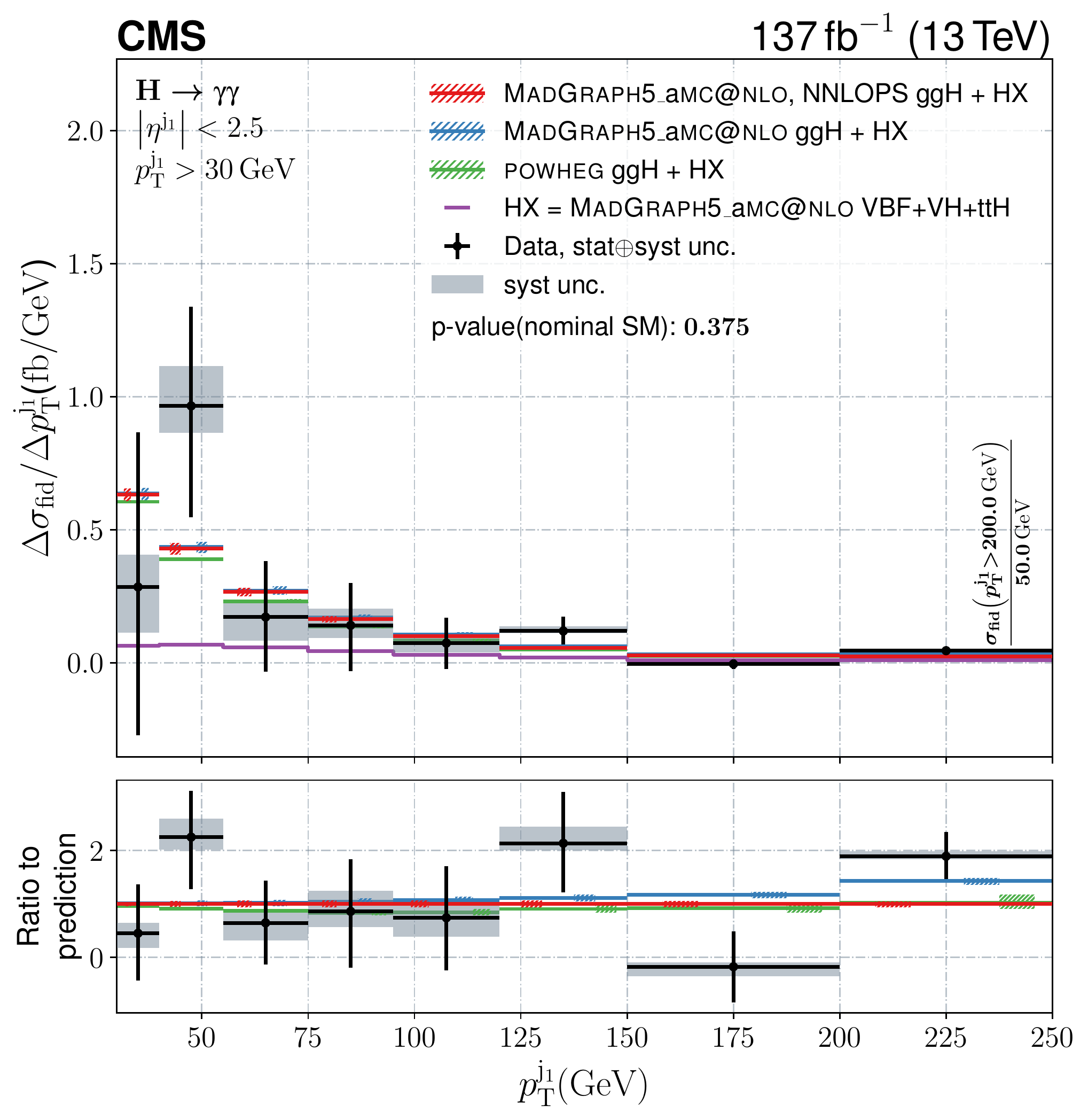}
  \includegraphics[width=0.49\textwidth]{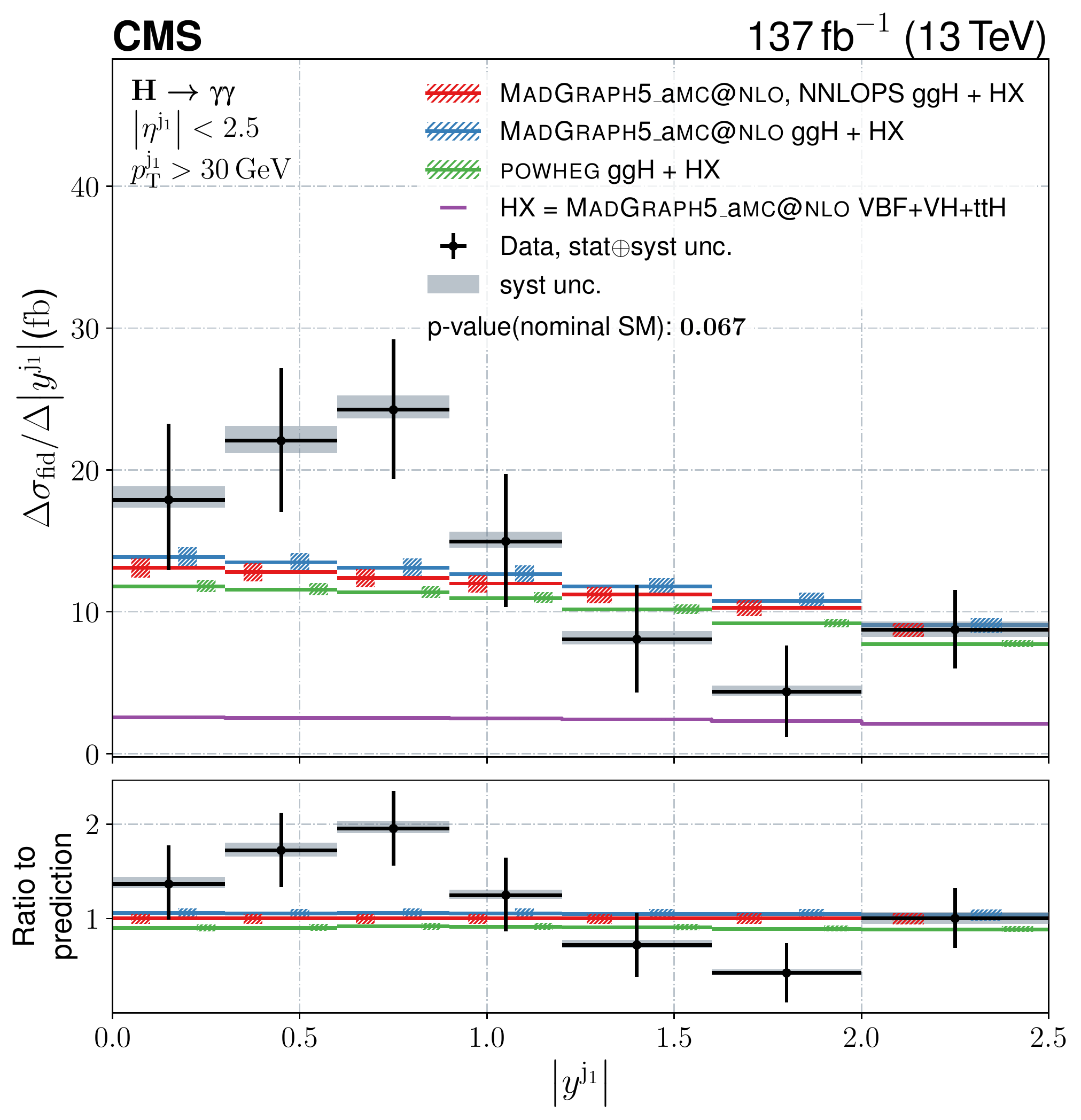}
  \caption{Differential fiducial cross section for $\abs{\phietaS}$, \taucj, $\ptjO$, and $\abs{y^{\mathrm{j}_{1}}}$. \resultsCaptionShort~The first bin in the upper right plot shows the cross section for $\taucj<15~\GeV$. This is marked in the plot together with the corresponding normalization.}
  \label{fig:resultsAbsPhiS}
\end{figure}

\begin{figure}[p]
\centering
  \includegraphics[width=0.49\textwidth]{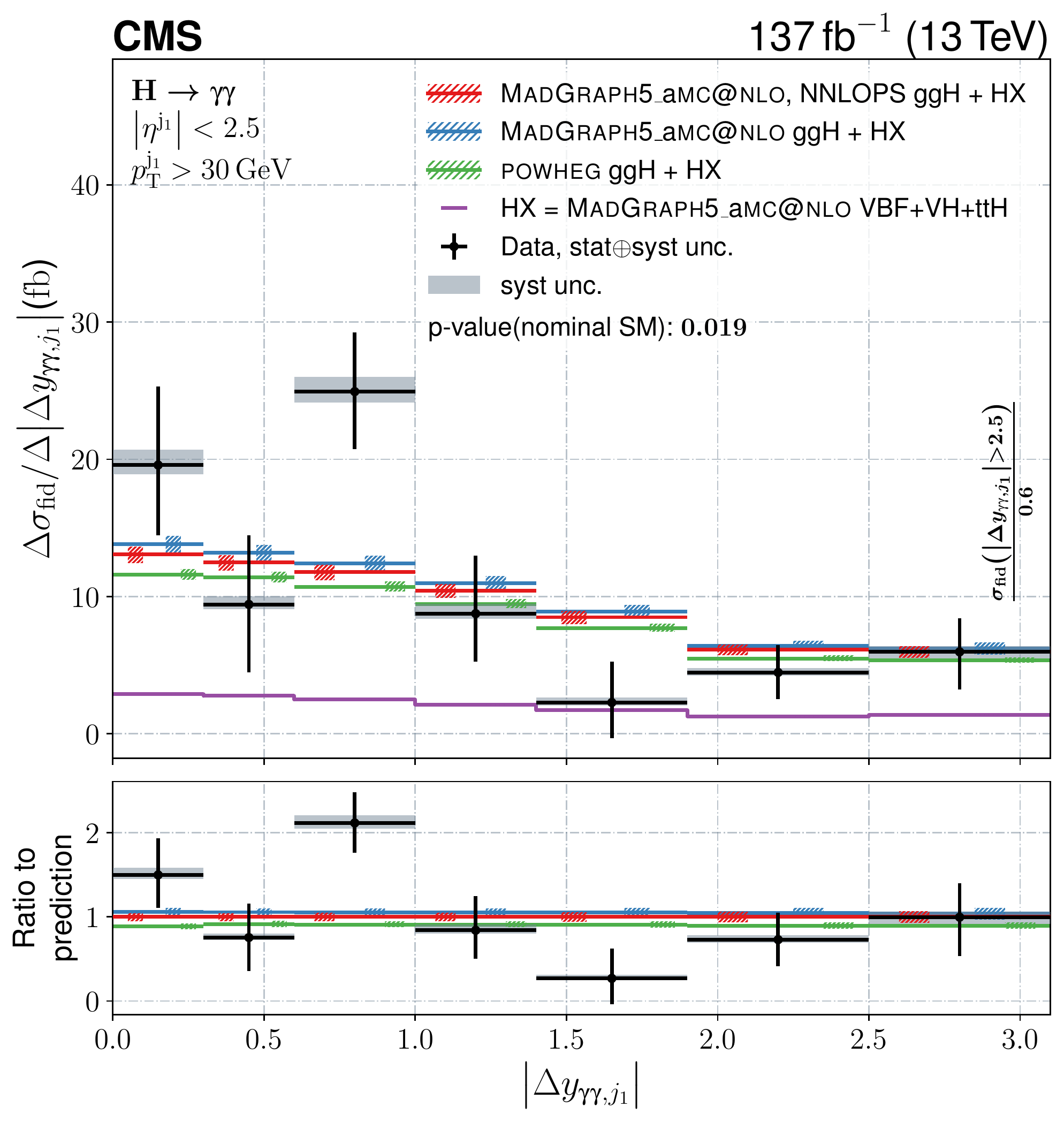}
  \includegraphics[width=0.49\textwidth]{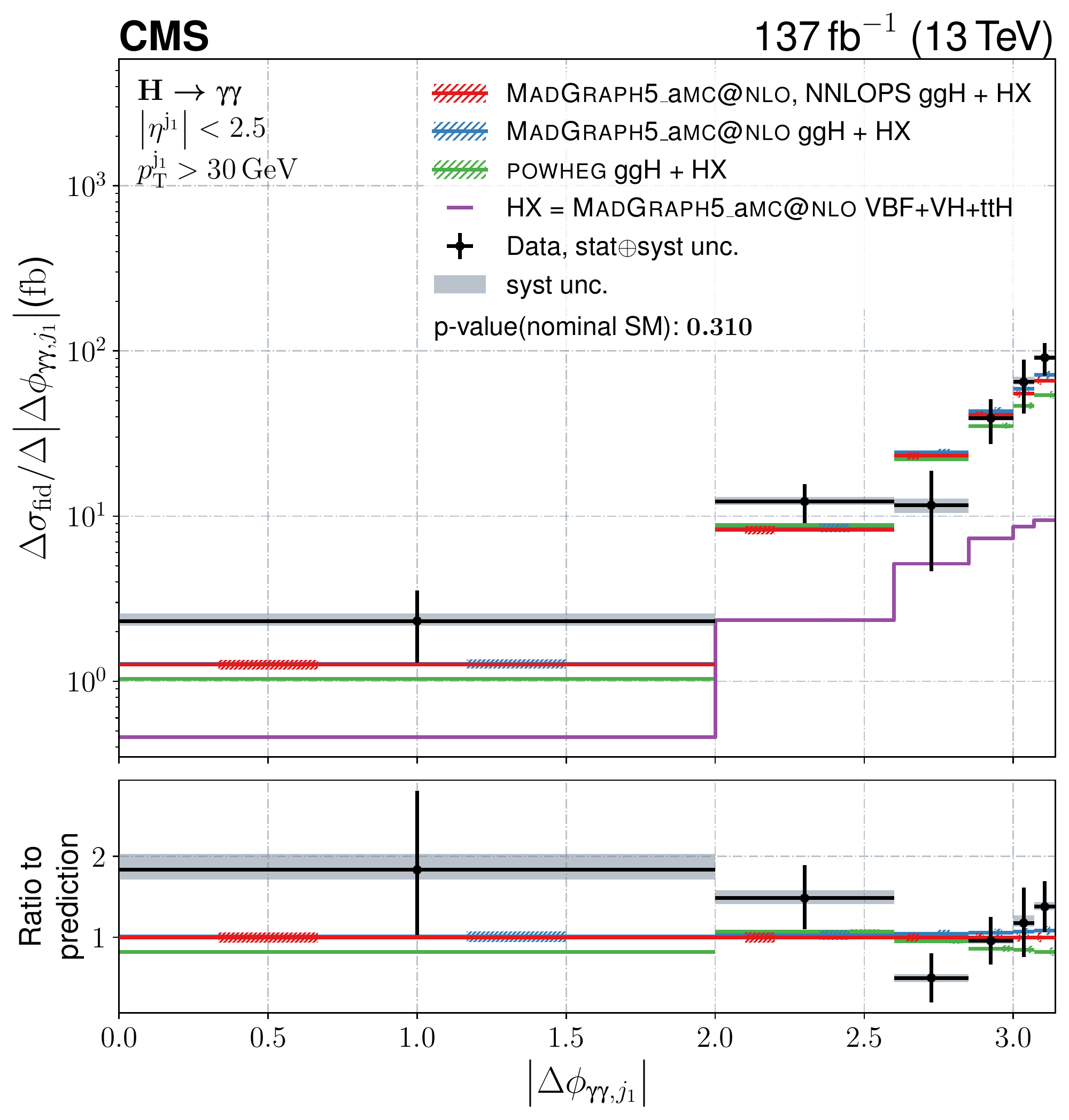}\\
  \includegraphics[width=0.49\textwidth]{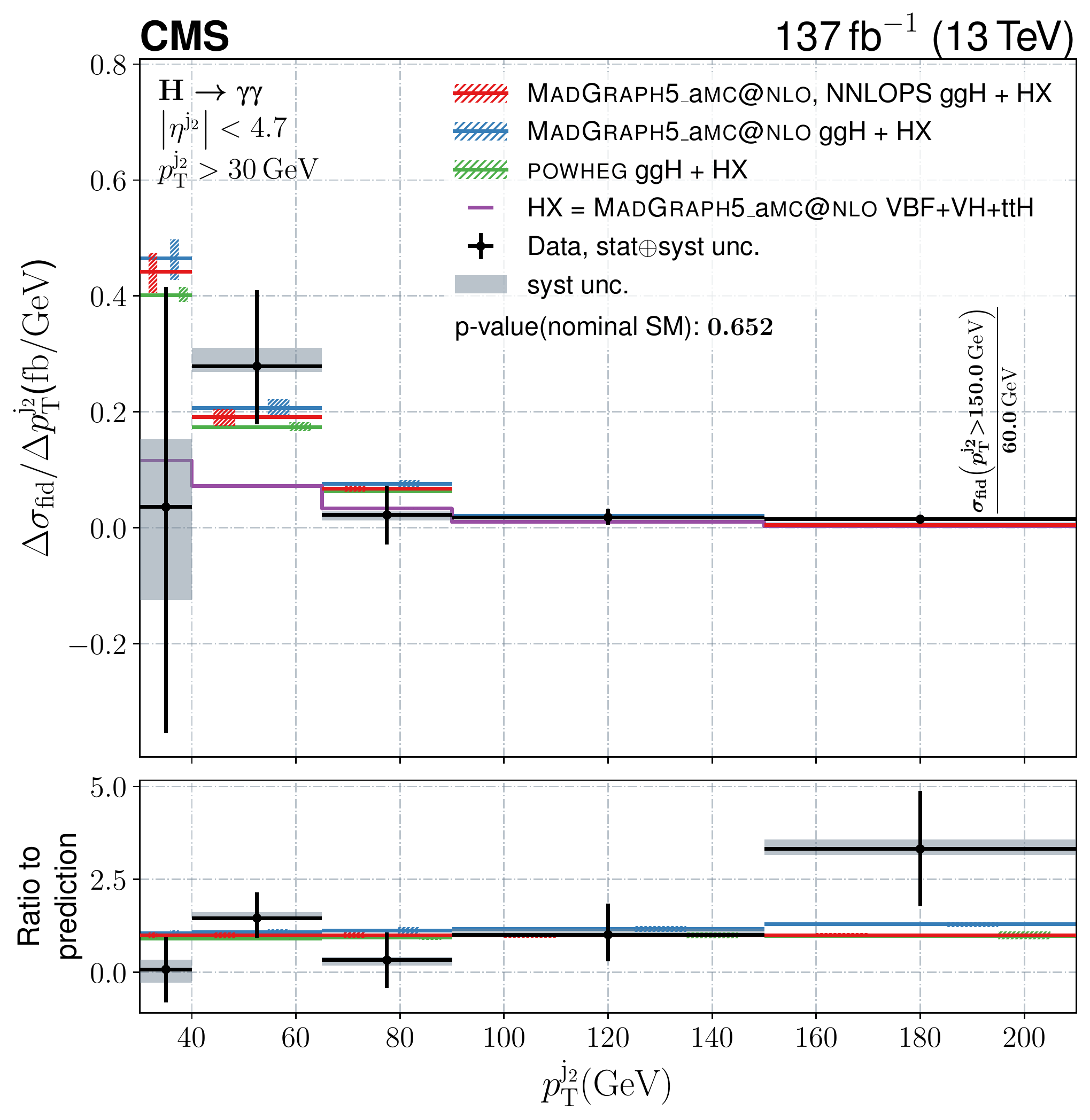}
  \includegraphics[width=0.49\textwidth]{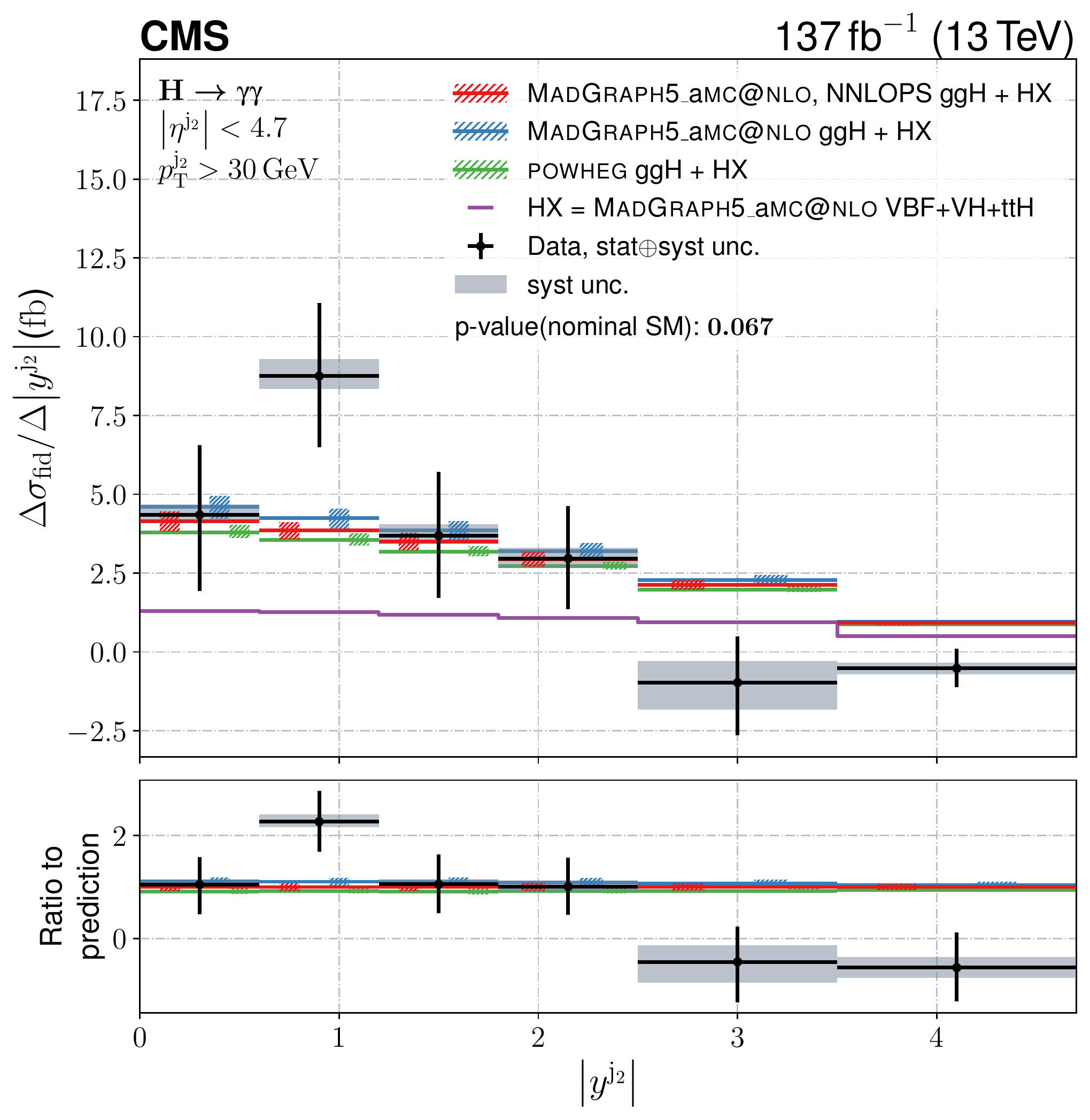}
  \caption{Differential fiducial cross sections for $\abs{\Delta y_{\PGg\PGg,\mathrm{j}_{1}}}$, $\abs{\Delta\phi_{\PGg\PGg,\mathrm{j}_{1}}}$, $\pt^{\mathrm{j}_{2}}$, and $\abs{y^{\mathrm{j}_{2}}}$. \resultsCaptionShort}
  \label{fig:resultsAbsDeltaRapidityGgJet2p50}
\end{figure}

\begin{figure}
\centering
  \includegraphics[width=0.49\textwidth]{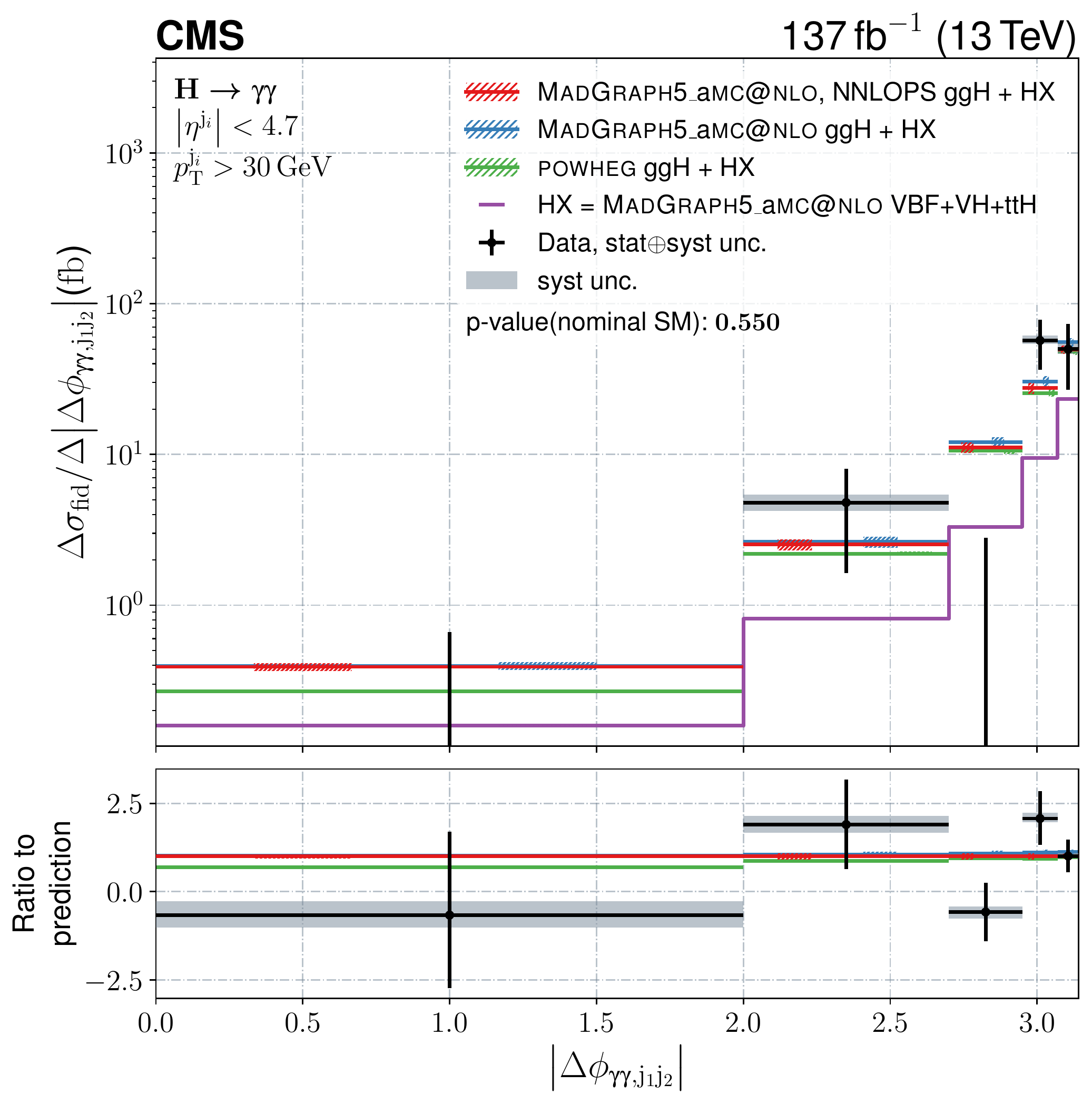}
  \includegraphics[width=0.49\textwidth]{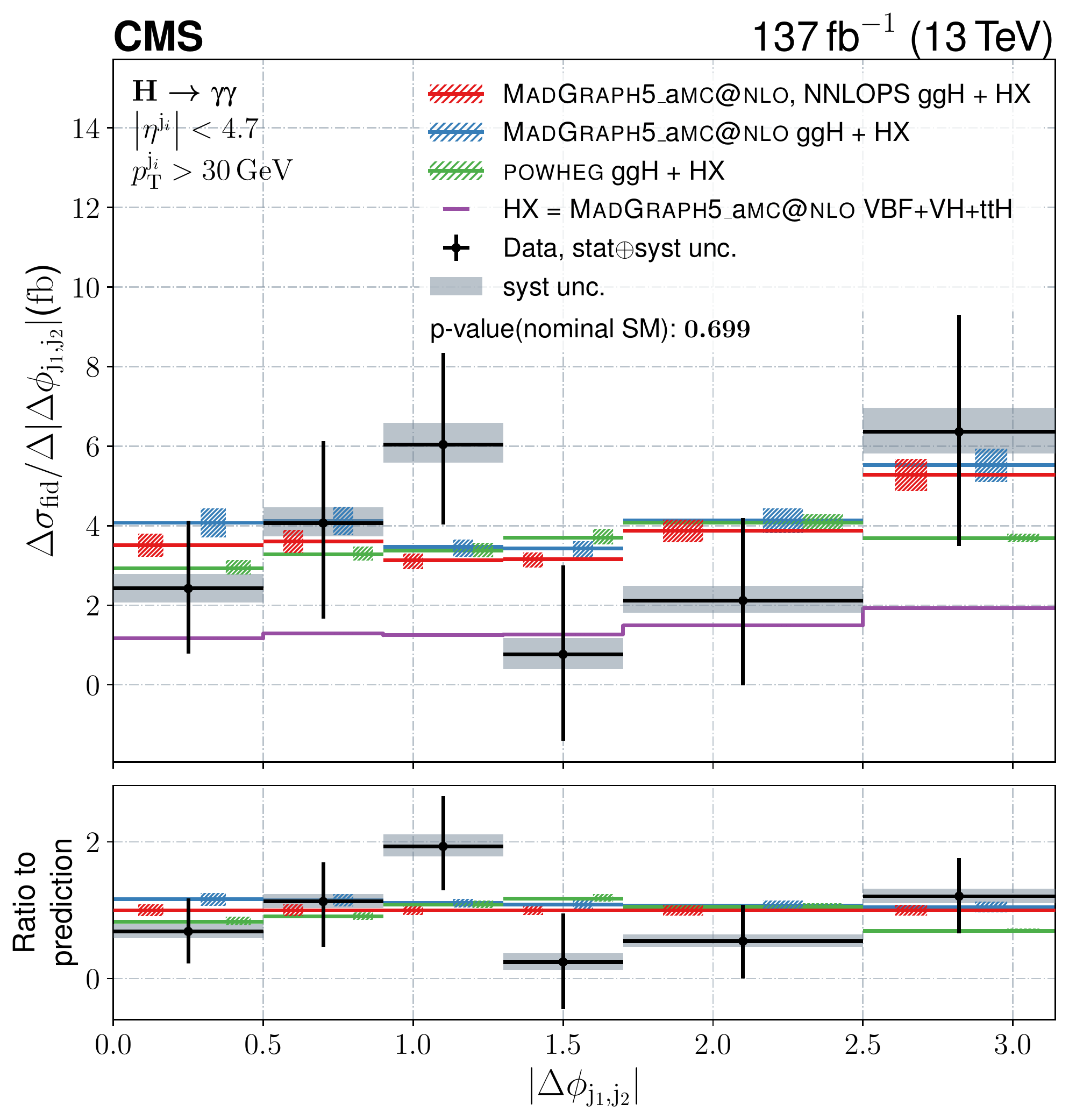}\\
  \includegraphics[width=0.49\textwidth]{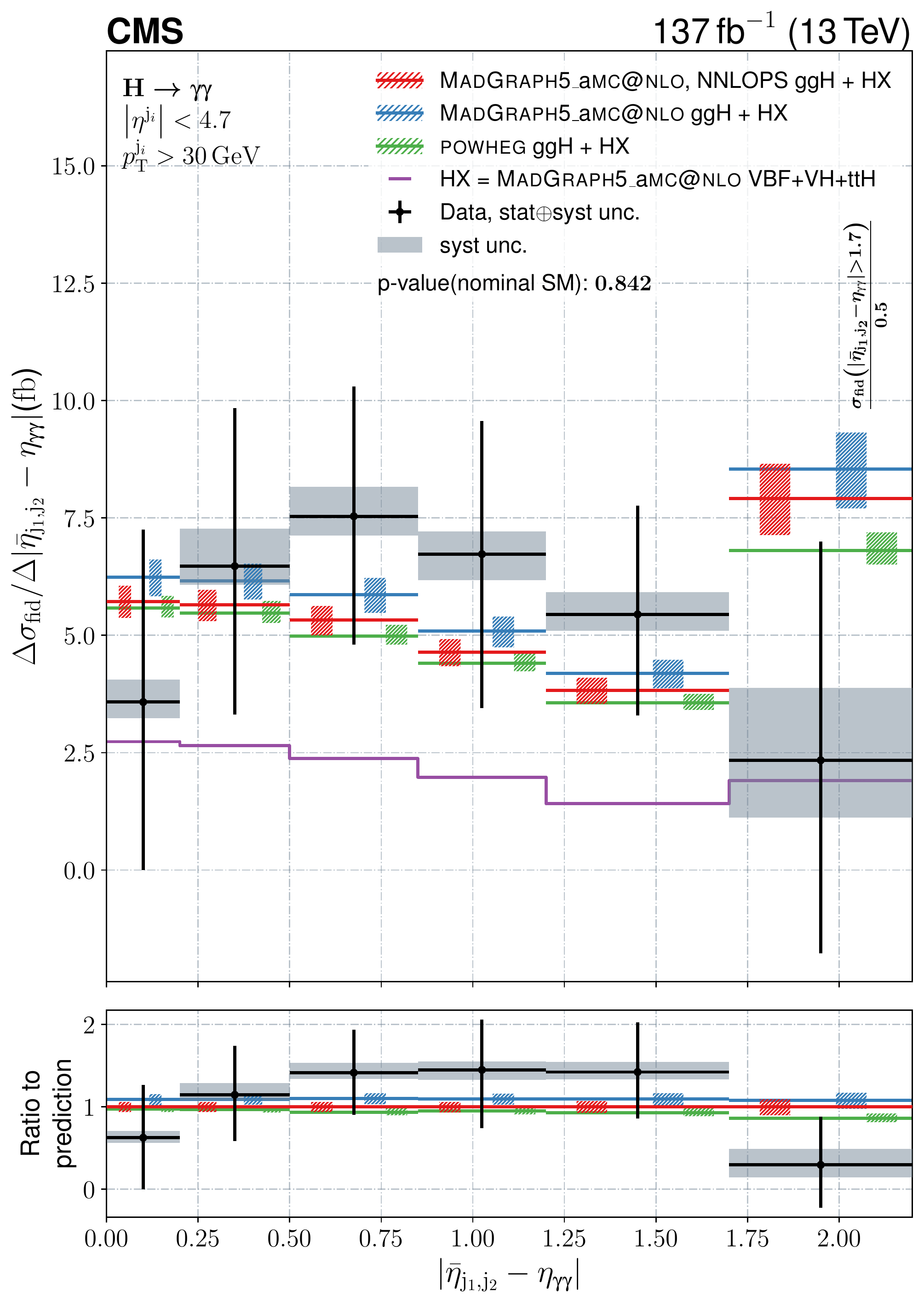}
  \includegraphics[width=0.49\textwidth]{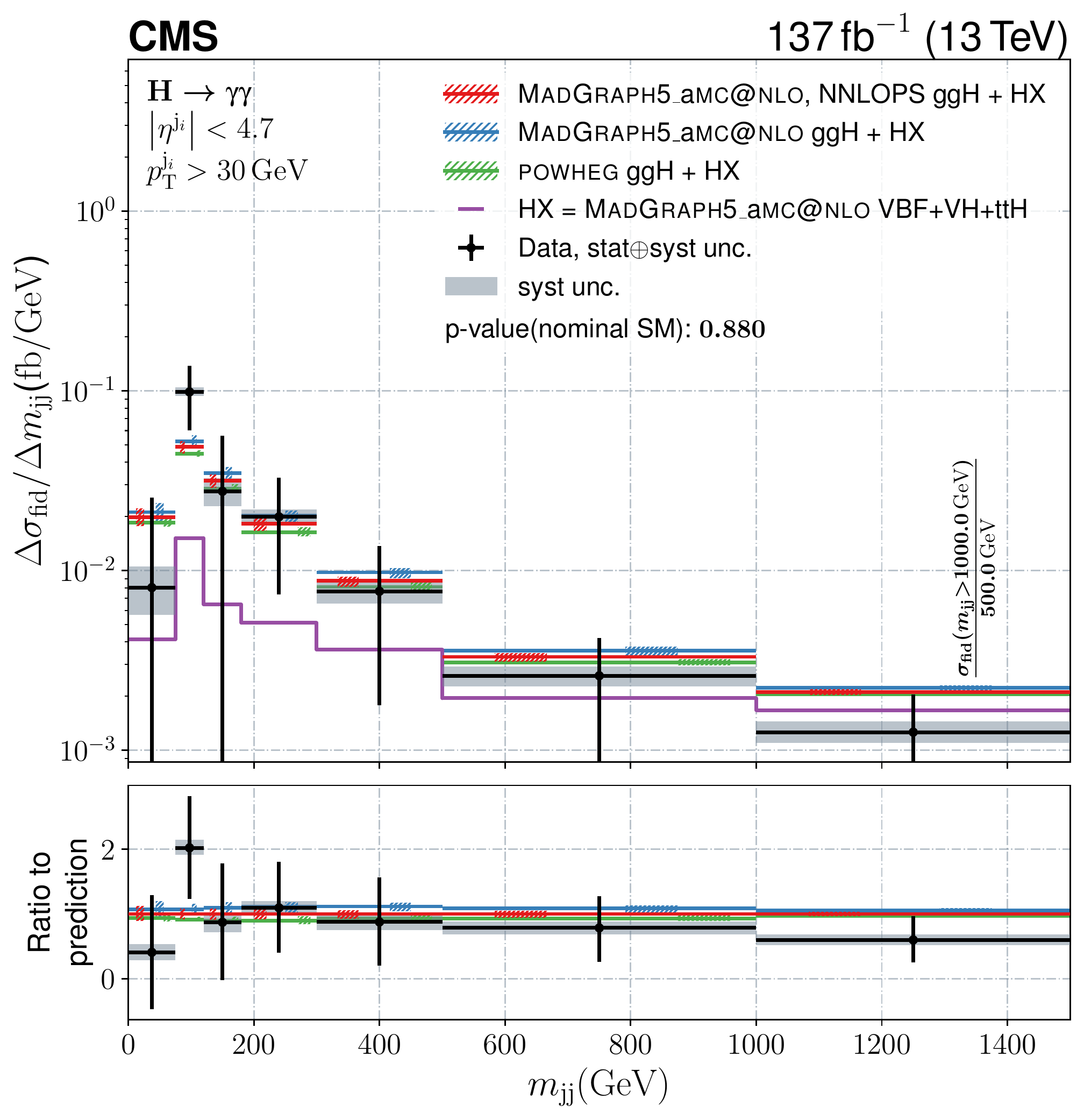}
  \caption{Differential fiducial cross sections for $\abs{\DphiggjOjT}$, $\abs{\DphijOjT}$, $\abs{\etaZepp}$, and $m_{\mathrm{jj}}$. \resultsCaptionShort}
  \label{fig:resultsAbsDeltaPhiGgJjJets4p7}
\end{figure}

\begin{figure}
\centering
  \includegraphics[width=0.49\textwidth]{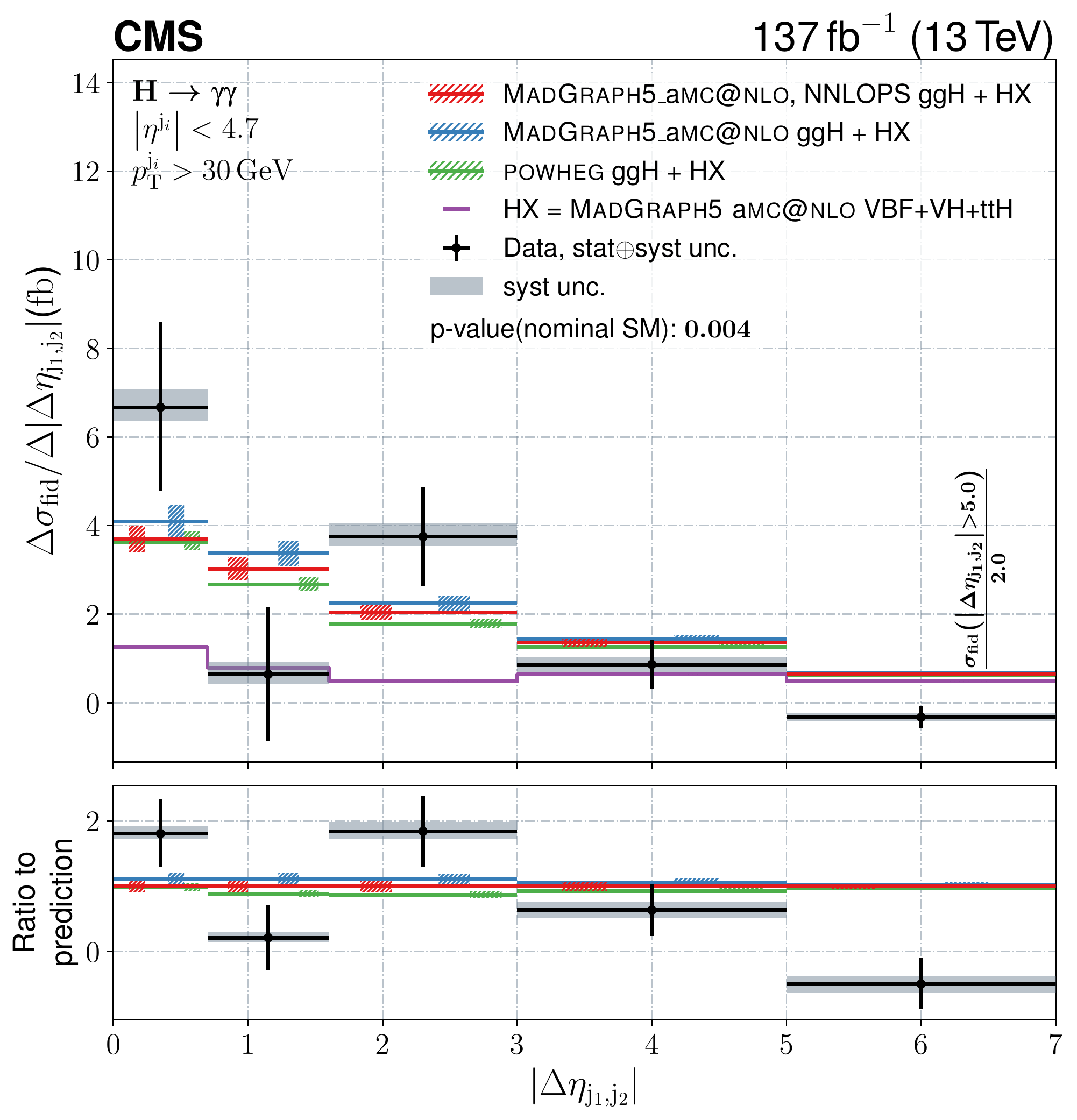}
  \includegraphics[width=0.49\textwidth]{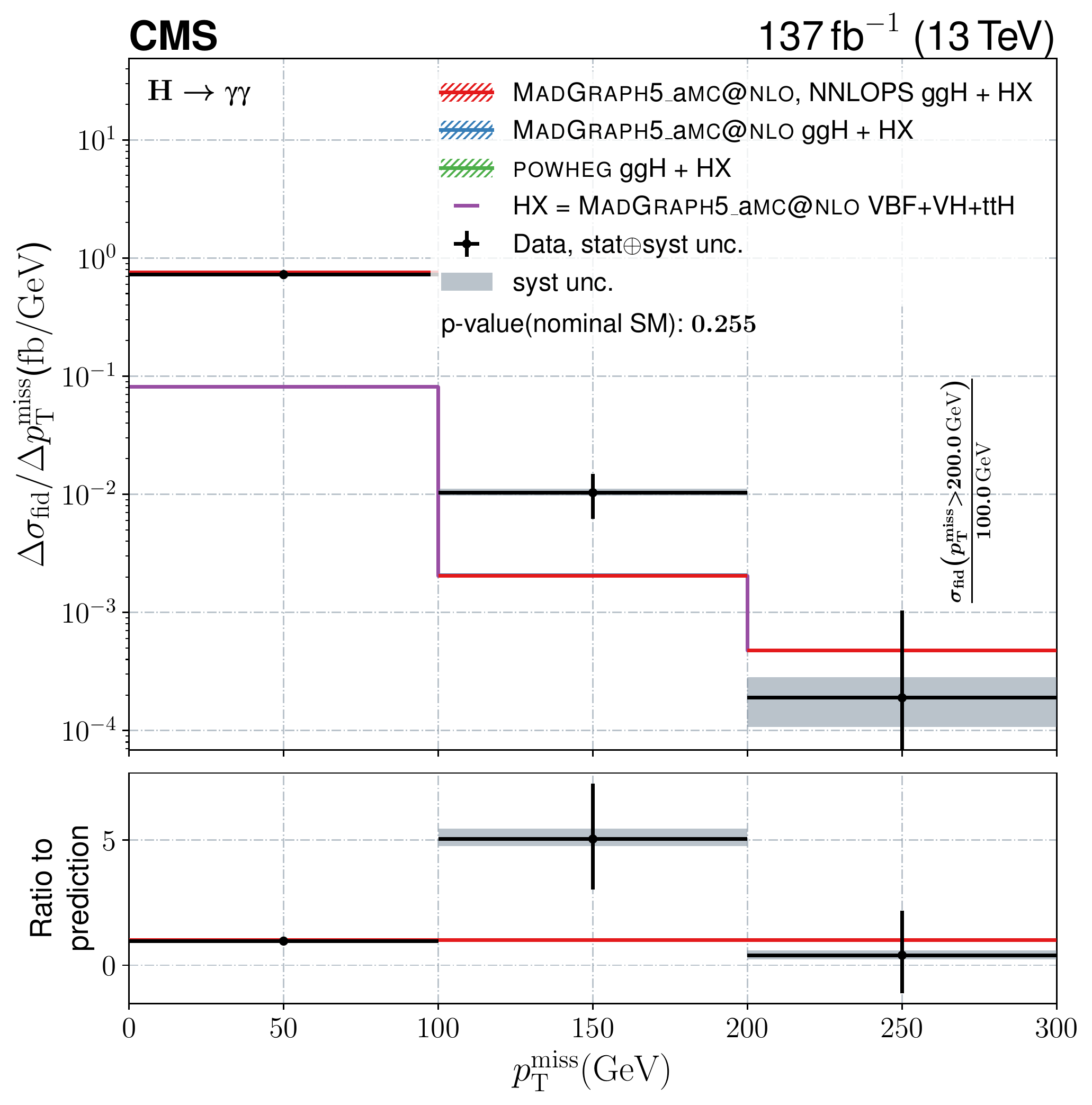}\\
  \includegraphics[width=0.49\textwidth]{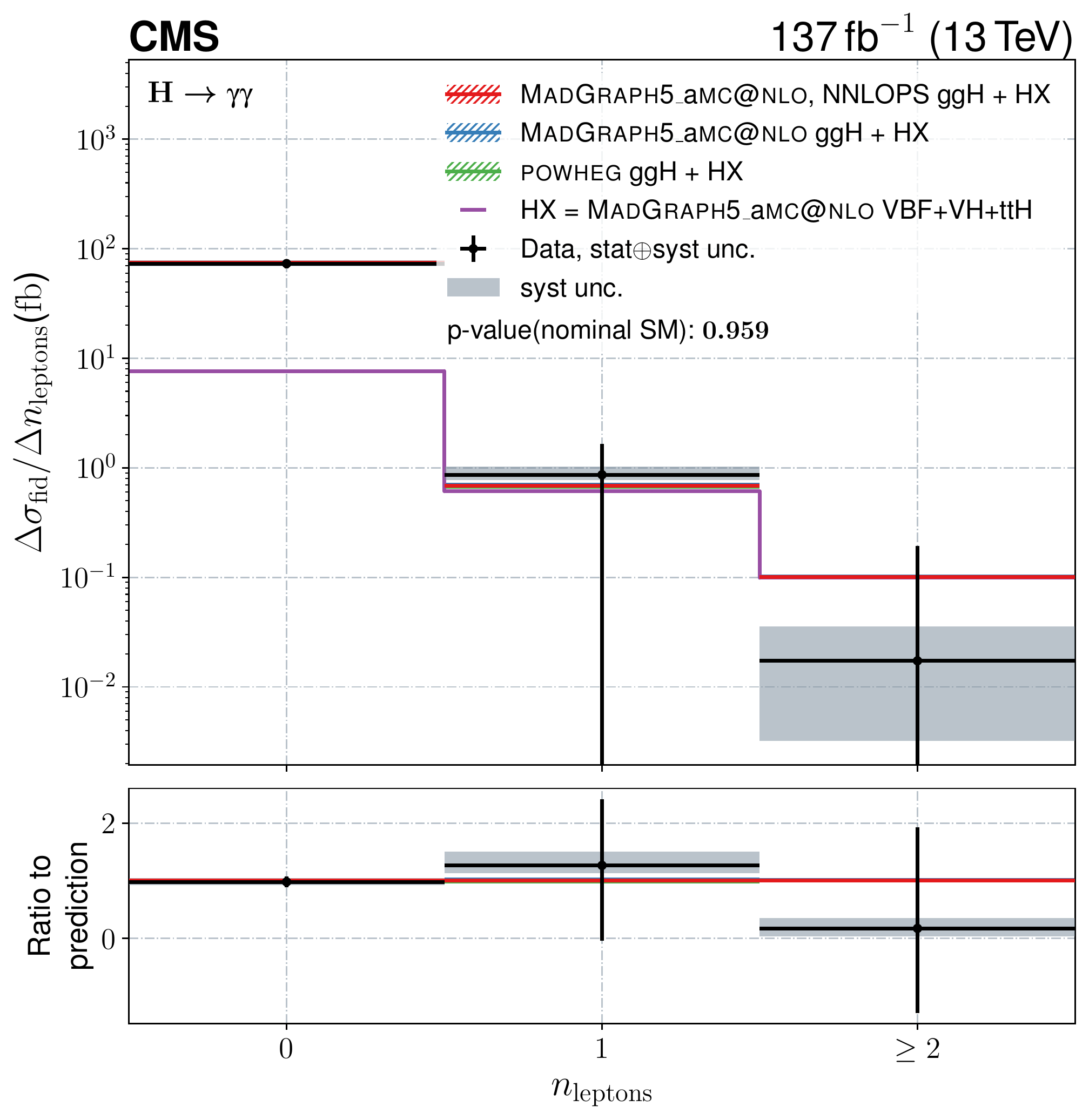}
  \includegraphics[width=0.49\textwidth]{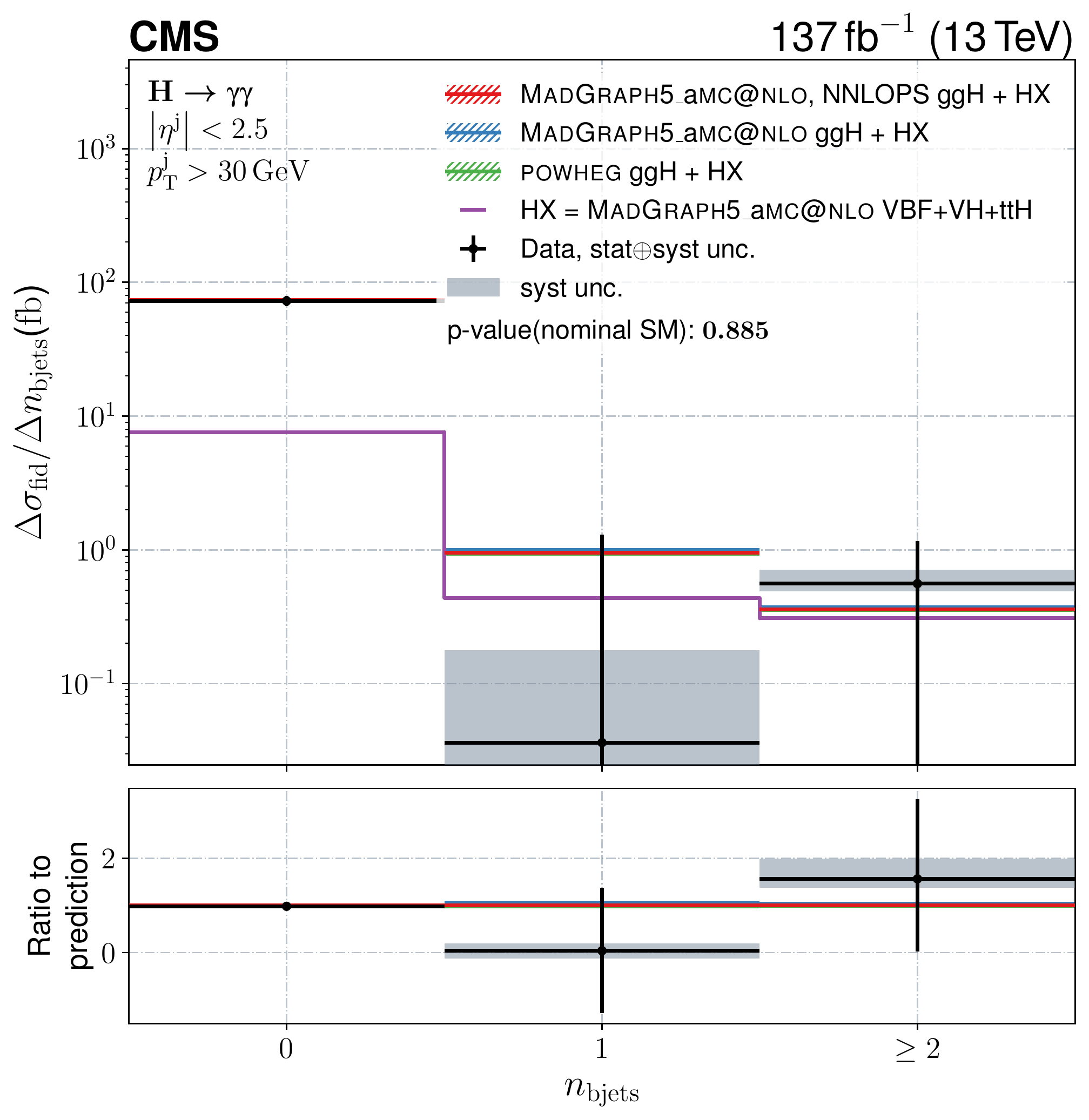}
  \caption{Differential fiducial cross sections for $\abs{\DetajOjT}$, $n_{\text{leptons}}$, \nbj, and \ptmiss. \resultsCaptionShort}
  \label{fig:resultsAbsDeltaEtaJ1J2Jets4p7}
\end{figure}

\begin{figure}
\centering
  \includegraphics[width=0.45\textwidth]{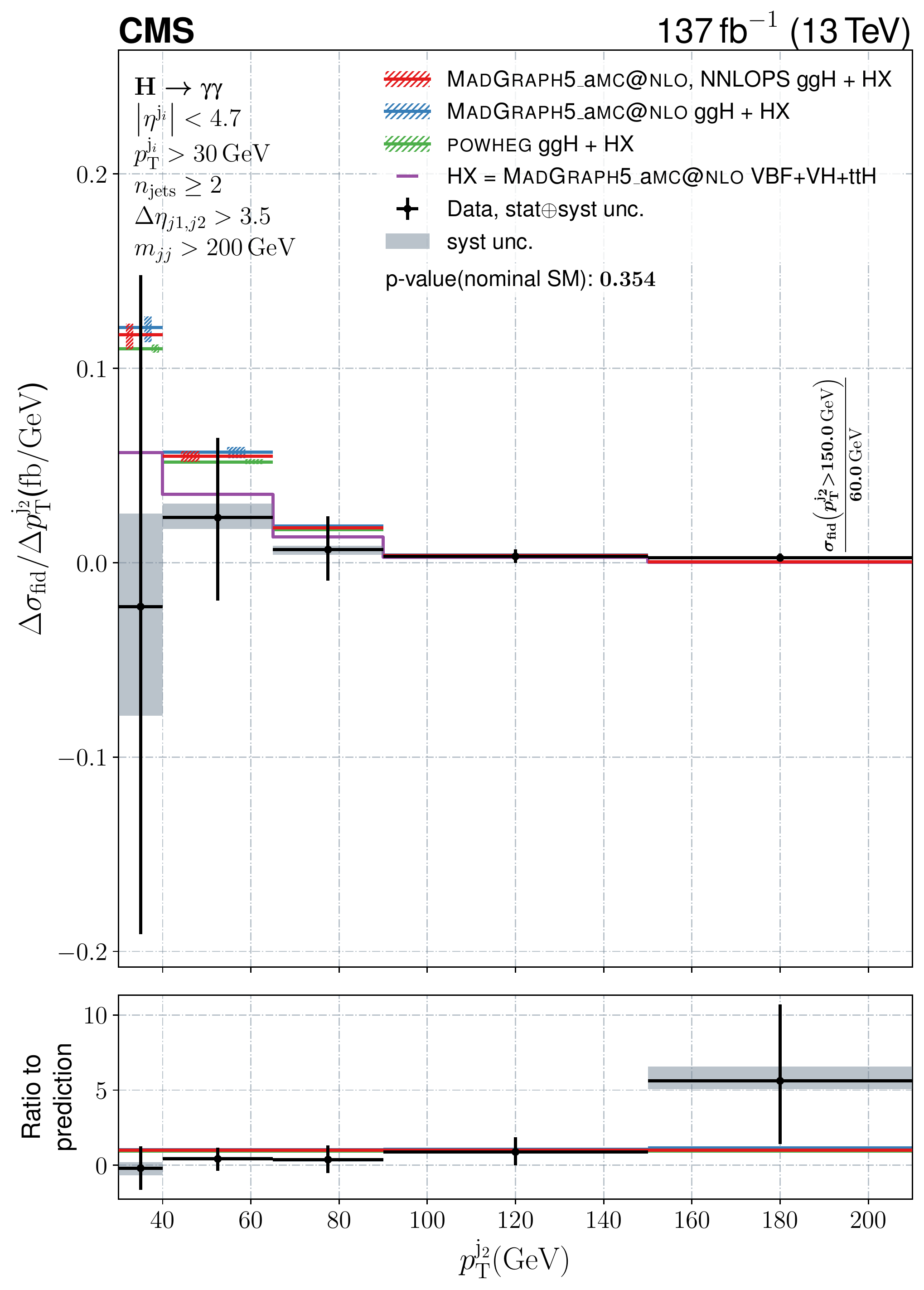}
  \includegraphics[width=0.45\textwidth]{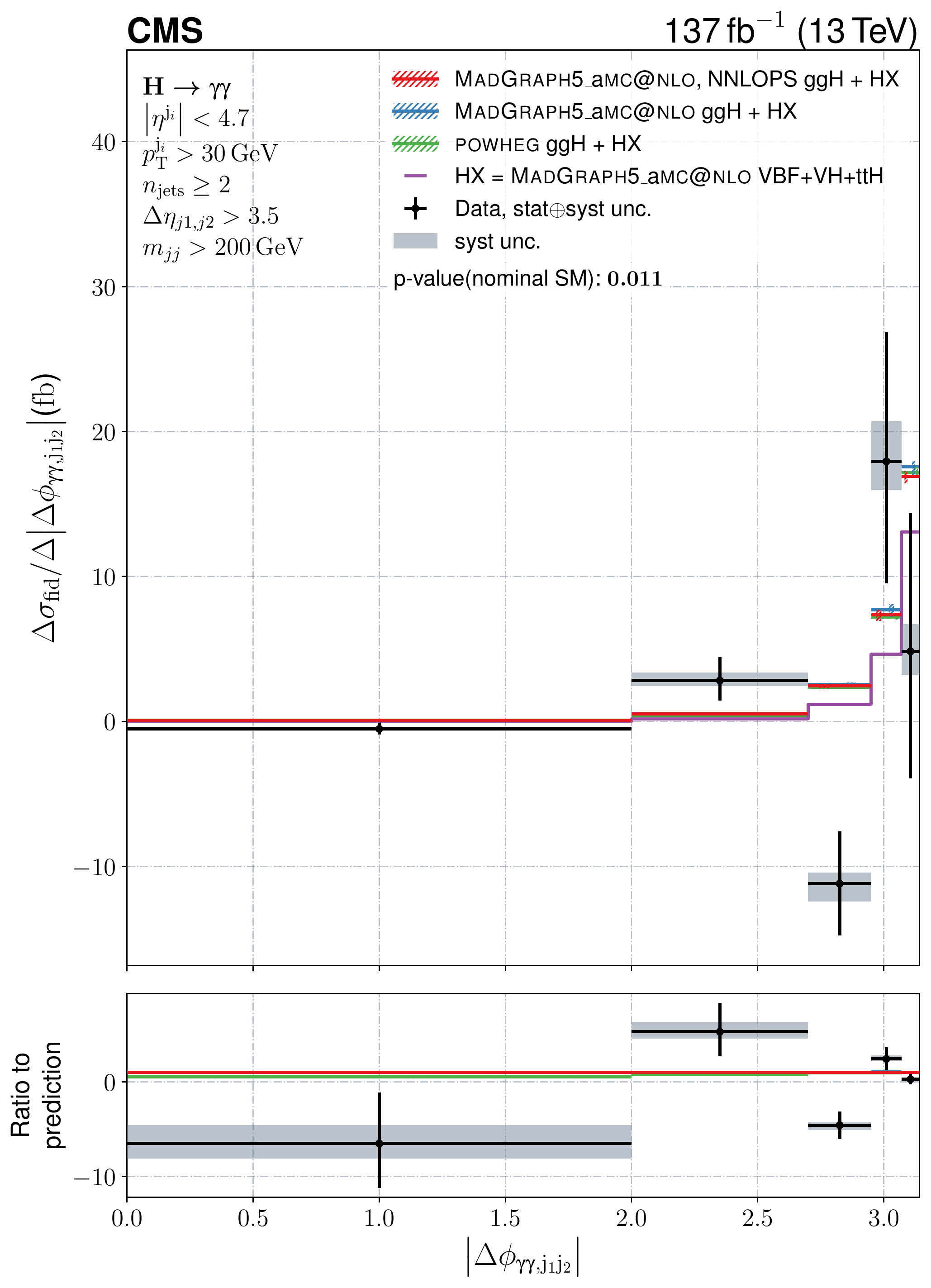}\\
  \includegraphics[width=0.45\textwidth]{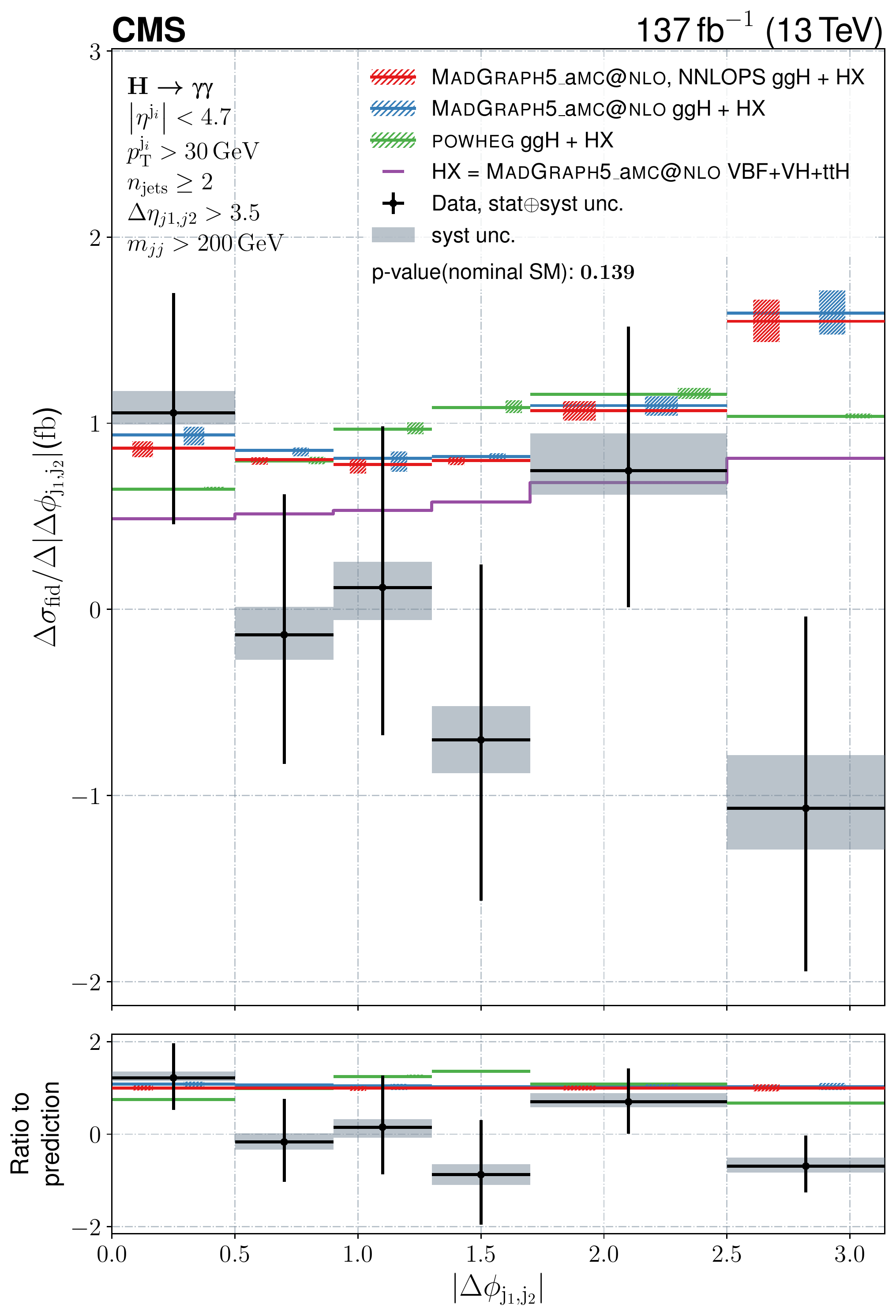}
  \includegraphics[width=0.45\textwidth]{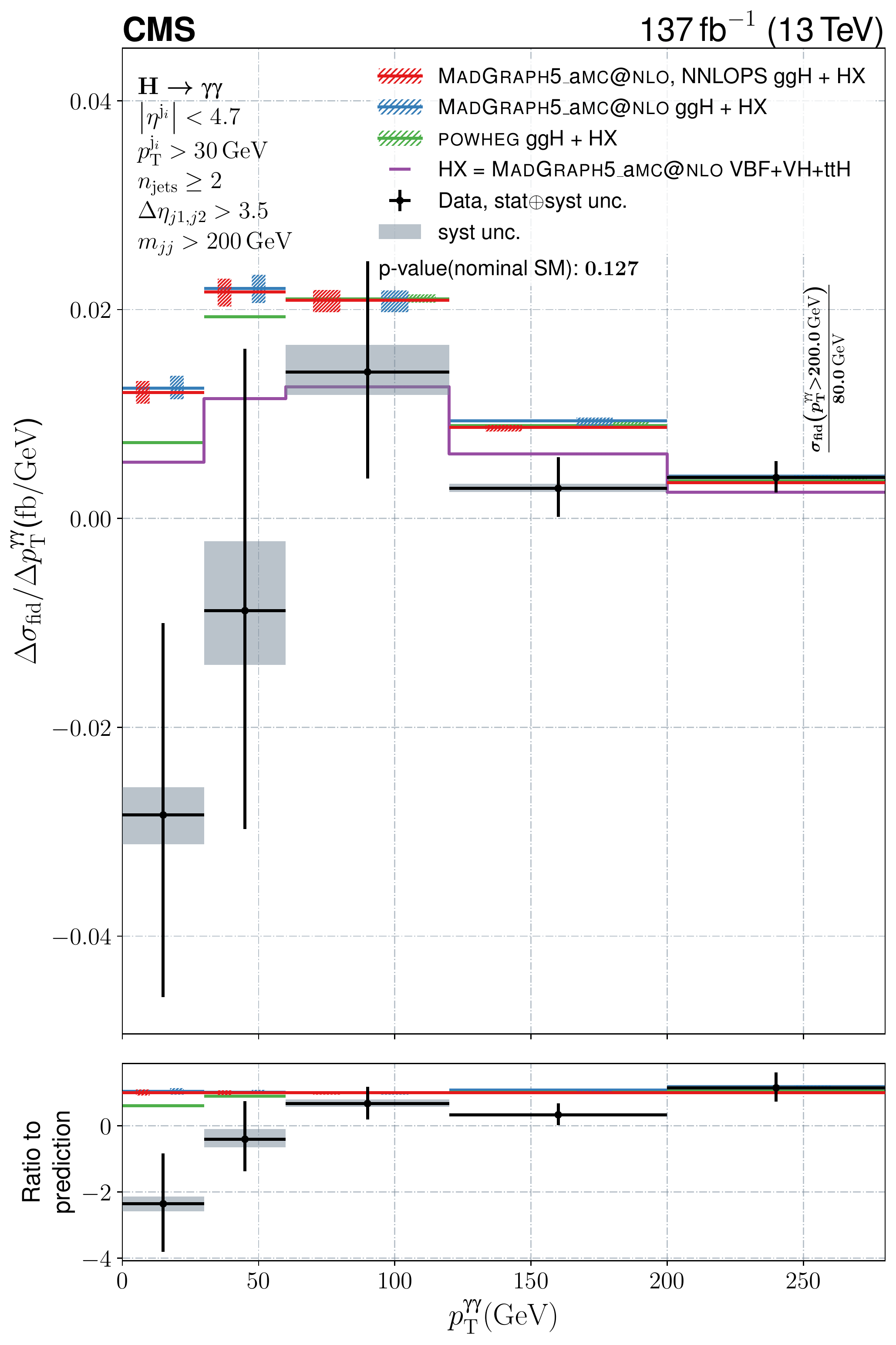}
  \caption{Differential fiducial cross sections for $\pt^{\mathrm{j}_{2}}$, $\abs{\DphiggjOjT}$, $\abs{\DphijOjT}$, and \ptgamgam in the VBF-enriched phase space region. \resultsCaptionShort}
  \label{fig:resultsJet4p7Pt1VBFlike}
\end{figure}

\begin{figure}
\centering
  \includegraphics[width=\textwidth]{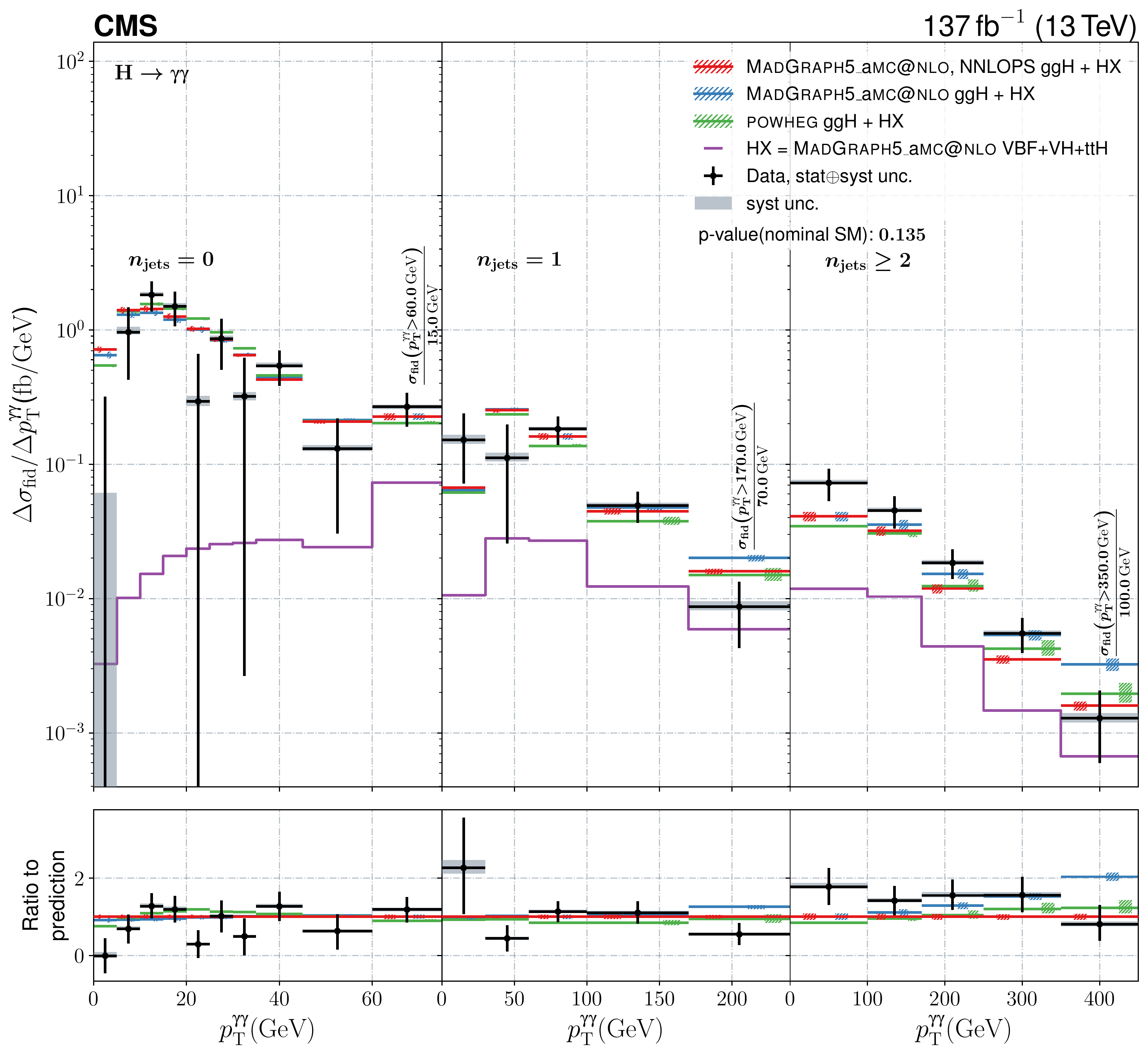}
  \caption{Double-differential fiducial cross section measured in bins of \ptgamgam and \njets. \resultsCaptionShortOne}
  \label{fig:resultsPtvsNjets}
\end{figure}

\begin{figure}
\centering
  \includegraphics[width=\textwidth]{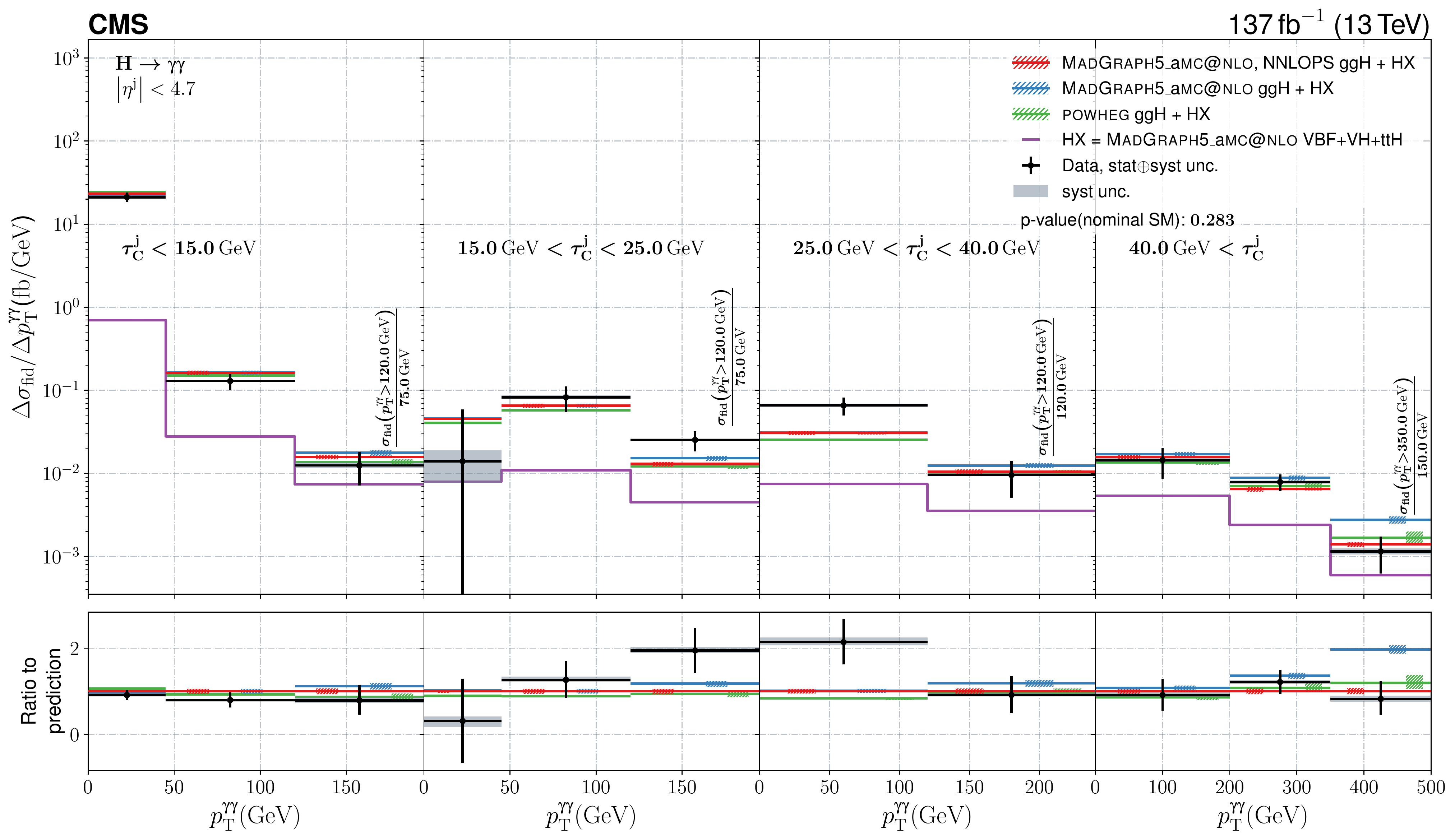}
  \caption{Double-differential fiducial cross section measured in bins of \ptgamgam and \taucj. \resultsCaptionShortOne}
  \label{fig:resultsPtvsTauC}
\end{figure}

The results for the differential fiducial cross section, with respect to all observables listed in Table~\ref{tab:obsBins}, are shown in Figs.~\ref{fig:resultsPtNjetsCosTsRapi}--\ref{fig:resultsJet4p7Pt1VBFlike}. The results of the double-differential cross section measurement with respect to \pt and \njets, and \pt and \taucj, are shown in Figs.~\ref{fig:resultsPtvsNjets} and~\ref{fig:resultsPtvsTauC}, respectively.
Each figure compares the measurement with the nominal prediction, as well as with several additional predictions. The $\PH\PX$ component, denoting the sum of the Higgs boson production cross sections from the \VBF, \VH, and \ttH production processes, is taken from the \MGvATNLO simulation and is common to the different SM predictions shown. The predicted cross section for the gluon-fusion production mode is taken from the \MGvATNLO simulation, with and without \NNLOPS reweighting, and the \POWHEG event generator \cite{POWHEG1,POWHEG2,POWHEG3,Powheg:ggH}, and added to the $\PH\PX$ component to obtain the total per-bin predictions shown. The uncertainty in the theoretical predictions only takes into account the variation in the predicted differential cross section shape coming from varying the set of PDF replicas, the renormalization and factorization scales, and \alpS. The uncertainty in the total Higgs boson production cross section and branching fraction is not taken into account for the results on the differential cross sections.

Overall, the differential cross section results agree within uncertainties with the nominal SM prediction. 
For each observable, a $p$-value is calculated using the test statistic given in Eq.~\eqref{form:testStat} evaluated at the 
SM point, where the \Hgg cross section is set to the nominal SM value in all particle-level bins, extracted using the \MGvATNLO simulation with the \NNLOPS reweighting for ggF. The $p$-value is then computed on the $\chi^{2}$ pdf with the number of degrees of freedom set to the number of particle-level bins for the respective observable. The cross section measured for the underflow bin is not reported on the results figure if not specified otherwise, but always taken into account for the calculation of the $p$-value. The observed $p$-values for the SM point are from 0.004 to 0.96, with the cross section measured as a function of \ptgamgam having a $p$-value of 0.24, and as a function of \njets a $p$-value of 0.69. For the fiducial cross section a $p$-value of 0.73 is observed for the SM point. The lowest $p$-value of 0.004 is seen for the difference between the $\eta$ coordinate of the leading and subleading jet $\abs{\DetajOjT}$, shown in Fig.~\ref{fig:resultsAbsDeltaEtaJ1J2Jets4p7}. The per-bin uncertainties for observables of the diphoton system range from 10\% to 40\%. For observables that involve the leading-\pt jet, uncertainties reach around 100\%. 

For the two-jet phase space for observables being calculated with the two leading-\pt jets, uncertainties are at the same level, because of the larger bin sizes. The cross section measurements as a function of \ptgamgam, \ptjO, $\abs{\DphiggjOjT}$, and $\abs{\DphijOjT}$ are also performed in the VBF-enriched phase space, where the per-bin uncertainties can reach 150\%. Overall, an under-fluctuation is observed for events that match the criteria for this phase space (see Fig.~\ref{fig:fidXSSummPlot}). A similar observation on the same data set has been reported in Ref.~\cite{HIG-19-015}. 

\clearpage

\section{Summary}\label{sec:summary}
The measurement of the fiducial inclusive Higgs boson (\PH) production cross section with the \Hgg decay mode has been presented. The fiducial phase space is defined by the ratio of the transverse momentum (\pt) of the leading (subleading) photon to diphoton invariant mass satisfying $\pt/m_{\gamgam}>1/3~(1/4)$, their pseudorapidity being within $\abs{\eta}<2.5$, and both photons being isolated. The production cross section for the Higgs boson decaying into two photons in the aforementioned phase space is measured to $\sigFid=73.4_{-5.9}^{+6.1}\fb$, in agreement with the theoretical prediction from the standard model (SM) of $75.4\pm4.1\fb$.

Furthermore, the $\Pp\Pp\to\PH+\PX$, \Hgg cross section in the fiducial phase space has been measured as a function of observables of the diphoton system, as well as several others involving properties of the leading-\pt and subleading-\pt jets. Observables corresponding to the number of jets, leptons, and b-tagged jets are included as well. For the first time using the CMS detector, the cross section has been measured as a function of the rapidity weighted jet-\pt (\taucj), using up to six additional jets in the event. The cross section as a function of a measure for the deviation from ``back-to-backness" $\abs{\phietaS}$ for the diphoton system has been measured for the first time using the \Hgg channel. Two double-differential cross section measurements have been performed: one in bins of \pt and the number of jets, the other in bins of \pt and \taucj. A selected set of differential measurements has been performed in a dedicated phase space enriched with events compatible with vector boson fusion Higgs boson production. Finally, the production cross section has been measured in three fiducial phase spaces loosely targeting the vector boson and \ttbar associated production modes.

Overall, the performed differential fiducial cross section measurements of the Higgs boson production in proton-proton collisions are found to be in agreement with the SM prediction within the uncertainties.

\begin{acknowledgments}
We congratulate our colleagues in the CERN accelerator departments for the excellent performance of the LHC and thank the technical and administrative staffs at CERN and at other CMS institutes for their contributions to the success of the CMS effort. In addition, we gratefully acknowledge the computing centres and personnel of the Worldwide LHC Computing Grid and other centres for delivering so effectively the computing infrastructure essential to our analyses. Finally, we acknowledge the enduring support for the construction and operation of the LHC, the CMS detector, and the supporting computing infrastructure provided by the following funding agencies: BMBWF and FWF (Austria); FNRS and FWO (Belgium); CNPq, CAPES, FAPERJ, FAPERGS, and FAPESP (Brazil); MES and BNSF (Bulgaria); CERN; CAS, MoST, and NSFC (China); MINCIENCIAS (Colombia); MSES and CSF (Croatia); RIF (Cyprus); SENESCYT (Ecuador); MoER, ERC PUT and ERDF (Estonia); Academy of Finland, MEC, and HIP (Finland); CEA and CNRS/IN2P3 (France); BMBF, DFG, and HGF (Germany); GSRI (Greece); NKFIH (Hungary); DAE and DST (India); IPM (Iran); SFI (Ireland); INFN (Italy); MSIP and NRF (Republic of Korea); MES (Latvia); LAS (Lithuania); MOE and UM (Malaysia); BUAP, CINVESTAV, CONACYT, LNS, SEP, and UASLP-FAI (Mexico); MOS (Montenegro); MBIE (New Zealand); PAEC (Pakistan); MES and NSC (Poland); FCT (Portugal); MESTD (Serbia); MCIN/AEI and PCTI (Spain); MOSTR (Sri Lanka); Swiss Funding Agencies (Switzerland); MST (Taipei); MHESI and NSTDA (Thailand); TUBITAK and TENMAK (Turkey); NASU (Ukraine); STFC (United Kingdom); DOE and NSF (USA).

\hyphenation{Rachada-pisek} Individuals have received support from the Marie-Curie programme and the European Research Council and Horizon 2020 Grant, contract Nos.\ 675440, 724704, 752730, 758316, 765710, 824093, 884104, and COST Action CA16108 (European Union); the Leventis Foundation; the Alfred P.\ Sloan Foundation; the Alexander von Humboldt Foundation; the Belgian Federal Science Policy Office; the Fonds pour la Formation \`a la Recherche dans l'Industrie et dans l'Agriculture (FRIA-Belgium); the Agentschap voor Innovatie door Wetenschap en Technologie (IWT-Belgium); the F.R.S.-FNRS and FWO (Belgium) under the ``Excellence of Science -- EOS" -- be.h project n.\ 30820817; the Beijing Municipal Science \& Technology Commission, No. Z191100007219010; the Ministry of Education, Youth and Sports (MEYS) of the Czech Republic; the Hellenic Foundation for Research and Innovation (HFRI), Project Number 2288 (Greece); the Deutsche Forschungsgemeinschaft (DFG), under Germany's Excellence Strategy -- EXC 2121 ``Quantum Universe" -- 390833306, and under project number 400140256 - GRK2497; the Hungarian Academy of Sciences, the New National Excellence Program - \'UNKP, the NKFIH research grants K 124845, K 124850, K 128713, K 128786, K 129058, K 131991, K 133046, K 138136, K 143460, K 143477, 2020-2.2.1-ED-2021-00181, and TKP2021-NKTA-64 (Hungary); the Council of Science and Industrial Research, India; the Latvian Council of Science; the Ministry of Education and Science, project no. 2022/WK/14, and the National Science Center, contracts Opus 2021/41/B/ST2/01369 and 2021/43/B/ST2/01552 (Poland); the Funda\c{c}\~ao para a Ci\^encia e a Tecnologia, grant CEECIND/01334/2018 (Portugal); the National Priorities Research Program by Qatar National Research Fund; MCIN/AEI/10.13039/501100011033, ERDF ``a way of making Europe", and the Programa Estatal de Fomento de la Investigaci{\'o}n Cient{\'i}fica y T{\'e}cnica de Excelencia Mar\'{\i}a de Maeztu, grant MDM-2017-0765 and Programa Severo Ochoa del Principado de Asturias (Spain); the Chulalongkorn Academic into Its 2nd Century Project Advancement Project, and the National Science, Research and Innovation Fund via the Program Management Unit for Human Resources \& Institutional Development, Research and Innovation, grant B05F650021 (Thailand); the Kavli Foundation; the Nvidia Corporation; the SuperMicro Corporation; the Welch Foundation, contract C-1845; and the Weston Havens Foundation (USA).
\end{acknowledgments}

\bibliography{auto_generated}

\providecommand{\href}[2]{#2}\begingroup\raggedright\begin{thebibliography}{10}%
\makeatletter
\providecommand{\hrefCMSnoop }[0]{\@secondoftwo}%
\makeatother
\providecommand{\doi}{\texttt{doi:}\begingroup \urlstyle{tt}\Url}

\bibitem{HiggsAtlas}
\hrefCMSnoop {}{{ATLAS Collaboration}, ``Observation of a new particle in the
  search for the standard model {Higgs} boson with the {ATLAS} detector at the
  {LHC}'',} \textit{ Phys. Lett. B} \textbf{ 716} (2012) 1,
  \href{http://dx.doi.org/10.1016/j.physletb.2012.08.020}{\doi{10.1016/j.physletb.2012.08.020}},
  \href{http://www.arXiv.org/abs/1207.7214}{\texttt{arXiv:1207.7214}}.

\bibitem{HiggsCMS}
\hrefCMSnoop {}{{CMS Collaboration}, ``Observation of a new boson at a mass of
  125 {GeV} with the {CMS} experiment at the {LHC}'',} \textit{ Phys. Lett. B}
  \textbf{ 716} (2012) 30,
  \href{http://dx.doi.org/10.1016/j.physletb.2012.08.021}{\doi{10.1016/j.physletb.2012.08.021}},
  \href{http://www.arXiv.org/abs/1207.7235}{\texttt{arXiv:1207.7235}}.

\bibitem{HiggsCMS1}
\hrefCMSnoop {}{{CMS Collaboration}, ``Observation of a new boson with mass
  near 125 {GeV} in {$\Pp\Pp$} collisions at {$\sqrt{s}$} = 7 and 8 {TeV}'',}
  \textit{ JHEP} \textbf{ 06} (2013) 081,
  \href{http://dx.doi.org/10.1007/JHEP06(2013)081}{\doi{10.1007/JHEP06(2013)081}},
  \href{http://www.arXiv.org/abs/1303.4571}{\texttt{arXiv:1303.4571}}.

\bibitem{ATLAS:2016neq}
\hrefCMSnoop {}{{ATLAS and CMS Collaborations}, ``Measurements of the {Higgs}
  boson production and decay rates and constraints on its couplings from a
  combined {ATLAS} and {CMS} analysis of the {LHC} {$\Pp\Pp$} collision data at
  $\sqrt{s}$ = 7 and 8 {TeV}'',} \textit{ JHEP} \textbf{ 08} (2016) 045,
  \href{http://dx.doi.org/10.1007/JHEP08(2016)045}{\doi{10.1007/JHEP08(2016)045}},
  \href{http://www.arXiv.org/abs/1606.02266}{\texttt{arXiv:1606.02266}}.

\bibitem{ATLAS:2014yga}
\hrefCMSnoop {}{{ATLAS Collaboration}, ``Measurements of fiducial and
  differential cross sections for {Higgs} boson production in the diphoton
  decay channel at {$\sqrt{s}=8$} {TeV} with {ATLAS}'',} \textit{ JHEP}
  \textbf{ 09} (2014) 112,
  \href{http://dx.doi.org/10.1007/JHEP09(2014)112}{\doi{10.1007/JHEP09(2014)112}},
  \href{http://www.arXiv.org/abs/1407.4222}{\texttt{arXiv:1407.4222}}.

\bibitem{CMS:2015qgt}
\hrefCMSnoop {}{{CMS Collaboration}, ``Measurement of differential cross
  sections for {Higgs} boson production in the diphoton decay channel in pp
  collisions at {$\sqrt{s}=8$} {TeV}'',} \textit{ Eur. Phys. J. C} \textbf{ 76}
  (2016) 13,
  \href{http://dx.doi.org/10.1140/epjc/s10052-015-3853-3}{\doi{10.1140/epjc/s10052-015-3853-3}},
  \href{http://www.arXiv.org/abs/1508.07819}{\texttt{arXiv:1508.07819}}.

\bibitem{ATLAS:2014xzb}
\hrefCMSnoop {}{{ATLAS Collaboration}, ``Fiducial and differential cross
  sections of {Higgs} boson production measured in the four-lepton decay
  channel in {$\Pp\Pp$} collisions at {$\sqrt{s}$} = 8 {TeV} with the {ATLAS}
  detector'',} \textit{ Phys. Lett. B} \textbf{ 738} (2014) 234,
  \href{http://dx.doi.org/10.1016/j.physletb.2014.09.054}{\doi{10.1016/j.physletb.2014.09.054}},
  \href{http://www.arXiv.org/abs/1408.3226}{\texttt{arXiv:1408.3226}}.

\bibitem{CMS:2015zpx}
\hrefCMSnoop {}{{CMS Collaboration}, ``Measurement of differential and
  integrated fiducial cross sections for {Higgs} boson production in the
  four-lepton decay channel in pp collisions at {$ \sqrt{s}=7 $} and 8
  {TeV}'',} \textit{ JHEP} \textbf{ 04} (2016) 005,
  \href{http://dx.doi.org/10.1007/JHEP04(2016)005}{\doi{10.1007/JHEP04(2016)005}},
  \href{http://www.arXiv.org/abs/1512.08377}{\texttt{arXiv:1512.08377}}.

\bibitem{ATLAS:2016vlf}
\hrefCMSnoop {}{{ATLAS Collaboration}, ``Measurement of fiducial differential
  cross sections of gluon-fusion production of {Higgs} bosons decaying to
  {$\PW\PW^{\ast}\rightarrow\Pe\Pgn\Pgm\Pgn$} with the {ATLAS} detector at
  {$\sqrt{s}=8$} {TeV}'',} \textit{ JHEP} \textbf{ 08} (2016) 104,
  \href{http://dx.doi.org/10.1007/JHEP08(2016)104}{\doi{10.1007/JHEP08(2016)104}},
  \href{http://www.arXiv.org/abs/1604.02997}{\texttt{arXiv:1604.02997}}.

\bibitem{CMS:2016ipg}
\hrefCMSnoop {}{{CMS Collaboration}, ``Measurement of the transverse momentum
  spectrum of the {Higgs} boson produced in pp collisions at {$ \sqrt{s}=8 $}
  {TeV} using {$\PH \to \PW\PW$} decays'',} \textit{ JHEP} \textbf{ 03} (2017)
  032,
  \href{http://dx.doi.org/10.1007/JHEP03(2017)032}{\doi{10.1007/JHEP03(2017)032}},
  \href{http://www.arXiv.org/abs/1606.01522}{\texttt{arXiv:1606.01522}}.

\bibitem{ATLAS:2017qey}
\hrefCMSnoop {}{{ATLAS Collaboration}, ``Measurement of inclusive and
  differential cross sections in the {$\PH \rightarrow \PZ\PZ^{\ast}
  \rightarrow 4\ell$} decay channel in {$\Pp\Pp$} collisions at {$\sqrt{s}=13$}
  {TeV} with the {ATLAS} detector'',} \textit{ JHEP} \textbf{ 10} (2017) 132,
  \href{http://dx.doi.org/10.1007/JHEP10(2017)132}{\doi{10.1007/JHEP10(2017)132}},
  \href{http://www.arXiv.org/abs/1708.02810}{\texttt{arXiv:1708.02810}}.

\bibitem{CMS:2017dib}
\hrefCMSnoop {}{{CMS Collaboration}, ``Measurements of properties of the
  {Higgs} boson decaying into the four-lepton final state in pp collisions at
  {$ \sqrt{s}=13 $} {TeV}'',} \textit{ JHEP} \textbf{ 11} (2017) 047,
  \href{http://dx.doi.org/10.1007/JHEP11(2017)047}{\doi{10.1007/JHEP11(2017)047}},
  \href{http://www.arXiv.org/abs/1706.09936}{\texttt{arXiv:1706.09936}}.

\bibitem{CMS:2018piu}
\hrefCMSnoop {}{{CMS Collaboration}, ``Measurements of {Higgs} boson properties
  in the diphoton decay channel in proton-proton collisions at {$\sqrt{s} =$}
  13 {TeV}'',} \textit{ JHEP} \textbf{ 11} (2018) 185,
  \href{http://dx.doi.org/10.1007/JHEP11(2018)185}{\doi{10.1007/JHEP11(2018)185}},
  \href{http://www.arXiv.org/abs/1804.02716}{\texttt{arXiv:1804.02716}}.

\bibitem{ATLAS:2018pgp}
\hrefCMSnoop {}{{ATLAS Collaboration}, ``Combined measurement of differential
  and total cross sections in the {$\PH \rightarrow \gamma \gamma$} and the
  {$\PH \rightarrow \PZ\PZ^{\ast} \rightarrow 4\ell$} decay channels at
  {$\sqrt{s} = 13$} {TeV} with the {ATLAS} detector'',} \textit{ Phys. Lett. B}
  \textbf{ 786} (2018) 114,
  \href{http://dx.doi.org/10.1016/j.physletb.2018.09.019}{\doi{10.1016/j.physletb.2018.09.019}},
  \href{http://www.arXiv.org/abs/1805.10197}{\texttt{arXiv:1805.10197}}.

\bibitem{deFlorian:2016spz}
\hrefCMSnoop {}{{LHC Higgs Cross Section Working Group}, ``Handbook of {LHC}
  {H}iggs cross sections: 4. {D}eciphering the nature of the {H}iggs sector'',}
  CERN Report CERN-2017-002-M, 2016.
\newblock
  \href{http://dx.doi.org/10.23731/CYRM-2017-002}{\doi{10.23731/CYRM-2017-002}},
  \href{http://www.arXiv.org/abs/1610.07922}{\texttt{arXiv:1610.07922}}.

\bibitem{ATLAS:2020rej}
\hrefCMSnoop {}{{ATLAS Collaboration}, ``{Higgs} boson production cross-section
  measurements and their {EFT} interpretation in the {$4\ell $} decay channel
  at {$\sqrt{s}=$}13 {TeV} with the {ATLAS} detector'',} \textit{ Eur. Phys. J.
  C} \textbf{ 80} (2020) 957,
  \href{http://dx.doi.org/10.1140/epjc/s10052-020-8227-9}{\doi{10.1140/epjc/s10052-020-8227-9}},
  \href{http://www.arXiv.org/abs/2004.03447}{\texttt{arXiv:2004.03447}}.
  [Errata: \DOI{10.1140/epjc/s10052-020-08644-x},
  \DOI{10.1140/epjc/s10052-021-09116-6}].

\bibitem{ATLAS:2020fcp}
\hrefCMSnoop {}{{ATLAS Collaboration}, ``Measurements of {$\PW\PH$} and
  {$\PZ\PH$} production in the {$\PH \rightarrow \PQb\PAQb$} decay channel in
  {$\Pp\Pp$} collisions at 13 {TeV} with the {ATLAS} detector'',} \textit{ Eur.
  Phys. J. C} \textbf{ 81} (2021) 178,
  \href{http://dx.doi.org/10.1140/epjc/s10052-020-08677-2}{\doi{10.1140/epjc/s10052-020-08677-2}},
  \href{http://www.arXiv.org/abs/2007.02873}{\texttt{arXiv:2007.02873}}.

\bibitem{ATLAS:2022yrq}
\hrefCMSnoop {}{{ATLAS Collaboration}, ``Measurements of {Higgs} boson
  production cross-sections in the~{$\PH\to\PGtp\PGtm$} decay channel in
  {$\Pp\Pp$} collisions at $\sqrt{s} $ = 13 {TeV} with the {ATLAS} detector'',}
  \textit{ JHEP} \textbf{ 08} (2022) 175,
  \href{http://dx.doi.org/10.1007/JHEP08(2022)175}{\doi{10.1007/JHEP08(2022)175}},
  \href{http://www.arXiv.org/abs/2201.08269}{\texttt{arXiv:2201.08269}}.

\bibitem{ATLAS:2022ooq}
\hrefCMSnoop {}{{ATLAS Collaboration}, ``Measurements of {Higgs} boson
  production by gluon$-$gluon fusion and vector-boson fusion using {$\PH\to
  \PW\PW^* \to \Pe\nu\mu\nu$} decays in {$\Pp\Pp$} collisions at
  {$\sqrt{s}=13\TeV$} with the atlas detector'',} 2022.
  \href{http://www.arXiv.org/abs/2207.00338}{\texttt{arXiv:2207.00338}}.
  Submitted to \textit{Phys. Rev.~D}.

\bibitem{ATLAS:2022vkf}
\hrefCMSnoop {}{{ATLAS Collaboration}, ``A detailed map of {Higgs} boson
  interactions by the {ATLAS} experiment ten years after the discovery'',}
  \textit{ Nature} \textbf{ 607} (2022) 52,
  \href{http://dx.doi.org/10.1038/s41586-022-04893-w}{\doi{10.1038/s41586-022-04893-w}},
  \href{http://www.arXiv.org/abs/2207.00092}{\texttt{arXiv:2207.00092}}.

\bibitem{ATLAS:2022tnm}
\hrefCMSnoop {}{{ATLAS Collaboration}, ``Measurement of the properties of
  {Higgs} boson production at {$\sqrt{s} = 13\TeV$} in the
  {$\PH\to\gamma\gamma$} channel using {$139\fbinv$} of {$\Pp\Pp$} collision
  data with the {ATLAS} experiment'',} 2022.
  \href{http://www.arXiv.org/abs/2207.00348}{\texttt{arXiv:2207.00348}}.
  Submitted to \textit{JHEP}.

\bibitem{CMS:2021ugl}
\hrefCMSnoop {}{{CMS Collaboration}, ``Measurements of production cross
  sections of the {Higgs} boson in the four-lepton final state in
  proton\textendash{}proton collisions at {$\sqrt{s} = 13\,\text {Te}\text {V}
  $}'',} \textit{ Eur. Phys. J. C} \textbf{ 81} (2021) 488,
  \href{http://dx.doi.org/10.1140/epjc/s10052-021-09200-x}{\doi{10.1140/epjc/s10052-021-09200-x}},
  \href{http://www.arXiv.org/abs/2103.04956}{\texttt{arXiv:2103.04956}}.

\bibitem{HIG-19-015}
\hrefCMSnoop {}{{CMS Collaboration}, ``Measurements of {Higgs} boson production
  cross sections and couplings in the diphoton decay channel at {$
  \sqrt{\mathrm{s}} $} = 13 {TeV}'',} \textit{ JHEP} \textbf{ 07} (2021) 027,
  \href{http://dx.doi.org/10.1007/JHEP07(2021)027}{\doi{10.1007/JHEP07(2021)027}},
  \href{http://www.arXiv.org/abs/2103.06956}{\texttt{arXiv:2103.06956}}.

\bibitem{ATLAS:2020wny}
\hrefCMSnoop {}{{ATLAS Collaboration}, ``Measurements of the {Higgs} boson
  inclusive and differential fiducial cross sections in the 4{$\ell$} decay
  channel at {$\sqrt{s}$} = 13 {TeV}'',} \textit{ Eur. Phys. J. C} \textbf{ 80}
  (2020) 942,
  \href{http://dx.doi.org/10.1140/epjc/s10052-020-8223-0}{\doi{10.1140/epjc/s10052-020-8223-0}},
  \href{http://www.arXiv.org/abs/2004.03969}{\texttt{arXiv:2004.03969}}.

\bibitem{ATLASHggDiff}
\hrefCMSnoop {}{{ATLAS Collaboration}, ``Measurements of the {Higgs} boson
  inclusive and differential fiducial cross-sections in the diphoton decay
  channel with {$\Pp\Pp$} collisions at {$\sqrt{s} = 13$} {TeV} with the
  {ATLAS} detector'',} 2022.
  \href{http://www.arXiv.org/abs/2202.00487}{\texttt{arXiv:2202.00487}}.
  Submitted to JHEP.

\bibitem{CMS:2020dvg}
\hrefCMSnoop {}{{CMS Collaboration}, ``Measurement of the inclusive and
  differential {Higgs} boson production cross sections in the leptonic {WW}
  decay mode at {$\sqrt{s} =$} 13 {TeV}'',} \textit{ JHEP} \textbf{ 03} (2021)
  003,
  \href{http://dx.doi.org/10.1007/JHEP03(2021)003}{\doi{10.1007/JHEP03(2021)003}},
  \href{http://www.arXiv.org/abs/2007.01984}{\texttt{arXiv:2007.01984}}.

\bibitem{HIG-20-015}
\hrefCMSnoop {}{{CMS Collaboration}, ``Measurement of the inclusive and
  differential {Higgs} boson production cross sections in the decay mode to a
  pair of {\ensuremath{\tau}} leptons in pp collisions at {$\sqrt{s}=13$}
  {TeV}'',} \textit{ Phys. Rev. Lett.} \textbf{ 128} (2022) 081805,
  \href{http://dx.doi.org/10.1103/PhysRevLett.128.081805}{\doi{10.1103/PhysRevLett.128.081805}},
  \href{http://www.arXiv.org/abs/2107.11486}{\texttt{arXiv:2107.11486}}.

\bibitem{HIG-17-025}
\hrefCMSnoop {}{{CMS Collaboration}, ``Measurement of inclusive and
  differential {Higgs} boson production cross sections in the diphoton decay
  channel in proton-proton collisions at {$\sqrt{s}=$} 13 {TeV}'',} \textit{
  JHEP} \textbf{ 01} (2019) 183,
  \href{http://dx.doi.org/10.1007/JHEP01(2019)183}{\doi{10.1007/JHEP01(2019)183}},
  \href{http://www.arXiv.org/abs/1807.03825}{\texttt{arXiv:1807.03825}}.

\bibitem{hepdata}
\hrefCMSnoop {}{}{HEPD}ata record for this analysis, 2022.
\newblock
  \href{http://dx.doi.org/10.17182/hepdata.132906}{\doi{10.17182/hepdata.132906}}.

\bibitem{CMSTRG}
\hrefCMSnoop {}{{CMS Collaboration}, ``The {CMS} trigger system'',} \textit{
  JINST} \textbf{ 12} (2017) P01020,
  \href{http://dx.doi.org/10.1088/1748-0221/12/01/P01020}{\doi{10.1088/1748-0221/12/01/P01020}},
  \href{http://www.arXiv.org/abs/1609.02366}{\texttt{arXiv:1609.02366}}.

\bibitem{Chatrchyan:2008zzk}
\hrefCMSnoop {}{{CMS Collaboration}, ``The {CMS} experiment at the {CERN}
  {LHC}'',} \textit{ JINST} \textbf{ 3} (2008) S08004,
  \href{http://dx.doi.org/10.1088/1748-0221/3/08/S08004}{\doi{10.1088/1748-0221/3/08/S08004}}.

\bibitem{CMSlumi2016}
\hrefCMSnoop {}{{CMS Collaboration}, ``Precision luminosity measurement in
  proton-proton collisions at {$\sqrt{s} =$} 13 {TeV} in 2015 and 2016 at
  {CMS}'',} \textit{ Eur. Phys. J. C} \textbf{ 81} (2021) 800,
  \href{http://dx.doi.org/10.1140/epjc/s10052-021-09538-2}{\doi{10.1140/epjc/s10052-021-09538-2}},
  \href{http://www.arXiv.org/abs/2104.01927}{\texttt{arXiv:2104.01927}}.

\bibitem{CMSlumi2017}
\href {https://cds.cern.ch/record/2621960}{{CMS Collaboration}, ``{CMS}
  luminosity measurement for the 2017 data-taking period at {$\sqrt{s} =
  13~\mathrm{TeV}$}'',} {CMS Physics Analysis Summary} CMS-PAS-LUM-17-004,
  2018.

\bibitem{CMSlumi2018}
\href {https://cds.cern.ch/record/2676164}{{CMS Collaboration}, ``{CMS}
  luminosity measurement for the 2018 data-taking period at {$\sqrt{s} =
  13~\mathrm{TeV}$}'',} {CMS Physics Analysis Summary} CMS-PAS-LUM-18-002,
  2019.

\bibitem{TAG-PROBE}
\hrefCMSnoop {}{{CMS Collaboration}, ``Measurements of inclusive {$\PW$} and
  {$\PZ$} cross sections in {$\Pp\Pp$} collisions at $\sqrt{s}=7$ {TeV}'',}
  \textit{ JHEP} \textbf{ 01} (2011) 080,
  \href{http://dx.doi.org/10.1007/JHEP01(2011)080}{\doi{10.1007/JHEP01(2011)080}},
  \href{http://www.arXiv.org/abs/1012.2466}{\texttt{arXiv:1012.2466}}.

\bibitem{EGMpaper}
\hrefCMSnoop {}{{CMS Collaboration}, ``Electron and photon reconstruction and
  identification with the {CMS} experiment at the {CERN} {LHC}'',} \textit{
  JINST} \textbf{ 16} (2021) P05014,
  \href{http://dx.doi.org/10.1088/1748-0221/16/05/P05014}{\doi{10.1088/1748-0221/16/05/P05014}},
  \href{http://www.arXiv.org/abs/2012.06888}{\texttt{arXiv:2012.06888}}.

\bibitem{AMCAT}
J.~Alwall\hrefCMSnoop {}{ {et~al.}, ``The automated computation of tree-level
  and next-to-leading order differential cross sections, and their matching to
  parton shower simulations'',} \textit{ JHEP} \textbf{ 07} (2014) 079,
  \href{http://dx.doi.org/10.1007/JHEP07(2014)079}{\doi{10.1007/JHEP07(2014)079}},
  \href{http://www.arXiv.org/abs/1405.0301}{\texttt{arXiv:1405.0301}}.

\bibitem{NNLOPS1}
\hrefCMSnoop {}{K.~Hamilton, P.~Nason, E.~Re, and G.~Zanderighi, ``{NNLOPS}
  simulation of {Higgs} boson production'',} \textit{ JHEP} \textbf{ 10} (2013)
  222,
  \href{http://dx.doi.org/10.1007/JHEP10(2013)222}{\doi{10.1007/JHEP10(2013)222}},
  \href{http://www.arXiv.org/abs/1309.0017}{\texttt{arXiv:1309.0017}}.

\bibitem{NNLOPS2}
\hrefCMSnoop {}{K.~Hamilton, P.~Nason, and G.~Zanderighi, ``{MINLO}:
  Multi-scale improved {NLO}'',} \textit{ JHEP} \textbf{ 10} (2012) 155,
  \href{http://dx.doi.org/10.1007/JHEP10(2012)155}{\doi{10.1007/JHEP10(2012)155}},
  \href{http://www.arXiv.org/abs/1206.3572}{\texttt{arXiv:1206.3572}}.

\bibitem{NNLOPS3}
\hrefCMSnoop {}{A.~Kardos, P.~Nason, and C.~Oleari, ``Three-jet production in
  {POWHEG}'',} \textit{ JHEP} \textbf{ 04} (2014) 043,
  \href{http://dx.doi.org/10.1007/JHEP04(2014)043}{\doi{10.1007/JHEP04(2014)043}},
  \href{http://www.arXiv.org/abs/1402.4001}{\texttt{arXiv:1402.4001}}.

\bibitem{Pythia82}
T.~Sj{\"o}strand\hrefCMSnoop {}{ {et~al.}, ``An introduction to {PYTHIA}
  {8.2}'',} \textit{ Comput. Phys. Commun.} \textbf{ 191} (2015) 159,
  \href{http://dx.doi.org/10.1016/j.cpc.2015.01.024}{\doi{10.1016/j.cpc.2015.01.024}},
  \href{http://www.arXiv.org/abs/1410.3012}{\texttt{arXiv:1410.3012}}.

\bibitem{CUETP8}
\hrefCMSnoop {}{{CMS Collaboration}, ``Event generator tunes obtained from
  underlying event and multiparton scattering measurements'',} \textit{ Eur.
  Phys. J. C} \textbf{ 76} (2016) 155,
  \href{http://dx.doi.org/10.1140/epjc/s10052-016-3988-x}{\doi{10.1140/epjc/s10052-016-3988-x}},
  \href{http://www.arXiv.org/abs/1512.00815}{\texttt{arXiv:1512.00815}}.

\bibitem{CP5}
\hrefCMSnoop {}{{CMS Collaboration}, ``Extraction and validation of a new set
  of {CMS} {PYTHIA8} tunes from underlying-event measurements'',} \textit{ Eur.
  Phys. J. C} \textbf{ 80} (2020) 4,
  \href{http://dx.doi.org/10.1140/epjc/s10052-019-7499-4}{\doi{10.1140/epjc/s10052-019-7499-4}},
  \href{http://www.arXiv.org/abs/1903.12179}{\texttt{arXiv:1903.12179}}.

\bibitem{SHERPA22}
\hrefCMSnoop {}{{Sherpa} Collaboration, ``Event generation with {Sherpa}
  {2.2}'',} \textit{ SciPost Phys.} \textbf{ 7} (2019) 034,
  \href{http://dx.doi.org/10.21468/SciPostPhys.7.3.034}{\doi{10.21468/SciPostPhys.7.3.034}},
  \href{http://www.arXiv.org/abs/1905.09127}{\texttt{arXiv:1905.09127}}.

\bibitem{ParticleFlow}
\hrefCMSnoop {}{{CMS Collaboration}, ``Particle-flow reconstruction and global
  event description with the {CMS} detector'',} \textit{ JINST} \textbf{ 12}
  (2017) P10003,
  \href{http://dx.doi.org/10.1088/1748-0221/12/10/P10003}{\doi{10.1088/1748-0221/12/10/P10003}},
  \href{http://www.arXiv.org/abs/1706.04965}{\texttt{arXiv:1706.04965}}.

\bibitem{HggMass}
\hrefCMSnoop {}{{CMS Collaboration}, ``A measurement of the {Higgs} boson mass
  in the diphoton decay channel'',} \textit{ Phys. Lett. B} \textbf{ 805}
  (2020) 135425,
  \href{http://dx.doi.org/10.1016/j.physletb.2020.135425}{\doi{10.1016/j.physletb.2020.135425}},
  \href{http://www.arXiv.org/abs/2002.06398}{\texttt{arXiv:2002.06398}}.

\bibitem{HIG-19-013}
\hrefCMSnoop {}{{CMS Collaboration}, ``Measurements of {$\mathrm{t\bar{t}}\PH$}
  production and the {CP} structure of the {Yukawa} interaction between the
  {Higgs} boson and top quark in the diphoton decay channel'',} \textit{ Phys.
  Rev. Lett.} \textbf{ 125} (2020) 061801,
  \href{http://dx.doi.org/10.1103/PhysRevLett.125.061801}{\doi{10.1103/PhysRevLett.125.061801}},
  \href{http://www.arXiv.org/abs/2003.10866}{\texttt{arXiv:2003.10866}}.

\bibitem{HIG-19-018}
\hrefCMSnoop {}{{CMS Collaboration}, ``Search for nonresonant {Higgs} boson
  pair production in final states with two bottom quarks and two photons in
  proton-proton collisions at {$ \sqrt{s} $} = 13 {TeV}'',} \textit{ JHEP}
  \textbf{ 03} (2021) 257,
  \href{http://dx.doi.org/10.1007/JHEP03(2021)257}{\doi{10.1007/JHEP03(2021)257}},
  \href{http://www.arXiv.org/abs/2011.12373}{\texttt{arXiv:2011.12373}}.

\bibitem{MTR}
\hrefCMSnoop {}{E.~Spyromitros-Xioufis, G.~Tsoumakas, W.~Groves, and
  I.~Vlahavas, ``Multi-target regression via input space expansion: treating
  targets as inputs'',} \textit{ Mach. Learn.} \textbf{ 104} (2016) 55,
  \href{http://dx.doi.org/10.1007/s10994-016-5546-z}{\doi{10.1007/s10994-016-5546-z}},
  \href{http://www.arXiv.org/abs/1211.6581}{\texttt{arXiv:1211.6581}}.

\bibitem{quantileReg}
\hrefCMSnoop {}{R.~Koenker and K.~F. Hallock, ``Quantile regression'',}
  \textit{ J. Econ. Perspect.} \textbf{ 15} (2001) 143,
  \href{http://dx.doi.org/10.1257/jep.15.4.143}{\doi{10.1257/jep.15.4.143}}.

\bibitem{scikit-learn}
\href {http://jmlr.org/papers/v12/pedregosa11a.html}{F.~Pedregosa {et~al.},
  ``{Scikit-learn}: Machine learning in {Python}'',} \textit{ J. Mach. Learn.
  Res.} \textbf{ 12} (2011) 2825,
  \href{http://www.arXiv.org/abs/1201.0490}{\texttt{arXiv:1201.0490}}.

\bibitem{xgboost}
\hrefCMSnoop {}{T.~Chen and C.~Guestrin, ``{XGBoost}: {A} scalable tree
  boosting system'',} in \textit{ Proc. 22nd ACM SIGKDD Int. Conf. Know.
  Discov. Data Min.}, p.~785.
\newblock 2016.
\newblock
  \href{http://dx.doi.org/10.1145/2939672.2939785}{\doi{10.1145/2939672.2939785}}.

\bibitem{AntiKt}
\hrefCMSnoop {}{M.~Cacciari, G.~P. Salam, and G.~Soyez, ``The {anti-\kt} jet
  clustering algorithm'',} \textit{ JHEP} \textbf{ 04} (2008) 063,
  \href{http://dx.doi.org/10.1088/1126-6708/2008/04/063}{\doi{10.1088/1126-6708/2008/04/063}},
  \href{http://www.arXiv.org/abs/0802.1189}{\texttt{arXiv:0802.1189}}.

\bibitem{fastjet}
\hrefCMSnoop {}{M.~Cacciari, G.~P. Salam, and G.~Soyez, ``{FastJet} user
  manual'',} \textit{ Eur. Phys. J. C} \textbf{ 72} (2012) 1896,
  \href{http://dx.doi.org/10.1140/epjc/s10052-012-1896-2}{\doi{10.1140/epjc/s10052-012-1896-2}},
  \href{http://www.arXiv.org/abs/1111.6097}{\texttt{arXiv:1111.6097}}.

\bibitem{DeepJet}
E.~Bols\hrefCMSnoop {}{ {et~al.}, ``Jet flavour classification using
  {DeepJet}'',} \textit{ JINST} \textbf{ 15} (2020) P12012,
  \href{http://dx.doi.org/10.1088/1748-0221/15/12/P12012}{\doi{10.1088/1748-0221/15/12/P12012}},
  \href{http://www.arXiv.org/abs/2008.10519}{\texttt{arXiv:2008.10519}}.

\bibitem{METperformance}
\hrefCMSnoop {}{{CMS Collaboration}, ``Performance of missing transverse
  momentum reconstruction in proton-proton collisions at {$\sqrt{s} =$} 13
  {TeV} using the {CMS} detector'',} \textit{ JINST} \textbf{ 14} (2019)
  P07004,
  \href{http://dx.doi.org/10.1088/1748-0221/14/07/P07004}{\doi{10.1088/1748-0221/14/07/P07004}},
  \href{http://www.arXiv.org/abs/1903.06078}{\texttt{arXiv:1903.06078}}.

\bibitem{CMS:2018rym}
\hrefCMSnoop {}{{CMS Collaboration}, ``Performance of the {CMS} muon detector
  and muon reconstruction with proton-proton collisions at {$\sqrt{s}=$} 13
  {TeV}'',} \textit{ JINST} \textbf{ 13} (2018) P06015,
  \href{http://dx.doi.org/10.1088/1748-0221/13/06/P06015}{\doi{10.1088/1748-0221/13/06/P06015}},
  \href{http://www.arXiv.org/abs/1804.04528}{\texttt{arXiv:1804.04528}}.

\bibitem{Collins:1977}
\hrefCMSnoop {}{J.~C. Collins and D.~E. Soper, ``Angular distribution of
  dileptons in high-energy hadron collisions'',} \textit{ Phys. Rev. D}
  \textbf{ 16} (1977) 2219,
  \href{http://dx.doi.org/10.1103/PhysRevD.16.2219}{\doi{10.1103/PhysRevD.16.2219}}.

\bibitem{Banfi:2010cf}
A.~Banfi\hrefCMSnoop {}{ {et~al.}, ``Optimisation of variables for studying
  dilepton transverse momentum distributions at hadron colliders'',} \textit{
  Eur. Phys. J. C} \textbf{ 71} (2011) 1600,
  \href{http://dx.doi.org/10.1140/epjc/s10052-011-1600-y}{\doi{10.1140/epjc/s10052-011-1600-y}},
  \href{http://www.arXiv.org/abs/1009.1580}{\texttt{arXiv:1009.1580}}.

\bibitem{boggia2017higgstools}
\hrefCMSnoop {}{M.~Boggia {et~al.}, ``The {HiggsTools} handbook: a beginners
  guide to decoding the {Higgs} sector'',} \textit{ J. Phys. G} \textbf{ 45}
  (2018) 065004,
  \href{http://dx.doi.org/10.1088/1361-6471/aab812}{\doi{10.1088/1361-6471/aab812}},
  \href{http://www.arXiv.org/abs/1711.09875}{\texttt{arXiv:1711.09875}}.

\bibitem{Gangal:2015}
\hrefCMSnoop {}{S.~Gangal, M.~Stahlhofen, and F.~J. Tackmann,
  ``Rapidity-dependent jet vetoes'',} \textit{ Phys. Rev. D} \textbf{ 91}
  (2015) 054023,
  \href{http://dx.doi.org/10.1103/PhysRevD.91.054023}{\doi{10.1103/PhysRevD.91.054023}},
  \href{http://www.arXiv.org/abs/1412.4792}{\texttt{arXiv:1412.4792}}.

\bibitem{Rainwater:1996}
\hrefCMSnoop {}{D.~L. Rainwater, R.~Szalapski, and D.~Zeppenfeld, ``Probing
  color singlet exchange in {$Z$} + two jet events at the {CERN LHC}'',}
  \textit{ Phys. Rev. D} \textbf{ 54} (1996) 6680,
  \href{http://dx.doi.org/10.1103/PhysRevD.54.6680}{\doi{10.1103/PhysRevD.54.6680}},
  \href{http://www.arXiv.org/abs/hep-ph/9605444}{\texttt{arXiv:hep-ph/9605444}}.

\bibitem{DiscreteProfilingMethod}
\hrefCMSnoop {}{P.~D. Dauncey, M.~Kenzie, N.~Wardle, and G.~J. Davies,
  ``Handling uncertainties in background shapes: the discrete profiling
  method'',} \textit{ JINST} \textbf{ 10} (2015) P04015,
  \href{http://dx.doi.org/10.1088/1748-0221/10/04/P04015}{\doi{10.1088/1748-0221/10/04/P04015}},
  \href{http://www.arXiv.org/abs/1408.6865}{\texttt{arXiv:1408.6865}}.

\bibitem{Asimov}
\hrefCMSnoop {}{G.~Cowan, K.~Cranmer, E.~Gross, and O.~Vitells, ``Asymptotic
  formulae for likelihood-based tests of new physics'',} \textit{ Eur. Phys. J.
  C} \textbf{ 71} (2011) 1554,
  \href{http://dx.doi.org/10.1140/epjc/s10052-011-1554-0}{\doi{10.1140/epjc/s10052-011-1554-0}},
  \href{http://www.arXiv.org/abs/1007.1727}{\texttt{arXiv:1007.1727}}.
[Erratum: \DOI{10.1140/epjc/s10052-013-2501-z}].

\bibitem{Butterworth:2015oua}
\hrefCMSnoop {}{J.~Butterworth {et~al.}, ``{PDF4LHC} recommendations for {LHC
  Run II}'',} \textit{ J. Phys. G} \textbf{ 43} (2016) 023001,
  \href{http://dx.doi.org/10.1088/0954-3899/43/2/023001}{\doi{10.1088/0954-3899/43/2/023001}},
  \href{http://www.arXiv.org/abs/1510.03865}{\texttt{arXiv:1510.03865}}.

\bibitem{NNPDF31}
\hrefCMSnoop {}{{NNPDF} Collaboration, ``Parton distributions from
  high-precision collider data'',} \textit{ Eur. Phys. J. C} \textbf{ 77}
  (2017) 663,
  \href{http://dx.doi.org/10.1140/epjc/s10052-017-5199-5}{\doi{10.1140/epjc/s10052-017-5199-5}},
  \href{http://www.arXiv.org/abs/1706.00428}{\texttt{arXiv:1706.00428}}.

\bibitem{NNPDF3}
\hrefCMSnoop {}{{NNPDF} Collaboration, ``Parton distributions for the {LHC} run
  {II}'',} \textit{ JHEP} \textbf{ 04} (2015) 040,
  \href{http://dx.doi.org/10.1007/JHEP04(2015)040}{\doi{10.1007/JHEP04(2015)040}},
  \href{http://www.arXiv.org/abs/1410.8849}{\texttt{arXiv:1410.8849}}.

\bibitem{MC2Hessian}
S.~Carrazza\hrefCMSnoop {}{ {et~al.}, ``An unbiased {Hessian} representation
  for {Monte Carlo} {PDFs}'',} \textit{ Eur. Phys. J. C} \textbf{ 75} (2015)
  369,
  \href{http://dx.doi.org/10.1140/epjc/s10052-015-3590-7}{\doi{10.1140/epjc/s10052-015-3590-7}},
\href{http://www.arXiv.org/abs/1505.06736}{\texttt{arXiv:1505.06736}}.

\bibitem{CMS:2014afl}
\hrefCMSnoop {}{{CMS Collaboration}, ``Observation of the diphoton decay of the
  {Higgs} boson and measurement of its properties'',} \textit{ Eur. Phys. J. C}
  \textbf{ 74} (2014) 3076,
  \href{http://dx.doi.org/10.1140/epjc/s10052-014-3076-z}{\doi{10.1140/epjc/s10052-014-3076-z}},
  \href{http://www.arXiv.org/abs/1407.0558}{\texttt{arXiv:1407.0558}}.

\bibitem{JetsInRun2}
\href {https://cds.cern.ch/record/2256875}{{CMS Collaboration}, ``Jet
  algorithms performance in {13} {TeV} data'',} {CMS Physics Analysis Summary}
  CMS-PAS-JME-16-003, 2017.

\bibitem{IntHggDipho}
\hrefCMSnoop {}{J.~Campbell, M.~Carena, R.~Harnik, and Z.~Liu, ``Interference
  in the {$\Pg\Pg\rightarrow \Ph \rightarrow \gamma\gamma$} on-shell rate and
  the {Higgs} boson total width'',} \textit{ Phys. Rev. Lett.} \textbf{ 119}
  (2017) 181801,
  \href{http://dx.doi.org/10.1103/PhysRevLett.119.181801}{\doi{10.1103/PhysRevLett.119.181801}},
  \href{http://www.arXiv.org/abs/1704.08259}{\texttt{arXiv:1704.08259}}.
  [Addendum: \DOI{10.1103/PhysRevLett.119.199901}].

\bibitem{POWHEG1}
\hrefCMSnoop {}{P.~Nason, ``A new method for combining {NLO} {QCD} with shower
  monte carlo algorithms'',} \textit{ JHEP} \textbf{ 11} (2004) 040,
  \href{http://dx.doi.org/10.1088/1126-6708/2004/11/040}{\doi{10.1088/1126-6708/2004/11/040}},
  \href{http://www.arXiv.org/abs/hep-ph/0409146}{\texttt{arXiv:hep-ph/0409146}}.

\bibitem{POWHEG2}
\hrefCMSnoop {}{S.~Frixione, P.~Nason, and C.~Oleari, ``Matching {NLO} {QCD}
  computations with parton shower simulations: the {\sc powheg} method'',}
  \textit{ JHEP} \textbf{ 11} (2007) 070,
  \href{http://dx.doi.org/10.1088/1126-6708/2007/11/070}{\doi{10.1088/1126-6708/2007/11/070}},
  \href{http://www.arXiv.org/abs/0709.2092}{\texttt{arXiv:0709.2092}}.

\bibitem{POWHEG3}
\hrefCMSnoop {}{S.~Alioli, P.~Nason, C.~Oleari, and E.~Re, ``A general
  framework for implementing {NLO} calculations in shower {Monte} {Carlo}
  programs: the {\sc powheg box}'',} \textit{ JHEP} \textbf{ 06} (2010) 043,
  \href{http://dx.doi.org/10.1007/JHEP06(2010)043}{\doi{10.1007/JHEP06(2010)043}},
  \href{http://www.arXiv.org/abs/1002.2581}{\texttt{arXiv:1002.2581}}.

\bibitem{Powheg:ggH}
\hrefCMSnoop {}{E.~Bagnaschi, G.~Degrassi, P.~Slavich, and A.~Vicini, ``{Higgs}
  production via gluon fusion in the {POWHEG} approach in the {SM} and in the
  {MSSM}'',} \textit{ JHEP} \textbf{ 02} (2012) 088,
  \href{http://dx.doi.org/10.1007/JHEP02(2012)088}{\doi{10.1007/JHEP02(2012)088}},
  \href{http://www.arXiv.org/abs/1111.2854}{\texttt{arXiv:1111.2854}}.

\end{thebibliography}\endgroup
\cleardoublepage \appendix\section{The CMS Collaboration \label{app:collab}}\begin{sloppypar}\hyphenpenalty=5000\widowpenalty=500\clubpenalty=5000
\cmsinstitute{Yerevan Physics Institute, Yerevan, Armenia}
{\tolerance=6000
A.~Tumasyan\cmsAuthorMark{1}\cmsorcid{0009-0000-0684-6742}
\par}
\cmsinstitute{Institut f\"{u}r Hochenergiephysik, Vienna, Austria}
{\tolerance=6000
W.~Adam\cmsorcid{0000-0001-9099-4341}, J.W.~Andrejkovic, T.~Bergauer\cmsorcid{0000-0002-5786-0293}, S.~Chatterjee\cmsorcid{0000-0003-2660-0349}, K.~Damanakis\cmsorcid{0000-0001-5389-2872}, M.~Dragicevic\cmsorcid{0000-0003-1967-6783}, A.~Escalante~Del~Valle\cmsorcid{0000-0002-9702-6359}, P.S.~Hussain\cmsorcid{0000-0002-4825-5278}, M.~Jeitler\cmsAuthorMark{2}\cmsorcid{0000-0002-5141-9560}, N.~Krammer\cmsorcid{0000-0002-0548-0985}, L.~Lechner\cmsorcid{0000-0002-3065-1141}, D.~Liko\cmsorcid{0000-0002-3380-473X}, I.~Mikulec\cmsorcid{0000-0003-0385-2746}, P.~Paulitsch, F.M.~Pitters, J.~Schieck\cmsAuthorMark{2}\cmsorcid{0000-0002-1058-8093}, R.~Sch\"{o}fbeck\cmsorcid{0000-0002-2332-8784}, D.~Schwarz\cmsorcid{0000-0002-3821-7331}, S.~Templ\cmsorcid{0000-0003-3137-5692}, W.~Waltenberger\cmsorcid{0000-0002-6215-7228}, C.-E.~Wulz\cmsAuthorMark{2}\cmsorcid{0000-0001-9226-5812}
% ...
\par}
\vskip\cmsinstskip
\dag:~Deceased\\
$^{1}$Also at Yerevan State University, Yerevan, Armenia\\
$^{2}$Also at TU Wien, Vienna, Austria\\
% ...
\end{sloppypar}
\end{document}